\documentclass{aa}
\usepackage{txfonts}
\usepackage{graphicx}
\usepackage{natbib}
\usepackage{longtable}
\usepackage{subeqnarray}
\usepackage{cases}
\usepackage{ulem}

\usepackage[colorlinks=true,citecolor=blue]{hyperref}

\begin{document}
\title{A 1.3\,cm spectral line study of the W33 region}
\author{Kadirya Tursun \inst{1,2}
\and Christian Henkel \inst{3,1}
\and Jarken Esimbek \inst{1,2}
\and Gang Wu \inst{1,2}
\and Dalei Li \inst{1,2}
\and Xindi Tang \inst{1,2}
\and Jianjun Zhou \inst{1,2}
\and T. L. Wilson \inst{3}
\and Friedrich Wyrowski \inst{3}
\and Rainer Mauersberger \inst{3}
\and Katharina Immer \inst{3,4}
\and B. Winkel \inst{3}
\and Yuxin He \inst{1,2}
\and Dongdong Zhou \inst{1,2}
\and Yingxiu Ma \inst{1,2}
\and Andrey M. Sobolev \inst{1,5}
\and Toktarkhan Komesh \inst{1,6,7}
\and Hailiang Shen \inst{1,2}
}
\institute{
State Key Laboratory of Radio Astronomy and Technology, Xinjiang Astronomical Observatory, CAS, 150 Science 1-Street, Urumqi, Xinjiang, 830011, P. R. China \\
e-mail: kadirya@xao.ac.cn, jarken@xao.ac.cn
\and
Xinjiang Key Laboratory of Radio Astrophysics, Urumqi 830011, PR China
\and
Max-Planck-Institut f\"ur Radioastronomie, Auf dem H\"ugel 69, D-53121 Bonn, Germany \\
e-mail: chenkel@mpifr-bonn.mpg.de
\and
Leiden Observatory, Leiden University, PO Box 9513, 2300 RA Leiden, The Netherlands
\and
Astronomical Observatory, Institute for Natural Sciences and Mathematics, Ural Federal University, 19 Mira street, Ekaterinburg, 620002, Russia
\and
Institute of Experimental and Theoretical Physics, Al-Farabi Kazakh National University, Almaty, 050040, Kazakhstan
\and
Energetic Cosmos Laboratory, Nazarbayev University, Astana 010000, Kazakhstan
}

\abstract
{At a distance of 2.4\,kpc, W33 is one of the most prolific sources of molecular line emission, and it is an excellent research target for a centimeter spectral line search. We carried out a 1.3\,cm spectral line survey in the frequency range 18--26\,GHz. The lines we identified include 44 radio recombination lines (RRLs) and 24 molecular lines, excluding transitions from the main isotopolog of NH$_3$. The RRLs are associated with the ionized gas from W33\,Main. Intensity ratios between RRL pairs with varying differences in the principal quantum number $n$\,(i.e., $\Delta n$) from the same element at adjacent frequencies agree with ratios expected under conditions of local thermodynamical equilibrium. In spite of a resulting helium-to-hydrogen abundance ratio (equal emitting volumes assumed) of (10.7$\pm$1.8)\%, which is consistent with expectations, helium shows broader turbulent line widths than hydrogen. The difference amounts to a few kilometers per second, hinting that the spatial distributions are slightly different. The molecular lines are attributed to nine different species (CH$_3$OH, HC$_3$N, SiS, c-C$_3$H$_2$, CH$_3$CN, NH$_2$D, HNCO, H$_2$O and CCS). Rotation temperatures and column densities were derived from CH$_3$OH transitions using rotational temperature diagram analysis. Maser emission produced by water vapor and methanol have been observed in W33\,Main, W33\,A, and W33\,B. Our survey discovered a CH$_{3}$OH\,(10$_{2,8}$--10$_{1,9}$\,E) maser in W33\,Main. Toward W33\,B1, the fractionated deuterium-to-hydrogen ratio (D/H) deduced from para-NH$_2$D/NH$_3$ is estimated to be $\lesssim$(1.0\,$\pm$\,0.2)$\times$10$^{-3}$. For the other molecular W33-hotspots, 3$\sigma$ upper limits are (5.0\,$\pm$\,0.4)$\times$10$^{-3}$. At linear scales of (0.5\,pc), fractional abundances and excitation temperatures do not reach values close to those in well-established hot cores, but higher-resolution measurements may alter this picture.}

\maketitle
\section{Introduction}
\label{sect:Introduction}
Interstellar and circumstellar gas condensations are widely found in various celestial environments, such as quiescent dark clouds, star-forming regions, ionized nebulae, stellar envelopes, pregalactic material, quasar absorption lines, and young supernova remnants. In these astronomical objects, the gas is characterized by a wide temperature range of about 10\,-\,7000\,K and a density range of about 10-10$^{11}$ cm$^{-3}$ (e.g., \citealt{1970A&A.....4..357R,1975ApJ...202...50D,1979A&A....71..205W,1986ApJ...300L..73K}). Considering this, it is clear that the transition lines of CO as probes do not cover the entire astrophysical complexity. Conditions required to excite different interstellar molecules or even different transitions of the same molecule are vary. Furthermore, many such clouds are associated with young massive stars, providing regions largely devoid of molecules but filled with ionized hydrogen and helium. Thus, only through the observations of multiline spectra can comprehensive information about the physical and chemical properties of the studied gas be obtained. Therefore, centimeter wave line observations can play an important role in analyzing the gas components forming new stars or being affected by stellar feedback.

Spectral line observations can be divided into the study of specific tracers (either selected atomic or molecular species) and into the systematic study of all spectral features in a well-defined frequency range. The latter is a very effective method for the study and analysis of the physical and chemical properties of astronomical objects (e.g., \citealt{2015A&A...581A..48Ga,2015A&A...574A..56Gb,2021A&A...656A..46M}). Lines at centimeter wavelengths are mostly optically thin, and molecular lines with sufficiently wide hyperfine  splitting can even be directly used to obtain opacities (e.g., \citealt{1983ARA&A..21..239H}), so analysis is easier than at millimeter wavelengths. Furthermore, the “line density” at this range is not as high as at millimeter wavelengths, reducing the confusion level.

\begin{table*}[t]
\centering
\begin{footnotesize}
\setlength{\tabcolsep}{5 pt}
\caption{Characteristics of the six  W33 clumps.}
\begin{tabular}{lccccccc}
\hline \hline
\vspace{4pt}
Source & R.A.\,(J2000) & Dec.\,(J2000) & Source Size &  $N(\rm H_{2})$ & Evol.\,Stage & Obs. Size \\
\vspace{2pt}
&($^{\rm h}$\,$^{\rm m}$\,$^{\rm s}$)&($^{\circ}$\,$^{\arcmin}$\,$^{\arcsec}$)& ($\arcsec$)/(pc)  & ($10^{23}$\,cm$^{-2}$) &   &    \\
\hline
W33\,Main  &18:14:13.50  &-17:55:47.0   & 58/0.7  &  4.6 $\pm$ 1.6   & Hot Core with H\,{\scriptsize II} region  & $380\arcsec\,\times\,320\arcsec$ \\
W33\,A     & 18:14:39.10 & -17:52:03.0  & 40/0.5  &  2.5 $\pm$ 0.6   & Hot Core\,?                                  & $40\arcsec\,\times\,40\arcsec$  \\
W33\,B     & 18:13:54.40 & -18:01:52.0  & 49/0.6  &   2.1 $\pm$ 0.5  & Hot Core\,?                                 & $120\arcsec\,\times\,120\arcsec$  \\
W33\,A1    &18:14:36.10  &-17:55:05.0   & 82/1.0  &  1.7 $\pm$ 0.6   & High-mass protostellar                    &  1 position  \\
W33\,B1    &18:14:07.10  &-18:00:45.0   & 102/1.2 & 0.5 $\pm$ 0.2    & High-mass protostellar                    & $40\arcsec\,\times\,40\arcsec$ \\
W33\,Main1 &18:14:25:00  &-17:53:58.0   & 75/0.9  &  0.9 $\pm$ 0.2   & High-mass protostellar                    & part of the W33\,Main map  \\
\hline
\end{tabular}
\label{table:1}
\tablefoot{Column\,1 identifies the source by name; columns\,2 and 3 denote the benchmark coordinates; column\,4 indicates the source size; column\,5 reports the peak H$_{2}$ column densities; column\,6 categorizes the evolutionary phase of the star-forming regions associated with W33, with data in columns\,4\,-\,6 taken from \cite{2014A&A...572A..63I}; column\,7 specifies the area observed in this survey, as illustrated in Fig.\,\ref{Fig:1}. The large W33\,Main area is also covering W33\,Main1. The sequence of sources is organized from the most evolved object, descending to smaller, less evolved structures.}
\end{footnotesize}
\end{table*}

Following the option to explore the physical and chemical properties of the interstellar medium through the analysis of spectral line surveys allows us to obtain an impartial snapshot of the studied gas. Nevertheless, despite the extensive legacy of influential large radio telescopes, there have only been a limited number of unbiased frequency surveys carried out within the centimeter wavelength domain. At centimeter wavelengths, anticipated spectroscopic signatures encompass radio recombination lines (RRLs) emanating from hydrogen, helium, and carbon as well as diverse kinds of molecular transitions (e.g., rotational, K-doublet or inversion lines), often from organic species. 

Here we provide a summary of prior studies conducted at centimeter wavelengths toward prominent galactic massive star-forming regions. A study of \citet{1993ApJS...86..211B} reported a spectral survey across a 4.4\,GHz range, from 17.6 to 22.0\,GHz, directed at the massive star-forming region W51, revealing numerous lines that were previously unidentified. Observations of the dark cloud TMC-1 have been conducted in specific frequency ranges of 4\,-\,6\,GHz and 8\,-\,10\,GHz by \citet{2004ApJ...610..329K}, in addition to a broader spectral scan from 9\,-\,50\,GHz by \citet{2004PASJ...56...69K}. \citet{2008ApJ...675L..85R} reported a series of targeted observations directed at the galactic center as part of their GBT PRIMOS project. More recently, two studies by \citet{2022ApJS..263...13L,2024ApJS..271....3L} reported spectral line surveys with frequency ranges of 26.1\,-\,35 GHz and 34.8\,-\,50 GHz, respectively, toward Orion\,KL using the Tianma radio telescope.

A few 1.3\,cm line surveys in the frequency range 17.9\,-\,26.2\,GHz obtained with 35$^{\prime \prime}$ to 50$^{\prime \prime}$ resolution have already been carried out with the 100\,m telescope at Effelsberg. The frequency range contains, for example, many lines of the symmetric top NH$_{3}$. \citet{2015A&A...581A..48Ga} surveyed Orion-KL from 17.9 to 26.2\,GHz with the 100\,m Effelsberg telescope, finding 261 spectral lines. The identified lines include 97 molecular transitions as well as almost twice as many radio recombination lines \citep{2015A&A...581A..48Ga}. The molecular lines are assigned to 13 different molecular species including rare isotopologues \citep{2015A&A...581A..48Ga}. A total of 23 molecular transitions from species already known to exist in Orion\,KL were detected for the first time in the interstellar medium. \citet{2015A&A...574A..56Gb} measured IRC\,+\,10216 from 17.9 to 26.2\,GHz with the Effelsberg 100\,m telescope, finding 78 spectral lines, and the authors showed that the centimeter wave range of the electromagnetic spectrum has an enormous potential for molecular line surveys.

\begin{figure}[t]
\vspace*{0.2mm}
\begin{center}
\includegraphics[trim={16.1cm 2.5cm 17.1cm 1.8cm}, clip, width=0.50\textwidth]{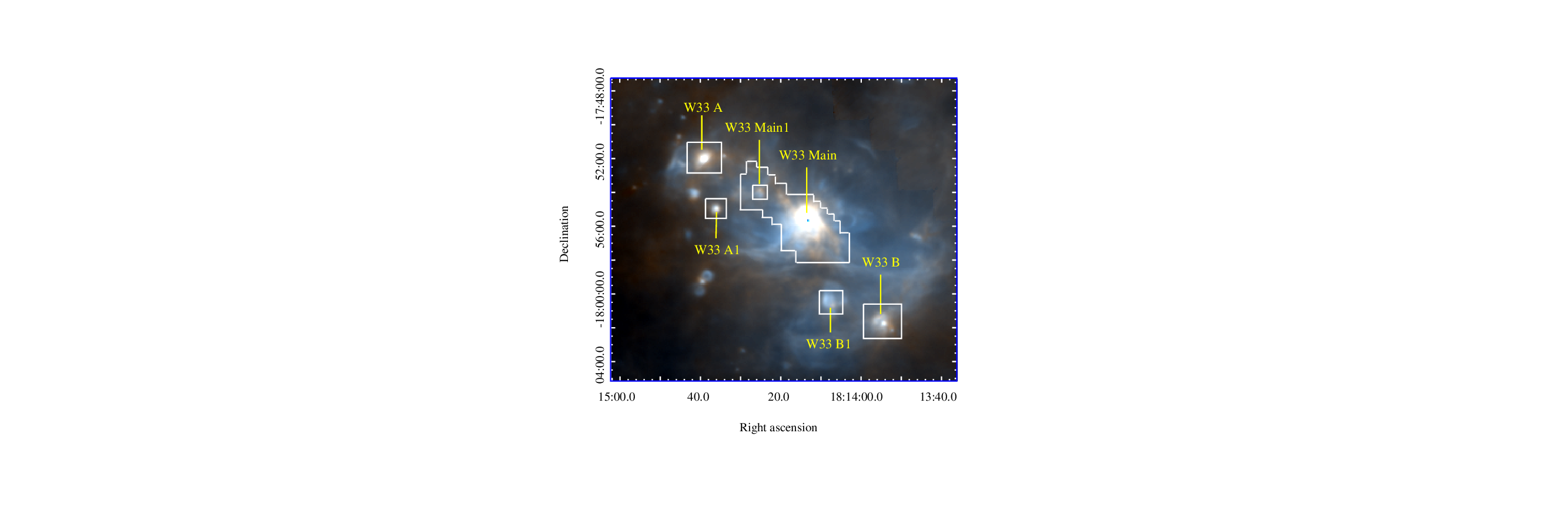}
\end{center}
\caption[]{Two-color image of the high-mass star-forming complex W33 and its surroundings (blue for 70\,$\mu$m, red for 160\,$\mu$m; all derived from $Herschel$ data, \citealt{2010PASP..122..314M}). The six boxes indicate the observed regions (adapted from \citealt{2022A&A...658A..34T}).}
\label{Fig:1}
\end{figure}

\begin{figure}[h]
\centering
\includegraphics[width=0.47\textwidth]{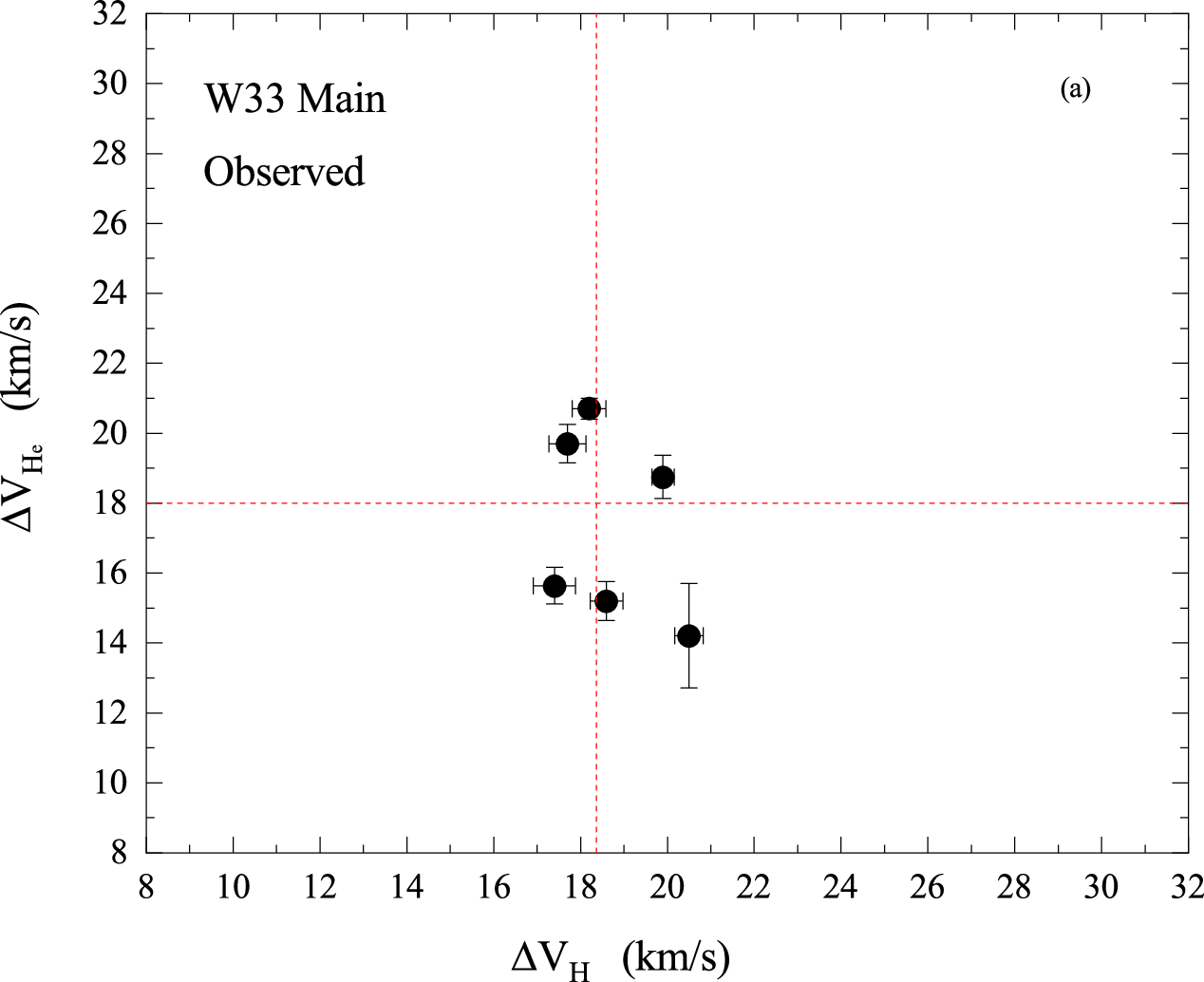}
\includegraphics[width=0.47\textwidth]{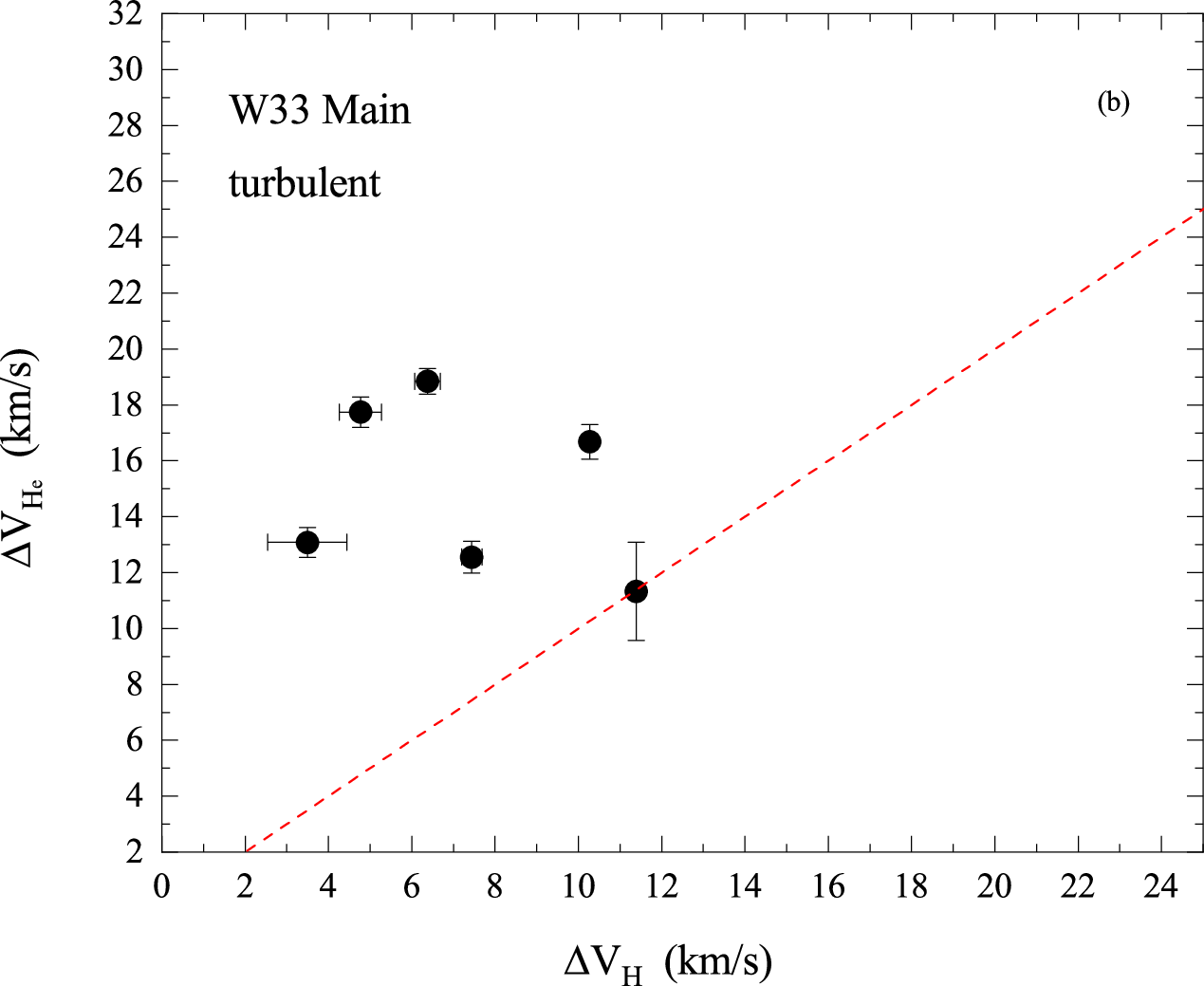}
\caption[]{(a) Correlation between the line widths of hydrogen (horizontal axis) and helium (vertical axis) RRLs. The dashed red lines denote the unweighted average line widths for the hydrogen and helium RRLs, respectively. (b) Correlation between the turbulent line widths as determined from hydrogen and helium RRLs. The dashed red line indicates the points at which the turbulent line widths for hydrogen and helium are equivalent.}
\label{Fig:2}
\end{figure}

W33, located at a distance of $\sim$2.4\,kpc \citep{2013A&A...553A.117I}, is a remarkable star-forming complex (see Fig.\,\ref{Fig:1}). It is both massive, with a mass exceeding $10^{5}$\,${\rm M}_{\odot}$, and bright, shining with a luminosity of $10^{6}$\,${\rm L}_{\odot}$  \citep{1984ApJ...283..573S}. Spanning an area of about 10\,pc, this complex encompasses a variety of regions, from quiescent infrared dark clouds to the intensely active infrared bright hotspots, which are linked to the presence of young massive stars (e.g., \citealt{2014A&A...572A..63I,2015ApJ...805..110M}). Maser emission from water and methanol have been observed in W33\,A and W33\,B as well as at the center of W33\,Main (e.g., \citealt{1977A&AS...30..145G,1981ApJ...250..621J,1986A&A...157..318M,1990ApJ...354..556H,2013A&A...553A.117I}). Additionally, OH masers have been detected in the regions of W33\,A and W33\,B \citep{1974ApJ...187...41W,1998MNRAS.297..215C}. 

In W33\,Main, \citet{1977ApJ...211..421D} observed a grouping of three infrared sources. Additionally, W33\,A is home to an infrared source exhibiting pronounced absorption lines at 3 and 10\,$\mu$m \citep{1977ApJ...211..421D,1978ApJ...226..863C,2010A&A...515A..45D}. Clusters of young stars are linked to W33\,Main, W33\,A, and W33\,B. Furthermore, W33\,Main shows intense radio continuum radiation \citep{1984ApJ...283..573S,2012PASP..124..939H,2022MNRAS.509.2234B}.

These sources represent varying stages of evolution (see Table\,\ref{table:1}). For instance, W33 A1, W33 B1, and W33 Main1 represent high-mass protostellar objects, among which W33 Main1 occupies an exceptionally early phase and lacks a notable heat provider. It may be that the other two objects have reached a warmer stage, suggesting they are more evolved. W33\,A and W33\,B have been identified as hot cores \citep{2014A&A...572A..63I}, and their intricate chemical compositions are significantly shaped by the recent evaporation of dust grain mantles. W33\,Main has reached a later stage of evolution and also contains an H\,{\scriptsize II} region that produces strong radio continuum emissions (e.g., \citealt{1984ApJ...283..573S}). Additionally, there are stellar clusters that have effectively dispersed the majority, if not all, of their surrounding molecular material (e.g., \citealt{2011ApJ...733...41M,2015ApJ...805..110M}). In this current investigation, our primary focus is on the regions W33\,Main, W33\,A, W33\,B, W33\,A1, W33\,B1, and W33\,Main1 (see Fig.\,\ref{Fig:1}). The regions we observed using  the Effelsberg 100\,m telescope are delineated by five distinct boxes in Fig.\,\ref{Fig:1}. Additionally, a comprehensive list of source parameters for the six W33 clumps is provided in Table\,\ref{table:1}.

Here, we systematically study the characteristics of W33 in the important $\lambda$\,$\sim$\,1.3\,cm (K-band) spectral range, which is known to contain a number of lines from prominent molecules such as H$_{2}$O, NH$_{3}$, CH$_{3}$OH, and the cyanopolyynes. The inversion line measurements of ammonia (NH$_{3}$) within the frequency range of 18–26 GHz, performed using the 100\,m Effelsberg telescope and directed toward W33, are detailed in \citet{2022A&A...658A..34T}. This article is organized as follows: In Sect.\,\ref{sect:Observation} we present our observations and data reduction. Results are highlighted in Sect.\,\ref{sect:Results}. A discussion is provided in Sect.\,\ref{sect:discussion}, and our main conclusions are summarized in Sect.\,\ref{sect:summary}. 

\section{Observations and data reduction}
\label{sect:Observation}
\subsection{Observations with the Effelsberg telescope}
\label{sect:Observation1}
The data were taken in January 2018 with the 100\,m Effelsberg telescope\footnote{The 100\,m telescope at Effelsberg is operated by the Max-Plank-Institut f\"ur Radioastronomie (MPIFR) on behalf of the Max-Plank-Gesellschaft (MPG).} located near Bonn, Germany. The measurements were performed using a dual-channel (LCP/RCP) K-band (17.9\,GHz-26.2\,GHz) HEMT receiver. The 5$\sigma$ noise level is $\sim$30\,mK for channels spanning 1\,km\,s$^{-1}$, given a system temperature of about $\sim$\,60\,K as measured on a $T^\ast_{A}$ scale. Four subbands\,--\,WFF4\,(17.9\,--\,20.4\,GHz), WFF3\,(20.0\,--\,22.5\,GHz), WFF2\,(21.6\,--\,24.1\,GHz) and WFF1\,(23.7\,--\,26.2\,GHz)\,--\,were used to concurrently span the full frequency range, ensuring an overlap of no less than 300\,MHz between neighboring subbands. Across the entire frequency spectrum, the beam size, as measured by the full width at half maximum (FWHM), ranged from 35$\arcsec$ (0.4\,pc) to 50$\arcsec$ (0.6\,pc) ($\sim40\arcsec$ (0.5\,pc) at 23\,GHz). 

The survey includes $\sim$20 hours of observation. The focus was adjusted at regular intervals, especially following the rise and the setting of the Sun. The pointing was calibrated hourly using a nearby reference source (primarily PKS\,1830--211), and it was determined to be precise to within 5$\arcsec$. The continuum emission from the intense source 3C\,286 was employed to calibrate the spectral line flux, with an assumed flux density of 2.5\,Jy at 22\,GHz, based on the standard set by \citet{1994A&A...284..331O}. The typical rms noise levels are approximately 0.04\,--\,0.21\,K for the 17.9\,--\,20.4\,GHz subband, 0.08\,--\,0.53\,K for the 20.0\,--\,22.5\,GHz subband, 0.07\,--\,0.60\,K for the 21.6\,--\,24.1\,GHz subband, and 0.06\,--\,0.35\,K for the 23.7\,--\,26.2\,GHz subband. The transformation ratio from Jy on a flux density scale ($S_{\nu}$) to K on the main beam brightness temperature scale ($T_{\rm mb}$) is $T_{\rm mb}$/$S_{\nu}$\,$\sim$\,1.7\,K\,Jy$^{-1}$ at 18.5\,GHz, 1.5\,K\,Jy$^{-1}$ at 22\,GHz, and 1.4\,K\,Jy$^{-1}$ at 23.7\,GHz. All velocity measurements are referenced to the Local Standard of Rest (LSR).

Prior to subsequent analysis, the above mentioned wide $\sim$2.5\,GHz subbands were split into smaller subbands of size $\sim$300\,MHz and then individually calibrated, accounting for atmospheric attenuation and elevation dependent telescope gain corrections. The calibration uncertainties are estimated to be $\pm$15\% (see, e.g., \citealt{2025A&A...700A.192A}).

Typically, each position was subjected to a total integration time of four minutes, alternating between on-source and off-source observations. Nevertheless, to manage possible fluctuations in pointing accuracy and calibration, the central continuum position of W33\,Main (see Table\,\ref{table:1}) was repeatedly monitored. All measurements were conducted using a position-switching method, where the offset positions were set at $900\arcsec$ in azimuth, alternating between the right and left sides. The maps were created with a $20\arcsec$ spacing, which guarantees complete sampling of the targeted sources.

\subsection{Data reduction}
\label{sect-2-2}
For data reduction, we utilized the GILDAS\footnote{http://www.iram.fr/IRAMFR/GILDAS/} software suite, which encompasses the CLASS and GREG tools. In the process of data reduction, it was observed that there were unavoidable flaws at the margins of the spectra; consequently, we discarded 100 channels from each end of the initial spectra. From each spectrum, which contained between 2000 to 3000 channels, baselines ranging from fourth to tenth order were removed to prevent line truncation at the boundaries of the subspectra. Following this, the subspectra were concatenated to reassemble the full spectrum. Due to the presence of time-varying radio frequency interference affecting some of the subspectra, channels exhibiting radio frequency interference signals were "flagged". Additionally, a small number of rogue channels were identified and subsequently removed.
We opted to use the "GAUSS" program to fit all the spectra. The integrated intensities $\int$$T_{\rm MB} {\rm d}\upsilon$, line center velocities $V_{\rm LSR}$, line widths $\Delta v$, and main beam brightness temperatures $T_{\rm MB}$ were derived using the "GAUSS" fitting method. 

\begin{figure*}[t]
\centering
\includegraphics[width=0.33\textwidth]{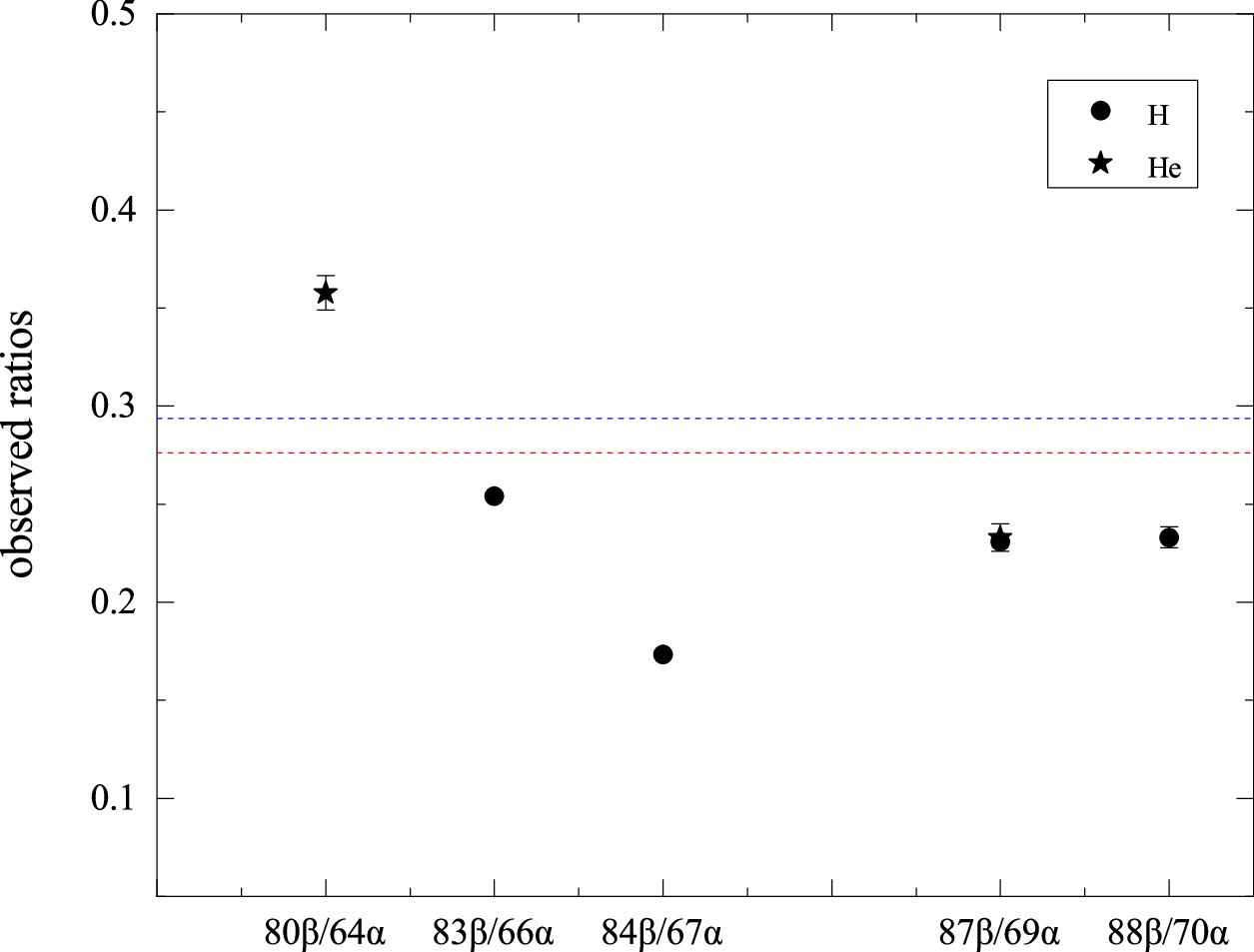}
\includegraphics[width=0.33\textwidth]{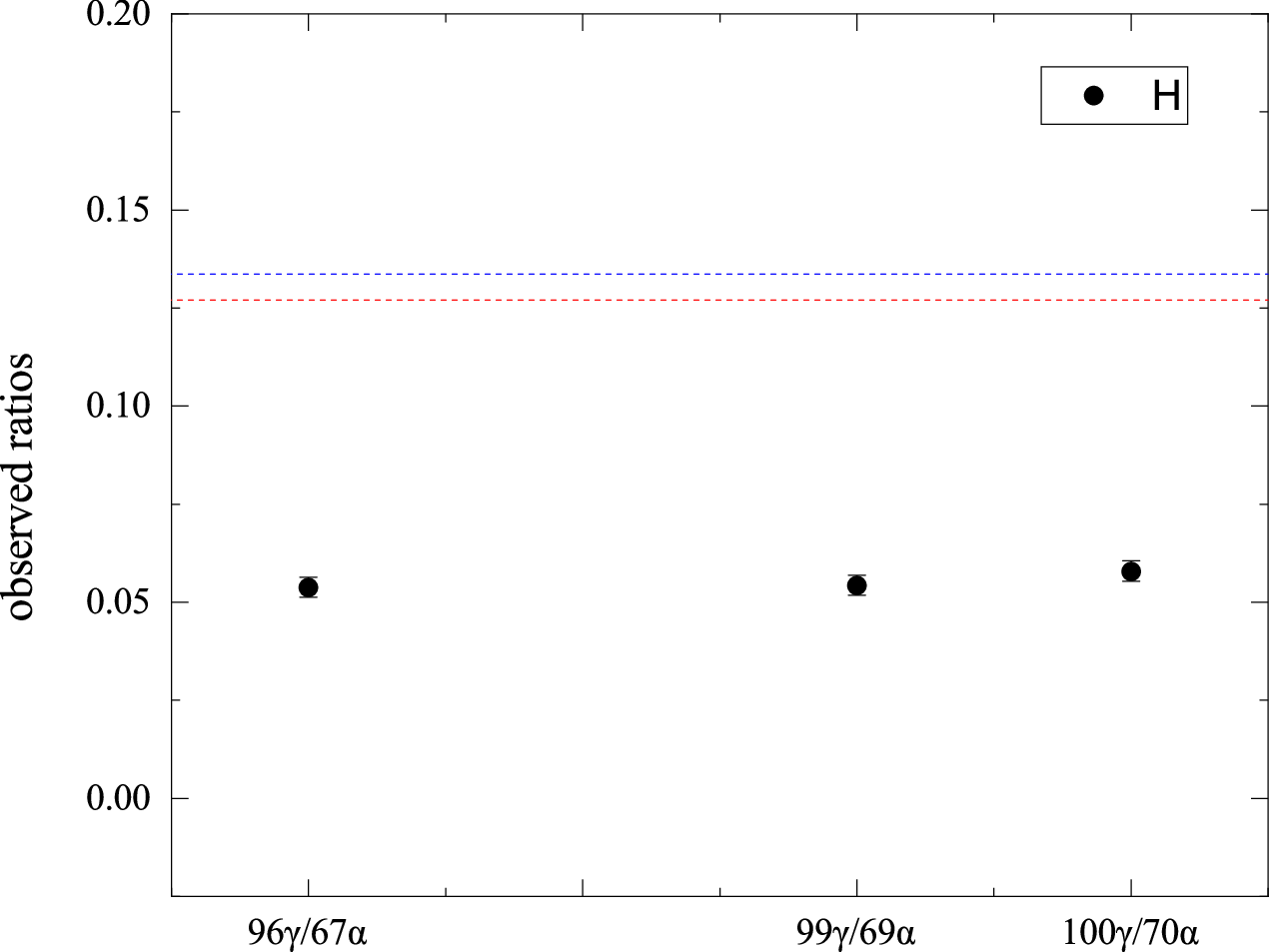}
\includegraphics[width=0.33\textwidth]{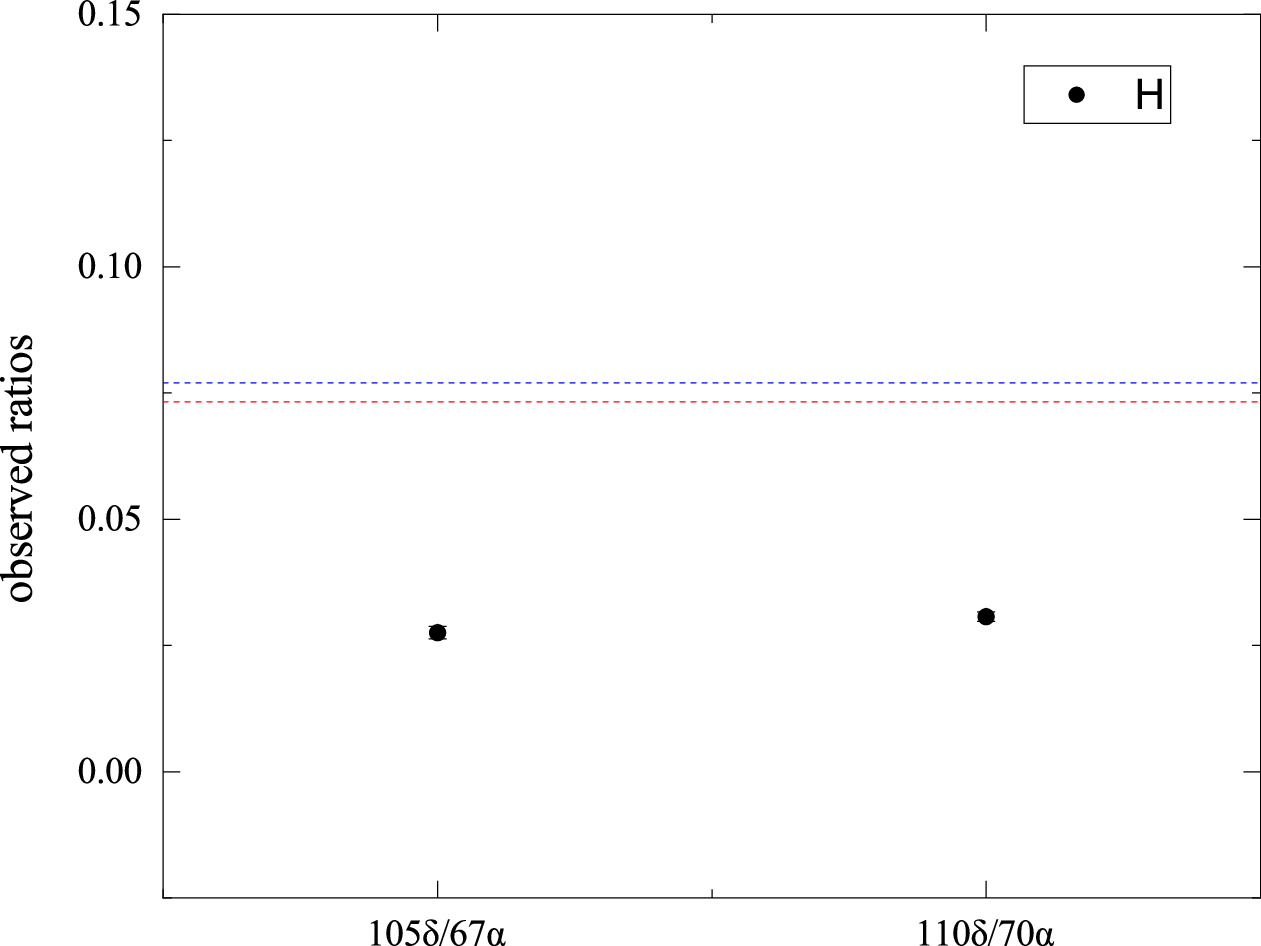}
\caption{{Analysis of the measured versus LTE ratios for RRLs in W33 Main. The actual ratios derived from hydrogen and helium recombination line data are indicated by circular and star-shaped symbols, respectively. The left panel shows a comparison of $\alpha$ and $\beta$, the central panel $\alpha$ and $\gamma$, and the right panel $\alpha$ and $\delta$ RRLs (see Sect.\,\ref{sect-3-1}). The theoretical LTE ratios are illustrated by dashed red lines in the \textit{left, middle, and right panels}, while the corresponding dashed blue lines represent the LTE ratios adjusted for departure coefficient corrections ($b_{n}$). Each ratio is displayed beneath its corresponding horizontal axis. For the majority of lines, the hydrogen and helium data overlap. Each data point has a small error, ranging from 4.3 $\times10^{-4}$ to 1.2 $\times10^{-3}$, with a mean error of about 2.2 $\times10^{-3}$.} 
\label{Fig:3}}
\end{figure*}

\section{Results and analysis}
\label{sect:Results}
\subsection{Overview}
\label{sect-3-0}
This study, similar to that of \citet{2022A&A...658A..34T}, is based on a total of 218 observed positions. Specifically, 182 data points were identified within the zone that includes W33\,Main and W33\,Main1. 9 positions were observed toward W33\,A, W33\,B was associated with 17 positions, and W33\,B1 was pinpointed by nine positions. Observations of W33\,A1 are part of the main map also encompassing W33\,Main. The sizes of the various maps displayed in Fig.\,\ref{Fig:1} are given in Table\,\ref{table:1}.

\subsection{Line identification}
\label{sect-3-1}
The identification of spectral lines was facilitated through the use of databases from JPL\footnote[3]{http://spec.jpl.nasa.gov}, CDMS\footnote[4]{http://www.astro.uni-koeln.de/cdms/catalog}, and Splatalogue\footnote[5]{http://www.splatalogue.online/\#/advanced}, in conjunction with the Lovas line list\footnote[6]{http://www.nist.gov/pml/data/micro/index.cfm} for astronomical spectroscopy \citep{1998JQSRT..60..883P,2004JPCRD..33..177L,2005JMoSt.742..215M}. Nevertheless, the rest frequencies for the RRLs are not present within these databases, necessitating their calculation using the Rydberg formula
\begin{equation}\label{redberg}
\nu_{n_2 \to n_1} = R\left(\frac{1}{n_1^2} - \frac{1}{n_2^2}\right), ~n_{1} < n_{2}, \Delta n = n_2 - n_1, 
\end{equation}
In this formula, $\Delta n$ represents the difference between the principal quantum numbers of the upper and lower states, with $n_{2}$ and $n_{1}$ being the quantum numbers for the respective states. The Rydberg constant, R, takes on the value of 3.28805129$\times 10^{15}$\,Hz for hydrogen and 3.28939118$\times 10^{15}$\,Hz for helium \citep{1968ApJS...16..143L,2009tra..book.....W}. The naming convention for RRLs is derived from $n_{1}$ and $\Delta n$. For instance, the designation H64$\alpha$ signifies a hydrogen RRL where $n_{1}$ is 64 and $n_{2}$ is 65, resulting in a $\Delta n$ of 1. The Greek letters are used to denote $\Delta n$ values, such that $\alpha$ indicates $\Delta n=1$, $\beta$ signifies $\Delta n=2$, and so forth.

A line is always considered genuine if it displays a 3$\sigma$ characteristic across more than three consecutive unprocessed channels. The sigma ($\sigma$) value corresponds to the product of the peak-to-rms ratio and the square root of the number of channels that significantly contribute to the signal feature. Lines that have been identified are presented in Table\,\ref{Tab:A.1}. 

Considering the intricate nature of line profiles in W33, various fitting techniques are utilized for our observed transitions. We employ a single Gaussian component to obtain the observed parameters of RRLs. For molecular lines that exhibit nuclear quadrupole hyperfine structure (such as HC$_{3}$N, CH$_{3}$CN, HNCO, and NH$_{3}$), we utilized the hyperfine fitting routines in CLASS to determine their observed properties (including optical depth). For most other molecular transitions, we used least-squares fits with Gaussian profiles. For lines that contain multiple distinct velocity components, our method involves using a spectrum of velocities and line widths for different components to distinguish their individual contributions. This approach is grounded in an overall understanding of the source. We utilize more than two Gaussian fittings to distinguish the relative contributions and derive the observed properties of the lines. For the remaining blended transitions, instead of separating the lines, we integrate the entire line profiles to establish upper limits on the integrated intensities of the distinct spatial subcomponents. 

All observed transitions show the well-known total velocity range of approximately $V_{\rm LSR}$\,$\sim$\,36\,km\,s$^{-1}$ to $V_{\rm LSR}$\,$\sim$\,58\,km\,s$^{-1}$, consistent with previous studies (e.g., \citealt{1972ApL....12..107G,1975ApL....16...29G,1983ApJ...265..791G,2013A&A...553A.117I,2014A&A...572A..63I}). However, the absorption and emission lines from specific sources fall into two distinct radial velocity groups: W33\,Main, W33\,A, W33\,Main1, W33\,A1,  W33\,B1 exhibit an observed range of 32 to 40\,km\,s$^{-1}$. W33\,B shows a notably different central radial velocity of  $\sim$58\,km\,s$^{-1}$ (see Figs.\,\ref{FigA.7} to \ref{FigA.12}). These two velocity regimes are consistent with the earlier CO observations reported by \cite{1983ApJ...265..791G}. Furthermore, the CO observations by \cite{1983ApJ...265..791G} reveal that both velocity components  are present in emission across the entire W33 complex.

Tables\,\ref{Tab:A.1} through\,\ref{Tab:A.5} in Appendix\,\ref{Appendix A} present the parameters of the observed lines for various species in the regions W33\,Main, W33\,A, W33\,B, W33\,A1,  W33\,B1, and W33\,Main1. Figs.\,\ref{FigA.1} to \ref{FigA.6} provide an overview of the 1.3\,cm spectral line survey for W33, in which 68 lines have been identified. Among the identified lines, there are 44 RRLs and 24 molecular lines. The RRLs, emanating from hydrogen and helium, are produced in the ionized medium of the W33 Main region, a portion of which falls within the scope of our central beam \citep{2022A&A...664A.140K}. The molecular lines pertain to nine different already previously identified species in the interstellar medium. In Figs.\,\ref{FigA.1} to \ref{FigA.6} of Appendix\,\ref{Appendix A} it is evident that the spectrum is primarily characterized by intense emissions from four distinct species: the exceptionally strong H$_{2}$O (6$_{1,6}$--5$_{2,3}$) maser transition, the CH$_{3}$OH maser lines, various metastable transitions of NH$_{3}$ (see Fig.\,2 to Fig.\,5 of \citealt{2022A&A...658A..34T}), and the H$\alpha$ RRL features. In Figs\,\ref{FigA.1} to \,\ref{FigA.6} each individual panel encompasses $\sim$500\,MHz, featuring a 10\,MHz overlap with the adjacent panels to ensure that spectral lines cut off in one panel are not similarly truncated in the next. Furthermore, Figs.\,\ref{FigA.7} to \ref{FigA.12} display enlarged views of all the identified spectral lines.

\begin{figure*}[t]
\centering
\includegraphics[width=0.49\textwidth]{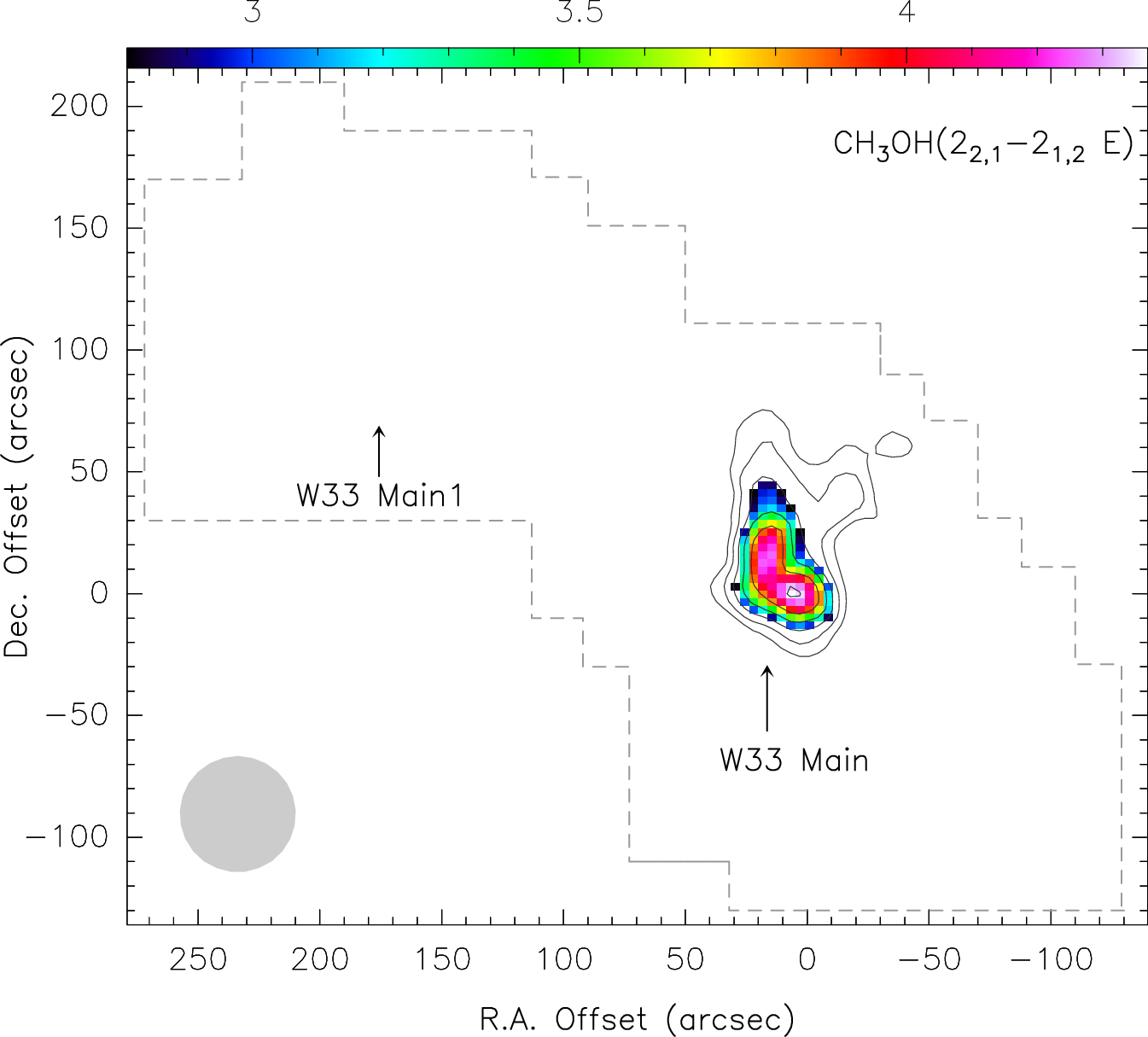}
\hspace{0.2cm}
\includegraphics[width=0.49\textwidth]{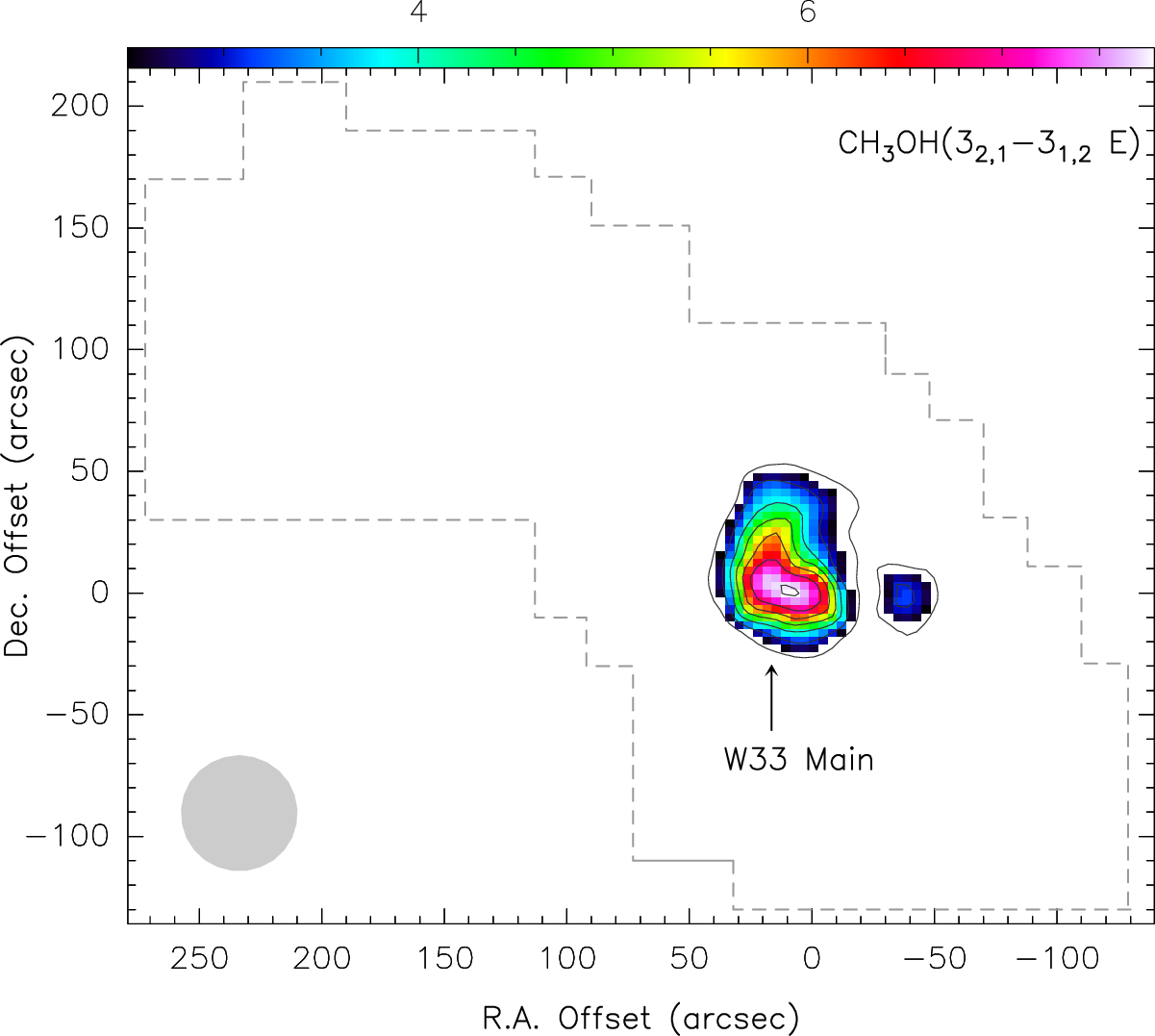}
\includegraphics[width=0.49\textwidth]{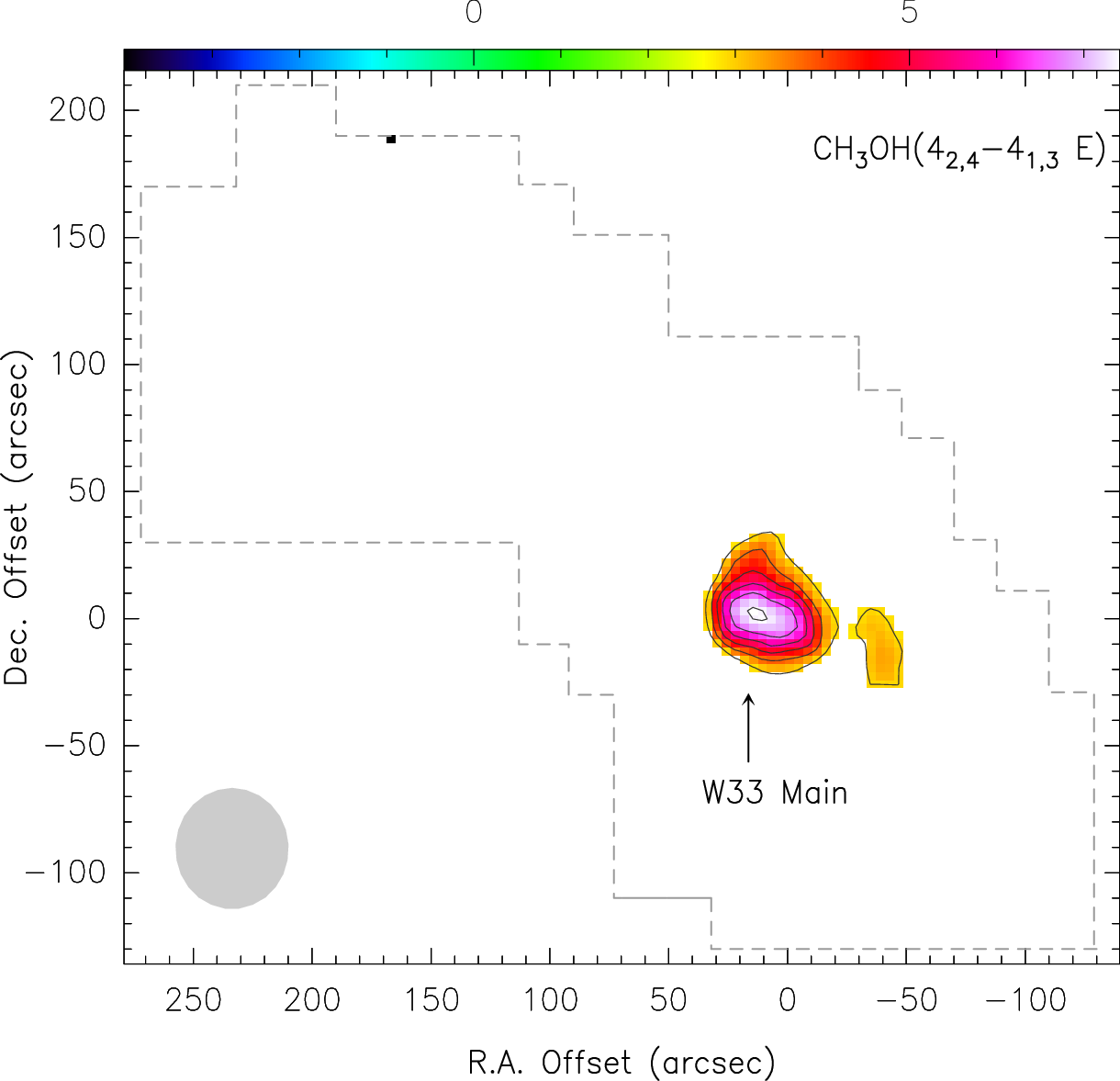}
\hspace{0.2cm}
\includegraphics[width=0.49\textwidth]{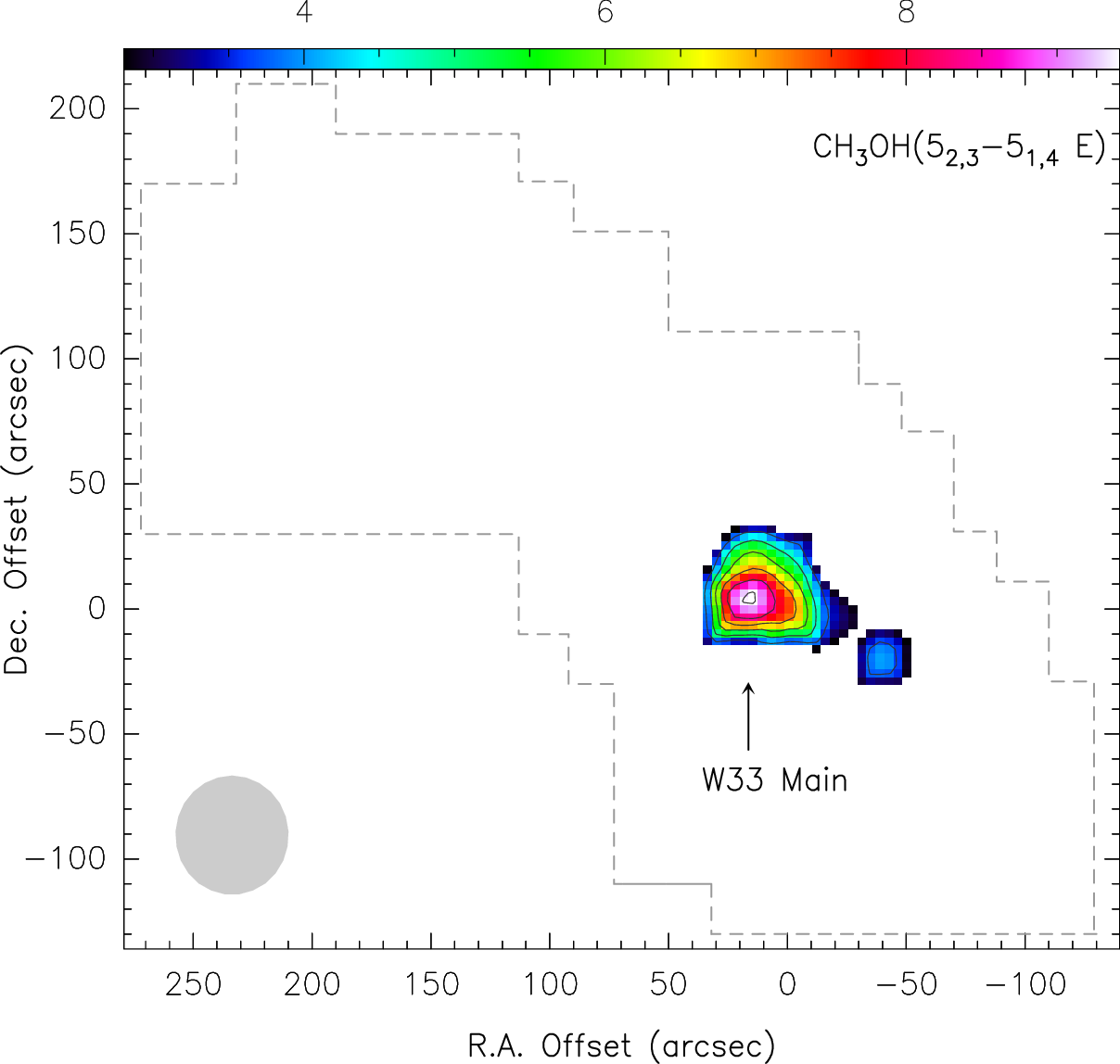}
\caption[]{Integrated intensity maps of CH$_3$OH\,(2$_{2,1}$--2$_{1,2}$ E) (\textit{top left}), CH$_3$OH\,(3$_{2,1}$--3$_{1,2}$ E) (\textit{top right}), CH$_3$OH\,(4$_{2,4}$--4$_{1,3}$ E) (\textit{bottom left}), and CH$_3$OH\,(5$_{2,3}$--5$_{1,4}$ E) (\textit{bottom right}) for the W33\,Main and W33\,Main1 regions. The reference position is R.A. 18:14:13.50 and DEC. -17:55:47.0 (J2000), corresponding to the location of W33\,Main. The integration range is 32\,$<$\,$V_{\rm LSR}$\,$<$\,40\,km\,s$^{-1}$. Contours begin at a level of 3.14\,K\,km\,s$^{-1}$ (3$\sigma$) on a main beam brightness temperature scale and increase in increments of 3.14\,K\,km\,s$^{-1}$. The unit of the color bars is kelvin kilometers per second. The boundaries of the mapped area are demarcated with dashed gray lines. The extent of the half-power beam width is represented by a gray-filled circle located in the bottom-left corners of the respective images. In this map, we only show W33 Main because the other regions are not extended point-like sources relative to the Effelsberg beam.}
\label{Fig:4}
\end{figure*}

\begin{figure*}[t]
\centering
\includegraphics[width=0.33\textwidth]{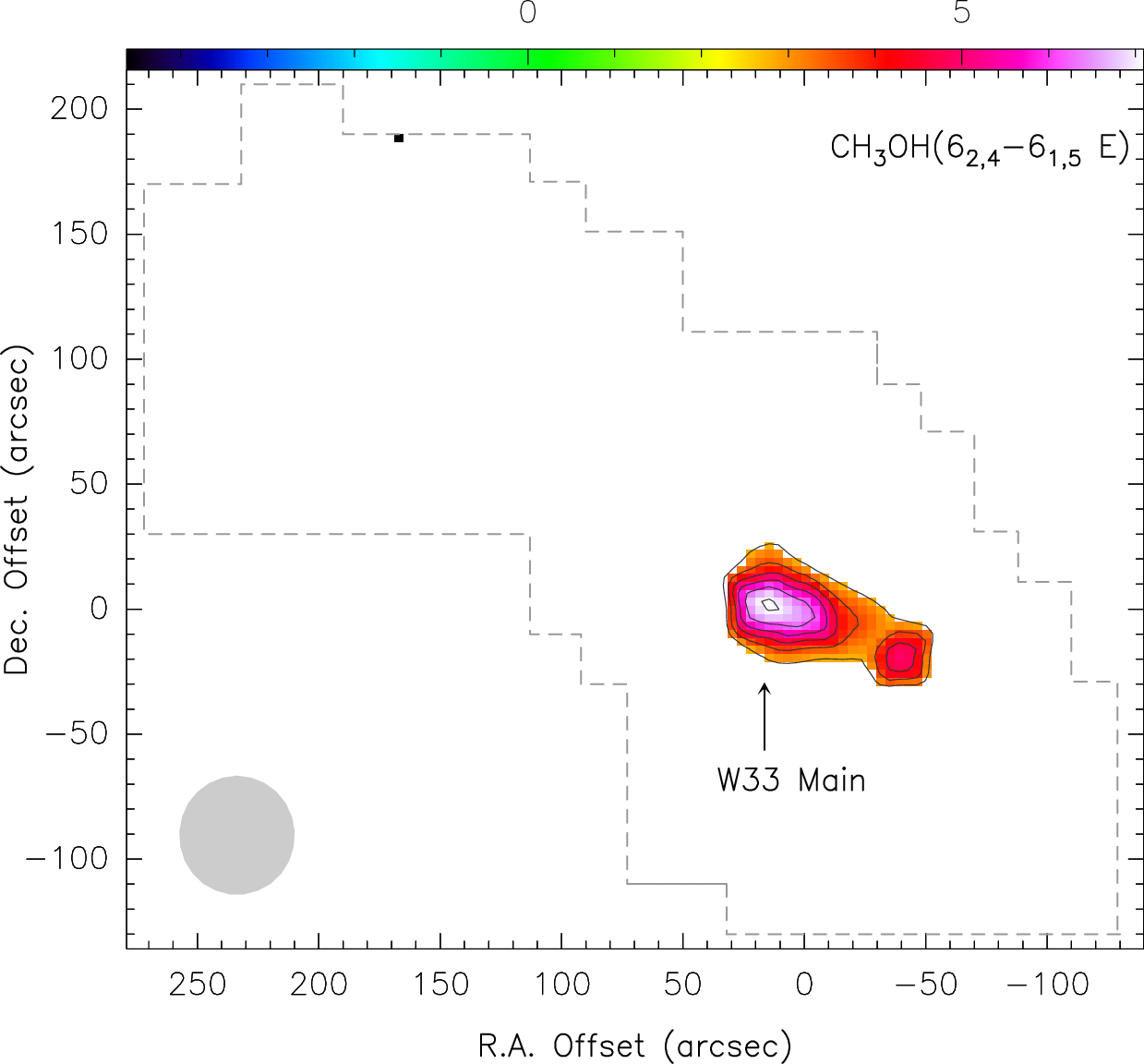}
\includegraphics[width=0.32\textwidth]{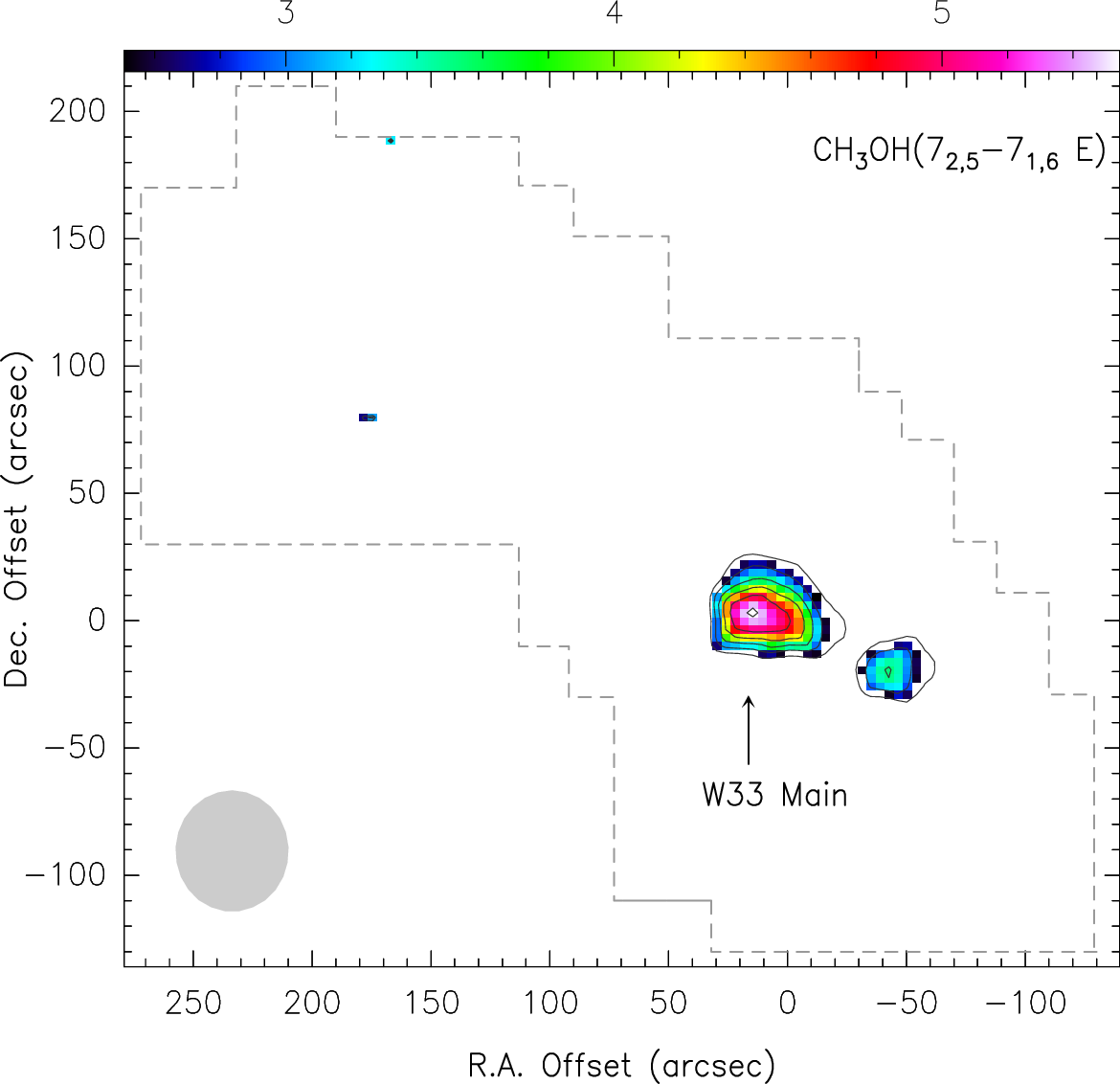}
\includegraphics[width=0.33\textwidth]{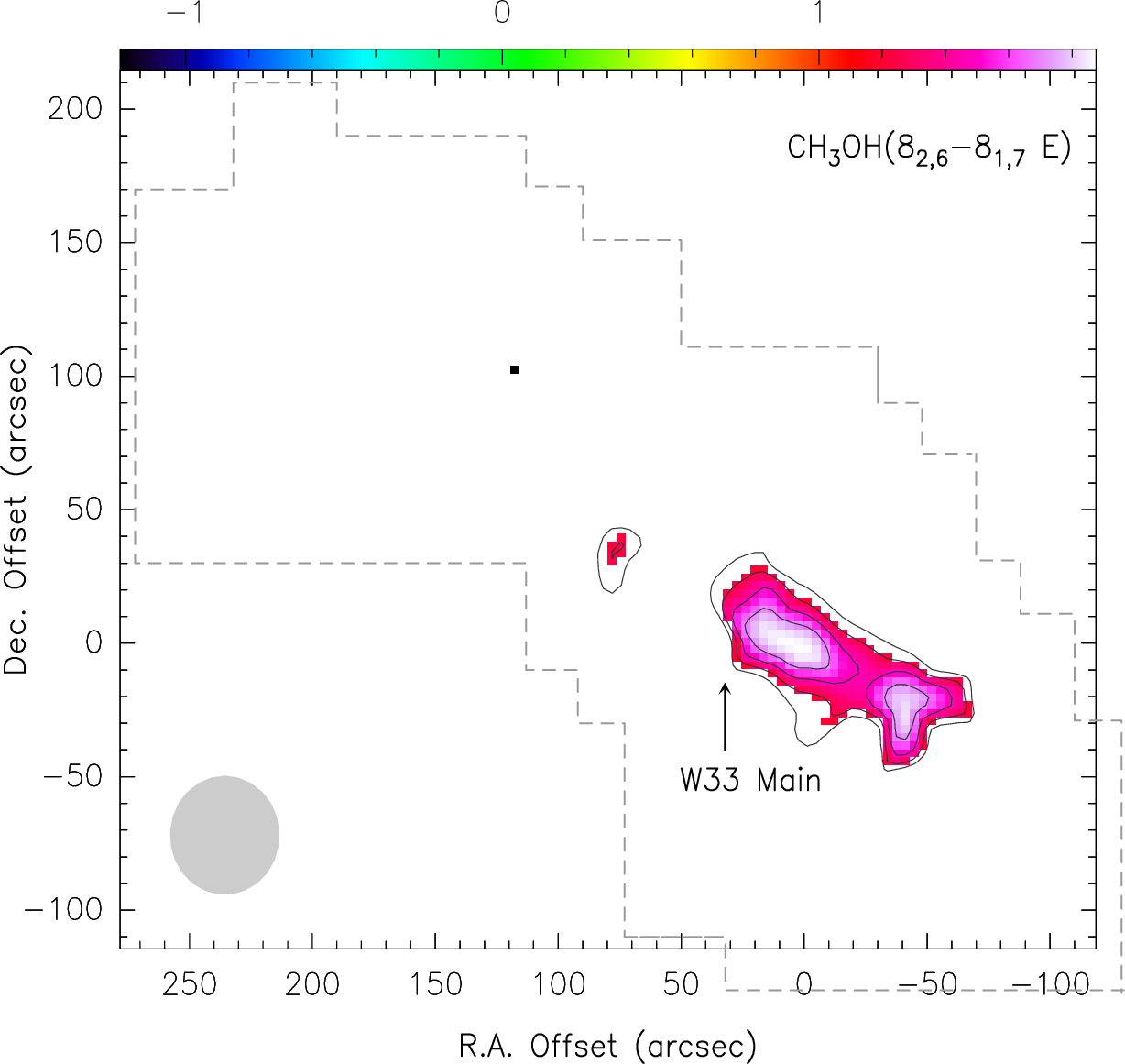}
\caption[]{Same as Figure\,\ref{Fig:4} but integrated intensities are presented for the CH$_3$OH\,(6$_{2,4}$--6$_{1,5}$ E) (\textit{left}), CH$_3$OH\,(7$_{2,5}$--7$_{1,6}$ E) (\textit{middle}), and CH$_3$OH\,(8$_{2,6}$--8$_{1,7}$ E) (\textit{right}) lines.} 
\label{Fig:5}
\end{figure*}

\subsection{Radio recombination lines}
\label{sect-3-2}
The spectrum is largely dominated by RRLs, as shown in Figs.\,\ref{FigA.1} to \ref{FigA.6} and Table\,\ref{Tab:A.2}. The RRL detections are associated with the W33\,Main H\,{\scriptsize II} region. Additionally, H69$\alpha$ was detected in W33\,Main1, which is related to a northeastern hotspot next to the W33\,Main H\,{\scriptsize II} region (see, e.g., Fig.1c of \citealt{1984ApJ...283..573S}). The offset between W33\,Main1 and the primary source W33\,Main is $(\Delta\alpha, \Delta\delta)$\,=\,($160\arcsec, 112\arcsec$). The 44 RRLs consists of 36 hydrogen RRLs, and 8 helium RRLs. The spectra for these are depicted in Figs.\,\ref{FigA.7}, \ref{FigA.8}, and \ref{FigA.12} , with the corresponding observed line parameters listed in Table\,\ref{Tab:A.2}. 29 hydrogen RRLs exhibiting $\Delta n \ge 2$ have been identified. Additionally, this survey has revealed hydrogen RRLs with $\Delta n$ reaching to 4.  

It is evident from Fig.\,\ref{Fig:2}a that the full widths to half maximum (FWHM) line widths of hydrogen RRLs are approximately 18.7\,km\,s$^{-1}$, which is wider than the corresponding line widths of helium RRLs, which are about 17.4\,km\,s$^{-1}$. The line broadening of RRLs is believed to be caused by thermal activity and turbulent motion, as discussed by \citet{2009tra..book.....W}. The observed line width, $\Delta \upsilon$, encompasses both thermal and turbulent components, $\Delta \upsilon = \sqrt{\Delta \upsilon_{\rm th}^{2}+\Delta \upsilon_{\rm tur}^{2}}$, with $\Delta \upsilon_{\rm th}$ representing the thermal component and $\Delta \upsilon_{\rm tur}$ denoting the width due to turbulent motion. Assuming an electron temperature of 6343\,$\pm$\,222\,K for the W33\,Main H{\scriptsize II} region, as reported by \citet{2022A&A...664A.140K}, and utilizing a Maxwell-Boltzmann velocity distribution, we calculated the thermal widths of hydrogen and helium to be 17.0\,km\,s$^{-1}$ and 8.6\,km\,s$^{-1}$, respectively. This was determined using the formula
\begin{equation}\label{eq-2}
\Delta \upsilon_{\text{th}} = 2\sqrt{2\ln 2} \Bigl( \frac{kT}{m} \Bigr)^{1/2}, 
\end{equation}
where k represents the Boltzmann constant, $T$ is the electron temperature, and $m$ denotes the mass of the particle. Hence, the measured average turbulent line widths for hydrogen and helium are calculated to be 7.3 $\pm$ 0.4\,km\,s$^{-1}$ and 15.0 $\pm$ 0.7\,km\,s$^{-1}$, respectively. Examining Fig.\,\ref{Fig:2}b, it becomes clear that the bulk of the data points reside above the red dashed line, representing the condition, where the turbulent line widths for both hydrogen and helium would be the same. This implies that the turbulent line width of helium could be marginally greater than that of hydrogen. The difference in turbulent line widths between helium and hydrogen ranges from 5.1 to 12.9\,km\,s$^{-1}$, with an average of 9.3\,$\pm$\,3.5\,km\,s$^{-1}$. This is likely due to spatially unresolved fine structure that cannot be disentangled with the current data and will require much higher angular resolution observations. 

Under the premise of local thermodynamical equilibrium (LTE), and utilizing the equations (6.24)--(6.27) presented by \citet{1972MNRAS.157..179B}, the intensity ratios for RRL pairs with varying $\Delta n$ in the W33\,Main H\,{\scriptsize II} region, originating from the same atomic species at adjacent frequencies (such as H88$\beta$/H70$\alpha$), were determined to be
\begin{equation}
\frac{T_{\Delta n_{1}}}{T_{\Delta n_{2}}} = \frac{\Delta n_{1}K(\Delta n_{1})}{\Delta n_{2}K(\Delta n_{2})} 
\end{equation} 
Here, $\Delta n_{1}$ and $\Delta n_{2}$ denote distinct $\Delta n$ values as outlined in equation (\ref{redberg}), and the corresponding $\Delta nK(\Delta n)$ values are provided in Table I from the work of  \citet{1972MNRAS.157..179B}. Fig.\,\ref{Fig:3} reveals that the observed ratios of H$\beta$/H$\alpha$,H$\gamma$/H$\alpha$,H$\delta$/H$\alpha$ are in agreement with those predicted by LTE ratios. Utilizing the table of departure coefficients ($b_{n}$) supplied by \citet{1979ApJS...39..633S}, we note that all $b_{n}$ values lie between 0.94 and 1.0, given an electron temperature of 6343\,$\pm$\,222\,K and an electron density of $1\times10^{4}$ cm$^{-3}$, utilizing the parameters set by \citet{1997A&A...327.1177W}. Hence, we propose that deviations from LTE conditions are negligible for the RRLs observed by us in W33\,Main.

When observed at centimeter wavelengths, compact H\,{\scriptsize II} regions frequently exhibit extremely broad radio recombination line profiles, exceeding the thermal line width by a factor of three to four. While this characteristic is not universal among all compact H\,{\scriptsize II} regions, those that display it are occasionally classified as broad recombination line objects \citep{1981A&A....93...48A,1994ApJ...437..697A,1994ApJ...428..670D,1995ApJ...444..765K,1999ApJ...520..162J,2002ApJ...568..754K,2006ApJ...637..850K}. The broadening of radio recombination lines in H\,{\scriptsize II} regions arises from several distinct factors: natural broadening due to the quantum uncertainty $\Delta$\,E of the energy level; thermal and microturbulent broadening caused by the small-scale motion of atoms and gas parcels; dynamical broadening, resulting from large-scale gas flows such as infall, rotation, and outflows; and pressure broadening, which is induced by high electron densities \citep{2008ApJ...672..423K,2012A&A...547L...3G,2017A&A...602A..37K,2025A&A...702A.107G}. We aim to estimate the respective contributions of these various broadening mechanisms.

For radio and (sub) millimeter recombination lines, natural broadening—caused by the energy uncertainty $\Delta$\,E of quantum states—is negligible. We calculated the thermal line widths of the hydrogen and helium radio recombination lines using Equation\,\ref{eq-2}. The results show that the thermal line width of the hydrogen recombination line is comparable to its observed line width ($\Delta \upsilon_{\rm obs}$\,=\,18.7\,km\,s$^{-1}$\,$\simeq$\,$\Delta \upsilon_{\rm th}$\,=\,17.0\,km\,s$^{-1}$), indicating that the broadening of the hydrogen line is dominated by thermal motion. In contrast, the observed line width of the helium recombination line is significantly larger than the thermal line width of the hydrogen line ($\Delta \upsilon_{\rm obs}$\,=\,17.4\,km\,s$^{-1}$\,$>$\,$\Delta \upsilon_{\rm th}$\,=\,8.6\,km\,s$^{-1}$), suggesting that the excess broadening is attributable to turbulent broadening. Turbulence is the most straightforward explanation for the nonthermal broadening. \citet{2008ApJ...672..423K} proposed that the turbulence in H\,{\scriptsize II} regions is driven by instabilities at the interface between the ionized gas and the surrounding molecular gas, providing a physical basis for the presence of turbulent widths. The observed spectral lines of hydrogen and helium radio recombination lines exhibit a narrow core and broad wing features, which is typically associated with dynamical broadening caused by stellar winds or outflows \citep{2012A&A...547L...3G}. \citet{2022A&A...664A.140K} report an electron density of 6.2\,$\times10^{3}$\,cm$^{-3}$ for W33\,Main (see Table\,5 of  \citealt{2022A&A...664A.140K}). With an electron density of $n$\,$<$\,$10^{4}$\,cm$^{-3}$, pressure broadening can be neglected; whereas when $n$\,$\sim$\,$10^{5}$\,--\,$10^{6}$\,cm$^{-3}$, pressure broadening must be taken into account in the K band. Therefore, in the W33\,Main region, pressure broadening can be neglected.

\subsection{Molecular lines}
\label{sect-3-3}
In this part of the study, we concentrate on the observed molecular lines in the six sources pertaining to W33 following Table\,\ref{table:1}.  This is based on the line parameters that were derived through Gaussian and hfs fitting methodologies outlined in Sect.\,\ref{sect-3-1}.

\subsubsection{NH$_{3}$ and NH$_{2}$D}
\label{sect-3-3-1}
In our band, we have identified six metastable and two non-metastable transitions of NH$_{3}$, as well as one NH$_{2}$D transition toward W33.
We observed metastable NH$_{3}$\,(1,1), (2,2), (3,3), (4,4), (5,5), and (6,6) transitions within W33\,Main, W33\,A, and W33\,B, as detailed in Figures\,2 and 3 of \citet{2022A&A...658A..34T}. We also observed the metastable NH$_{3}$\,(1,1), (2,2), (3,3), and (4,4) transitions in W33\,A1 and W33\,Main1 (see Fig.\,4 of \citealt{2022A&A...658A..34T}). In W33\,B1, only the NH$_{3}$\,(1,1), (2,2), and (3,3) transitions were detected (see Fig.\,5 of \citealt{2022A&A...658A..34T}). The non-metastable NH$_{3}$\,(2,1) and (3,2) lines are detected toward the central positions of W33\,Main, W33\,A, and W33\,B (see the right panel of Fig.\,3 of \citealt{2022A&A...658A..34T}). An NH$_{2}$D\,(3$_{1,3}$s--3$_{0,3}$a) transition line was only tentatively detected in W33\,B1 (see Figs.\,\ref{FigA.12}). The spectral line parameters for the NH$_{2}$D\,(3$_{1,3}$s--3$_{0,3}$a) line are presented in Table\,\ref{Tab:A.5}, while a NH$_{2}$D/NH$_{3}$ ratio is calculated in Sect.\,\ref{4.3}.

Figure\,\ref{FigB.1} displays the integrated intensity maps of the NH$_3$\,(1,1), (2,2), and (3,3) emission for the main hyperfine groups in the W33\,Main and W33\,Main1 regions. This includes absorption toward the H\,{\scriptsize II} region of W33 Main. Intensities were integrated over the LSR velocity ($V_{\rm LSR}$) range of 32 to 40 \,km\,s$^{-1}$. The NH$_3$\,(1,1) emission is spatially extended and traces the denser molecular structure. NH$_3$\,(2,2) is detected over a slightly smaller area, while the NH$_3$\,(3,3) distribution is even more compact. In each panel, the half-power beam width is depicted by a gray circle in the lower-left corner. The dashed gray lines indicate the boundaries of the mapped region.

\subsubsection{H$_{2}$O}
\label{sect-3-3-2}
We have detected the H$_{2}$O\,(6$_{1,6}$--5$_{2,3}$) line in the W33 region.  The H$_{2}$O\,(6$_{1,6}$--5$_{2,3}$) maser shows multiple velocity components within the three W33\,Main, W33\,A, and W33\,B clumps (see Figs.\,\ref{FigA.9}, \ref{FigA.10}, and \ref{FigA.11}). For W33\,Main, the H$_{2}$O emission line exhibits a three-component velocity structure with $V_{\rm LSR}$\,$\sim$\,33\,km\,s$^{-1}$, $V_{\rm LSR}$\,$\sim$\,35\,km\,s$^{-1}$, and $V_{\rm LSR}$\,$\sim$\,38\,km\,s$^{-1}$.  In W33\,A, the H$_{2}$O emission line shows five distinct velocity components at  $V_{\rm LSR}$\,$\sim$\,35\,km\,s$^{-1}$, $V_{\rm LSR}$\,$\sim$\,38\,km\,s$^{-1}$, $V_{\rm LSR}$\,$\sim$\,44\,km\,s$^{-1}$, $V_{\rm LSR}$\,$\sim$\,46\,km\,s$^{-1}$, and $V_{\rm LSR}$\,$\sim$\,49\,km\,s$^{-1}$. Additionally, the H$_{2}$O emission in W33\,B displays a double-peaked velocity profile, with peaks centered at $V_{\rm LSR}$\,$\sim$\,56\,km\,s$^{-1}$ and $V_{\rm LSR}$\,$\sim$\,60\,km\,s$^{-1}$. The peak main beam brightness temperatures for W33\,Main, W33\,A, and W33\,B are greater than 5\,K, 40\,K, and 200\,K, respectively, at their corresponding peak positions in the W33 region, as listed in Table\,\ref{table:1}. A comparison of the line parameters with those obtained from previous observations (e.g., \citealt{1977A&AS...30..145G,1981ApJ...250..621J,2015A&A...575A..49C}) reveals that the line shapes and peak main beam brightness temperatures are in good agreement. The spectral line parameters for the H$_{2}$O\,(6$_{1,6}$--5$_{2,3}$) transition lines are detailed in Table\,\ref{Tab:A.5}.

\subsubsection{Cyanopolyynes: HC$_{3}$N, HC$_{5}$N, HC$_{7}$N} 
\label{sect-3-3-3}
Cyanopolyynes have been extensively researched and are commonly found in a variety of astronomical settings, including the late-type carbon star IRC +10216 and the dense cold cores   encountered, for example, in Taurus and Aquila \citep[e.g.,][]{1979ApJ...233L.141M,1980ApJ...242L..87B,1983ApJ...270..589B,2000A&AS..142..181C,2004PASJ...56...69K,2013A&A...559A..91L,2024ApJS..271...45T} as well as in Chameleon molecular clouds \citep{2019MNRAS.488..495W}. To provide an idea what to expect in W33, we note that \citet{2018ApJ...854..133T} used the Nobeyama 45\,m radio telescope (with a beam size of 37$\arcsec$) to observe 17 high-mass starless cores (HMSCs) and 35 high-mass protostellar objects (HMPOs) in HC$_{3}$N and HC$_{5}$N emission. Among these, HC$_{3}$N was detected in 15 HMSCs and 28 HMPOs, while HC$_{5}$N was identified in 5 HMSCs and 14 HMPOs. In our survey, we detected the HC$_{3}$N\,(2–1) line at 18.2\,GHz toward W33\,Main and W33\,A1 (see Figs.\,\ref{FigA.9} and \ref{FigA.11}). Toward W33\,A, we detected the HC$_{3}$N\,(2–1) line, as well as the HC$_{5}$N\,(7–6) and HC$_{5}$N\,(8–7) transitions (see Fig.\,\ref{FigA.10}). We also detected the HC$_{3}$N\,(2–1), HC$_{5}$N\,(7–6), (8–7), (9–8), and tentatively the HC$_{7}$N\,(19–18) transitions in W33\,B (see Fig.\,\ref{FigA.11}). Toward W33\,Main1, we also detected HC$_{3}$N\,(2–1) emission. Additionally, we observed the HC$_{5}$N\,(7–6) and HC$_{5}$N\,(9–8) transitions at 18.6\,GHz and 23.9\,GHz, respectively (see Fig.\,\ref{FigA.12}). The HC$_{5}$N\,(8–7) line was not detected in W33\,Main1 due to contamination by Radio Frequency Interference (FRI). Spectral line parameters of the HC$_{n}$N transition lines are given in Table\,\ref{Tab:A.4}.

\begin{figure*}[t]
\centering
\includegraphics[width=0.23\textwidth]{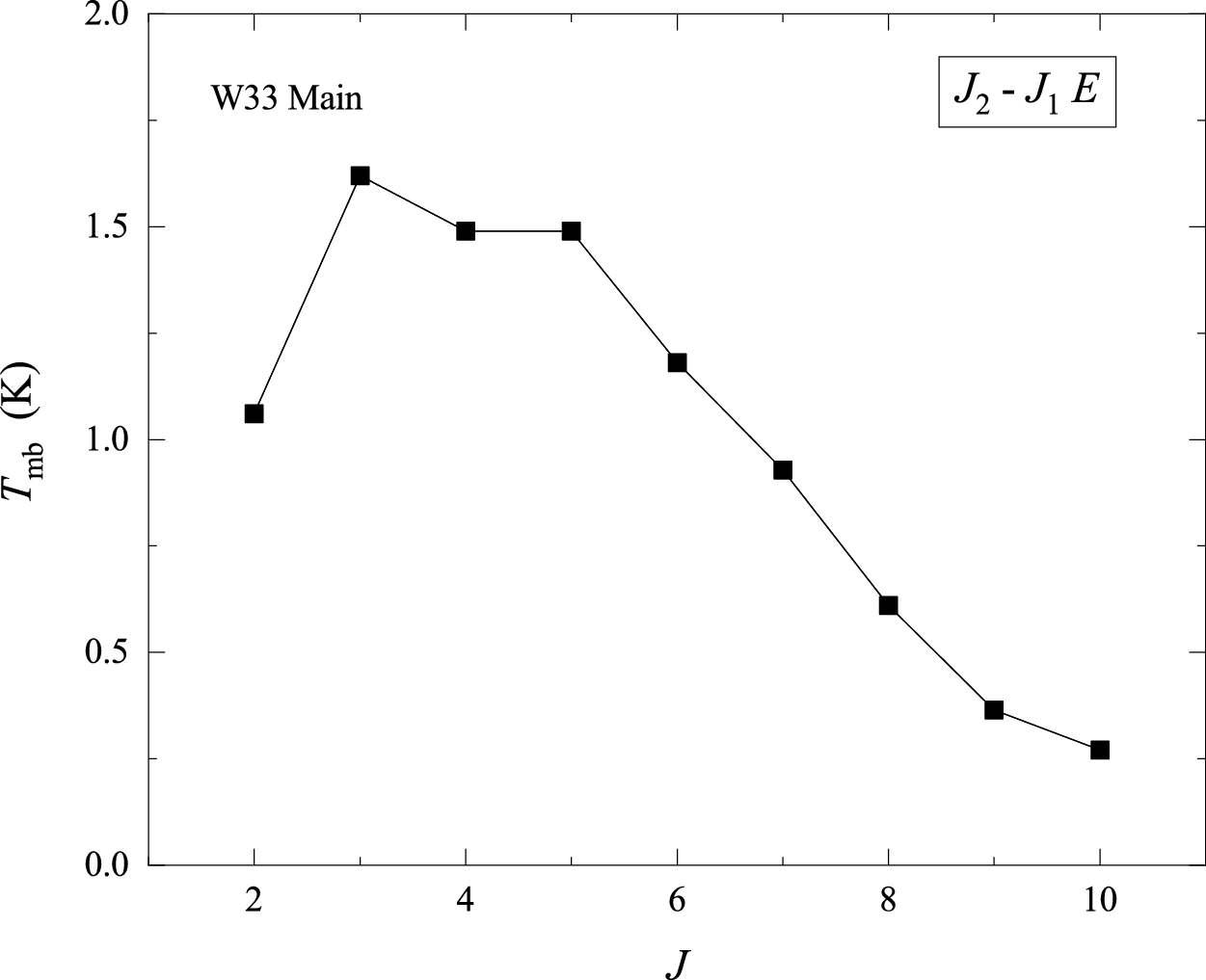}
\hspace{0.2cm}
\includegraphics[width=0.23\textwidth]{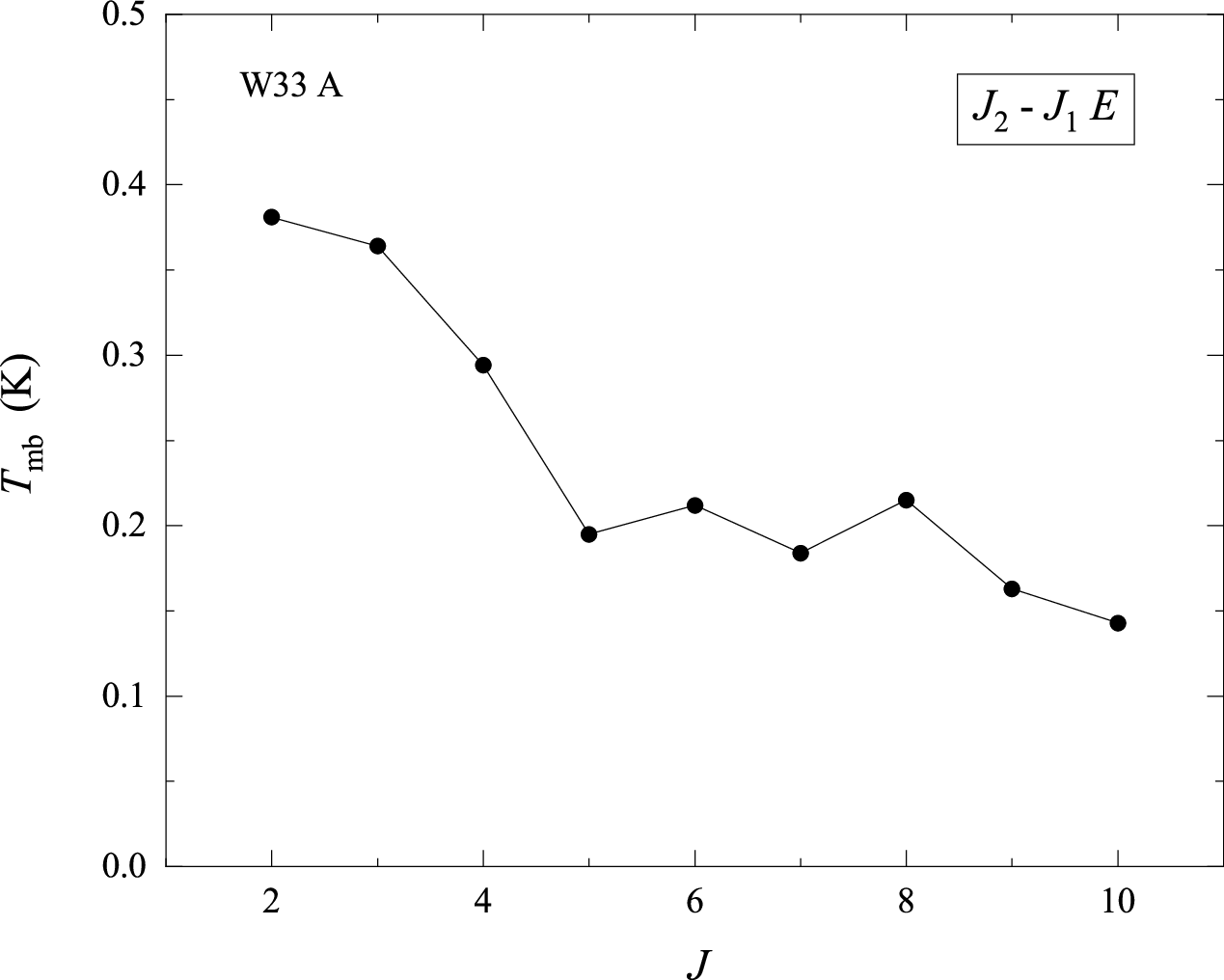}
\includegraphics[width=0.23\textwidth]{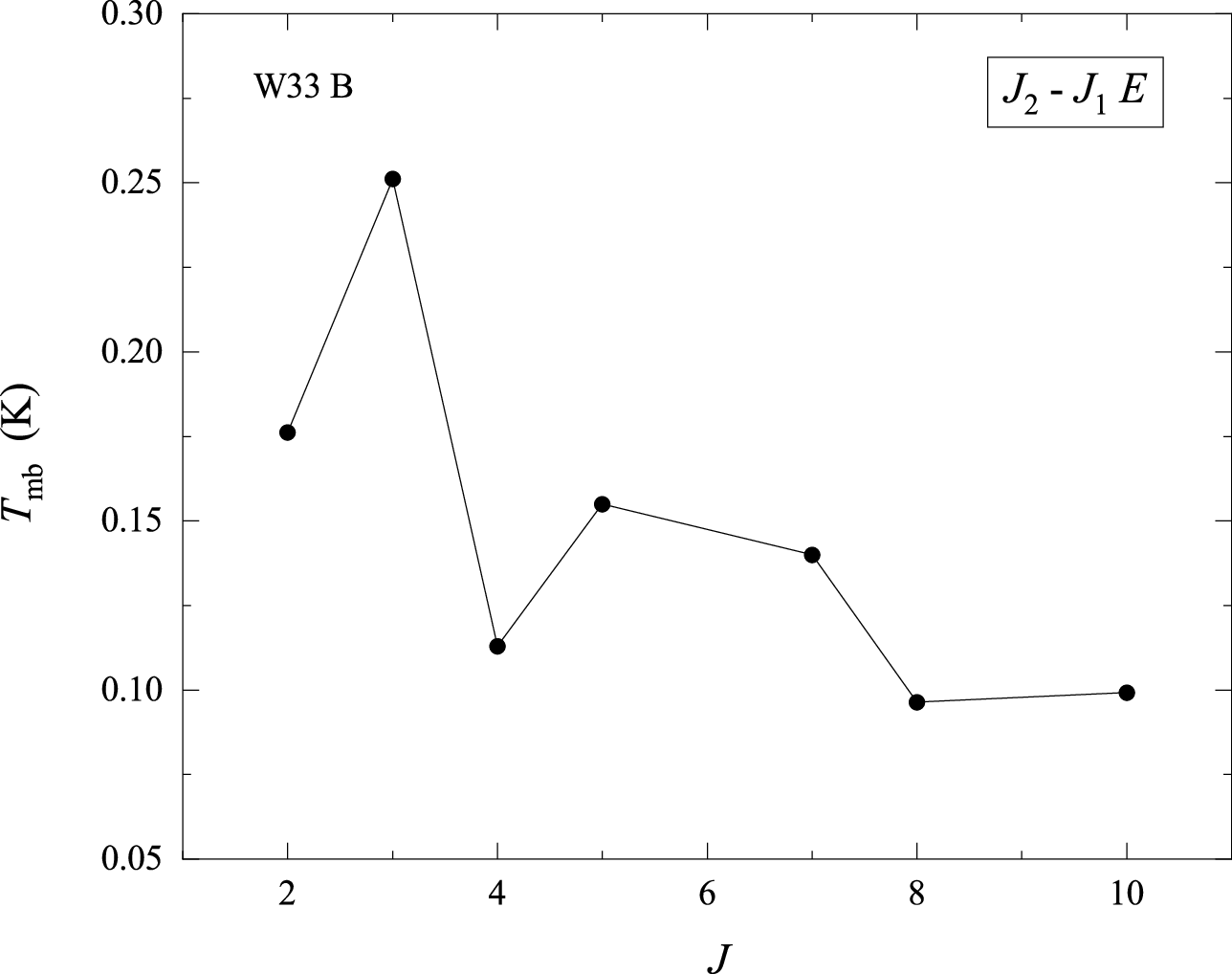}
\hspace{0.2cm}
\includegraphics[width=0.23\textwidth]{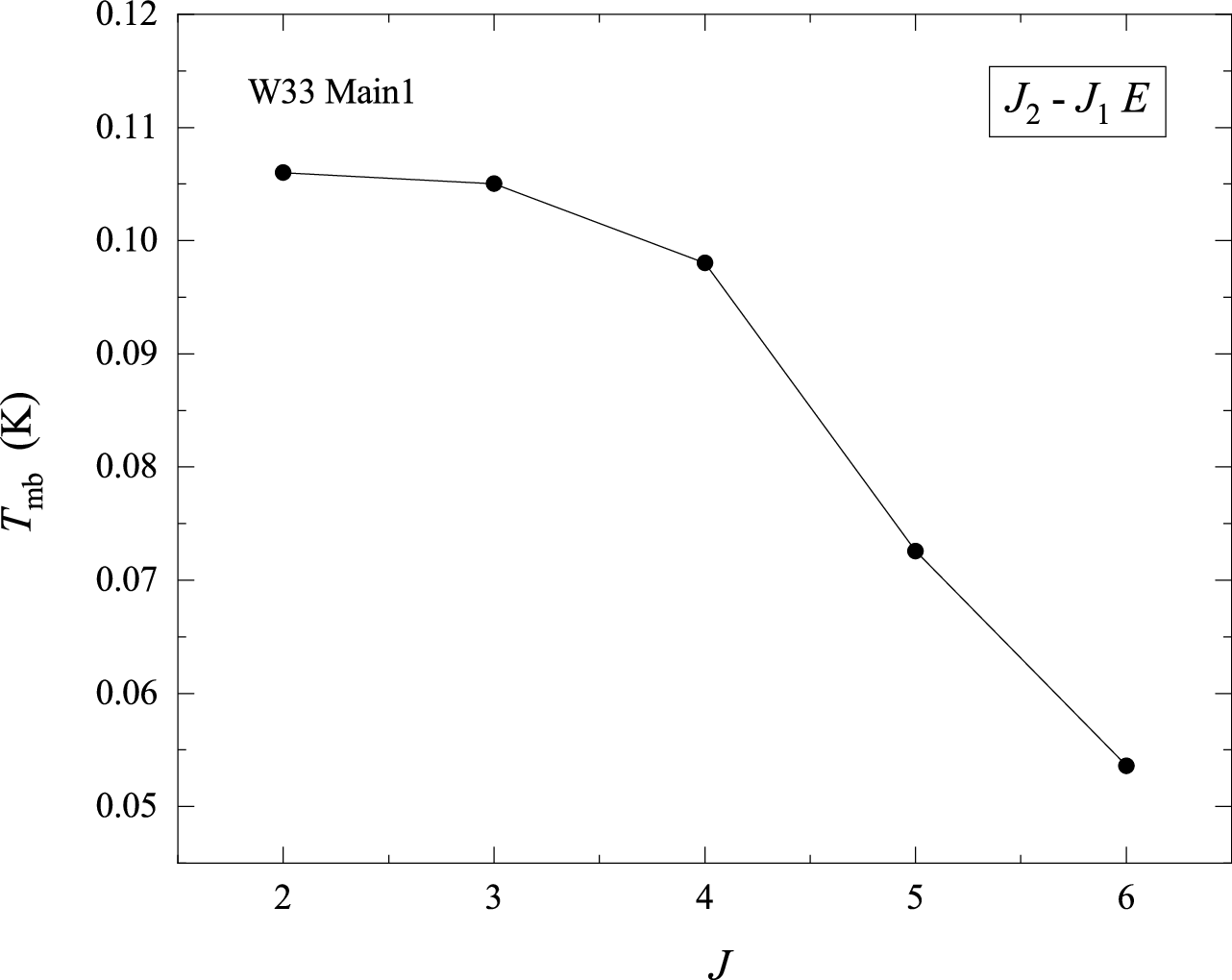}
\caption[]{Observed peak intensities of CH$_{3}$OH ($J_{2}-J_{1}$~$E$) transitions plotted against the rotational quantum number $J$. The panels are arranged as follows: \textit{Top left} for  W33\,Main, \textit{top right} for W33\,A, \textit{bottom left} for W33\,B, and \textit{bottom right} for W33\,Main1.}
\label{Fig:6}
\end{figure*}

\subsubsection{Methanol}
\label{sect-3-3-4}
In total, 218 positions were observed in an approximately fully sampled mode (spacing: $20\arcsec$; see Sect.\,\ref{sect:Observation1}). Ten to 15 positions revealed a detection at a 3$\sigma$ level in case of the CH$_3$OH\,(2$_{2,1}$--2$_{1,2}$ E), CH$_3$OH\,(3$_{2,1}$--3$_{1,2}$ E), CH$_3$OH\,(4$_{2,4}$--4$_{1,3}$ E), CH$_3$OH\,(5$_{2,3}$--5$_{1,4}$ E), CH$_3$OH\,(6$_{2,4}$--6$_{1,5}$ E), CH$_3$OH\,(7$_{2,5}$--7$_{1,6}$ E), and CH$_3$OH\,(8$_{2,6}$--8$_{1,7}$ E) lines. Figs.\,\ref{Fig:4} and \ref{Fig:5} display the integrated intensity maps for the different CH$_3$OH transition lines, covering a velocity interval of 32\,$<$\,$V_{\rm LSR}$\,$<$\,40\,km\,s$^{-1}$. Figs.\,\ref{Fig:4} and \ref{Fig:5} indicate that inside of W33\,Main the CH$_3$OH transition lines can only be identified within the central compact area. Each panel features the half-power beam width represented by a gray circle in the bottom-left corners of the images. The boundaries of the mapped area are demarcated by gray dashed lines.

In this survey, we find ten CH$_{3}$OH transitions toward W33\,Main, including the nine $J_{2}-J_{1}$~$E$ methanol transitions with quantum number $J$ ranging from $J$\,=\,2 to 10. From Fig.\,\ref{Fig:6} (\textit{top-left} panel), our measurements toward W33\,Main show that the peak intensities of these transitions increase with principle quantum number $J$ from 2 to 3 and decrease with $J$ from 3 to 10. In the W33\,Main region, E-type methanol masers have been detected in the transitions $J_{2}-J_{1}$ for $J$\,=\,2--9 (e.g., \citealt{1986A&A...157..318M}), which include the lines (2$_{2,1}$--2$_{1,2}$ E) to (9$_{2,7}$--9$_{1,8}$ E). \citet{1986A&A...157..318M} reported that the W33\,Main source produces strong, narrow-line emission in these transitions, which they identified as masers. In our survey, we have discovered a CH$_{3}$OH\,(10$_{2,8}$--10$_{1,9}$ E) maser. We identify this as a maser for two primary reasons. First, the line profile has a line width similar to that of the confirmed maser transitions with $J$\,$\textless$\,10. Second, the level placements for the $J$\,=\,10 states, as seen in methanol energy level schemes (for example, in \citealt{2016A&A...592A..31L}), are very similar to those of the $J$\,$\textless$\,10 transitions that are known to be masers. This interpretation is supported by Fig.\,3 in \citet{2016A&A...592A..31L}, where the $J=10$ levels are connected by red dotted lines, indicating that the associated transition has previously been identified as a maser. Maser emission in the CH$_{3}$OH\,(10$_{2,8}$--10$_{1,9}$ E) transition has been detected in Orion \citep{1998ApJ...498..763S,2016A&A...592A..31L}.

Toward W33\,A, we detected nine CH$_{3}$OH\,($J_{2}-J_{1}$~$E$) transitions. Our measurements of nine $J_{2}-J_{1}$~$E$ transitions show that there is a steep decay in intensity from $J$\,=\,2 to 5 followed by a shallow one at higher $J$ (see Fig.\,\ref{Fig:6} \textit{top-right} panel). For W33\,B, we detected seven CH$_{3}$OH\,($J_{2}-J_{1}$~$E$) transitions. Our measurements show that the peak intensities of these transitions increase with principle quantum number $J$ from 2 to 3 to then decay at higher $J$ (see Fig.\,\ref{Fig:6} \textit{bottom-left} panel). Toward W33\,Main1, we find six CH$_{3}$OH transitions. The figure including five $J_{2}-J_{1}$~$E$ lines shows that the peak intensities decrease with principal quantum number $J$ from 2 to 6 (see Fig.\,\ref{Fig:6} \textit{bottom-right} panel). The spectral line parameters for the CH$_{3}$OH transition lines are provided in Table\,\ref{Tab:A.3}.

In W33\,A, W33\,B, and W33\,Main1, the CH$_{3}$OH\,($J_{2}-J_{1}$~$E$) line emission is not strong, with peak main-beam brightness temperatures of $T_{\rm mb}$\,$\textless$\,0.50\,K (see Table\,\ref{Tab:A.3}). Investigating the properties of all $J_{2}-J_{1}$ E-type transitions in W33\,A, W33\,B, and W33\,Main1, we use the maserdb.net\footnote[7]{https://maserdb.net/} database and previous studies. The $J$\,=\,5, $J_{2}-J_{1}$ E-type methanol maser transition at 24.959\,GHz was already detected in the W33\,B ($G12.68-0.18$) region (e.g., \citealt{2017ApJS..230...22T}) and was judged as emission resulting from inverted level populations. Motivated by the question of whether the other CH$_{3}$OH\,($J_{2}-J_{1}$~$E$) transitions in W33\,A, W33\,B, and W33\,Main1 are also maser lines, we calculated their excitation temperatures ($T_{\rm ex}$) and optical depths ($\tau$) using the RADEX radiative transfer code (as implemented by \citealt{2007A&A...468..627V}). We determined the methanol $T_{\rm ex}$ and $\tau$ using independent values of H$_{2}$ volume density ($n_{\rm H_{2}}$) and kinetic temperature ($T_{\rm kin}$) for each source in W33. For W33\,Main, $n_{\rm H_{2}}$\,=\,1.2\,$\times$\,10$^{4}$, $T_{\rm kin}$\,=\,30\,K. For W33\,A, $n_{\rm H_{2}}$\,=\,2.6\,$\times$\,10$^{4}$, $T_{\rm kin}$\,=\,16\,K. For W33\,B, $n_{\rm H_{2}}$\,=\,1.1\,$\times$\,10$^{4}$, $T_{\rm kin}$\,=\,13\,K. Toward W33\,Main1, $n_{\rm H_{2}}$\,=\,1.8\,$\times$\,10$^{4}$, $T_{\rm kin}$\,=\,13\,K. These values are based on NH$_3$ results from \citet{2022A&A...658A..34T}. The results indicate that all obtained $T_{\rm ex}$ and $\tau$ values are negative (see Table\,\ref{Tab2}). The negative $T_{\rm ex}$ of the different transitions show variations ranging from -7.1\,K to -1.2\,K with differences between most transitions amounting to approximately 2\,K. We conclude that all these CH$_{3}$OH\,($J_{2}-J_{1}$~$E$) transitions are masers arising from population inversion. This interpretation is supported by the following evidence. The entire set of CH$_{3}$OH\,$J_{2}-J_{1}$ transitions is either inverted or not, depending on the relative abundance of the $k$\,=1 and 2 stacks, as these ladders have a unique abundance ratio \citep{2016A&A...592A..31L}. With all observed $J_{2}-J_{1}$ E-type transitions in W33\,Main being identified as masers (see the previous paragraph and \citealt{1986A&A...157..318M}), we conclude that the CH$_{3}$OH\,$J_{2}-J_{1}$ emission in W33\,A, W33\,B, and W33\,Main1 are also arising from inverted level populations.

\subsubsection{Other molecules}
\label{sect-3-3-5}
\paragraph{HNCO:} The emission from isocyanic acid HNCO\,(1$_{0,1}$\,--\,0$_{0,0}$) at 21.9 GHz  is detected in W33\,Main, W33\,A, and W33\,B, as shown in Figs.\,\ref{FigA.9}, \ref{FigA.10}, and \ref{FigA.11}. Table\,\ref{Tab:A.5} lists the spectral line parameters of the HNCO transition. HNCO serves as an effective tracer for the evolutionary state of starbursts and the presence of shocks \citep{2009ApJ...694..610M}. For instance, the abundance of HNCO varies by nearly two orders of magnitude between young and old starbursts, making it an excellent diagnostic tool for assessing their evolutionary status \citep{2009ApJ...694..610M}. The abundance of HNCO in galaxies is primarily dominated by two factors: shocks in the early stages of massive cloud formation and strong UV fields in the evolved stages \citep{2009ApJ...694..610M}. The shocks in W33\,Main may originate from the swept-up gas of the H\,{\scriptsize II} region in W33\,Main. To determine the HNCO\,(1$_{0,1}$\,--\,0$_{0,0}$) column density and abundance in W33\,Main, we used the RADEX radiative transfer code \citep{2007A&A...468..627V}. We adopted physical parameters from \citet{2022A&A...658A..34T} ($T_{\rm kin}$\,=\,30\,K, $n_{\rm H_{2}}$\,=\,1.2\,$\times$\,10$^{4}$\,cm$^{-3}$). Using the observed line width ($\Delta v$\,=\,9.1\,km\,s$^{-1}$) and main beam brightness temperature ($T_{\rm mb}$\,=\,0.04\,K; see Table\,\ref{Tab:A.5}) of the HNCO\,(1$_{0,1}$\,--\,0$_{0,0}$) transition, the model yields a column density of $N$\,=\,2.3\,$\times$\,10$^{13}$\,cm$^{-2}$. The resulting fractional abundance of HNCO relative to H$_2$, defined as $\chi$\,(CH$_{3}$OH)\,=\,$N$(CH$_{3}$OH)\,/\,$N$(H$_{2}$), is 4.9\,$\times$\,10$^{-11}$ at the peak positions of the W33\,Main. 

\paragraph{CH$_{3}$CN:} In our survey, we find the CH$_{3}$CN\,($1_{0}-0_{0}$) emission line toward W33\,Main, W33\,A, W33\,B, and W33\,Main1. The results of our CH$_{3}$CN observations are displayed in  Figs.\,\ref{FigA.9}, \ref{FigA.10}, \ref{FigA.11}, and \ref{FigA.12}, with the Gaussian line parameters summarized in Table\,\ref{Tab:A.5}. CH$_{3}$CN was first detected in Orion\,KL and Sagittarius\,B2 \citep{1975ApJ...198L..81S}. CH$_{3}$CN serves as a tracer for hot cores in high-mass star-forming regions \citep{2017A&A...602A..59C}, and it has also been identified as an indicator of dense gas and shocks, offering a more comprehensive perspective on chemical abundances \citep{2015A&A...579A.101A}.

\paragraph{c-C$_{3}$H$_{2}$:} The c-C$_{3}$H$_{2}$ (cyclopropenylidene) molecule, first detected in space toward Sgr\,B2 by \citet{1985ApJ...299L..63T}, is a small cyclic hydrocarbon commonly observed in interstellar environments. Following its initial discovery, c-C$_{3}$H$_{2}$ has been observed across various astrophysical environments, such as diffuse and dense clouds, protostellar regions, planetary nebulae, and even extragalactic sources \citep{1985ApJ...298L..61M,1989AJ.....97.1403M,1987A&A...181L..19C,1988A&A...206..108C,1989A&A...209..382C,2001ApJ...552..168F,2009ApJ...704L.108T,2012ApJ...753L..28L}, and it ranks among the most abundant molecules within the interstellar medium \citep{2004A&A...417..135T}. The c-C$_{3}$H$_{2}$ molecule can be used to trace the chemical evolution of carbon isotope ratios across the Galaxy, and it also serves as an important indicator for revealing the initial chemical conditions during star formation \citep{1989AJ.....97.1403M}. In our band, we have detected the c-C$_{3}$H$_{2}$\,(1$_{1,}$$_{0}$-1$_{0,}$$_{1}$) transition at 18.3\,GHz in all six W33 regions: W33\,Main, W33\,A, W33\,B, W33\,A1, W33\,B1, and W33\,Main1 (see Figs.\,\ref{FigA.9}, \ref{FigA.10}, \ref{FigA.11}, and \ref{FigA.12}). Additionally, we observed a single c-C$_{3}$H$_{2}$\,(2$_{2,}$$_{0}$-2$_{2,}$$_{1}$) absorption line at 21.6\,GHz toward W33\,Main (see Fig.\,\ref{FigA.9}). The results of the Gaussian fits are shown in Table\,\ref{Tab:A.5}. The (1$_{1,}$$_{0}$-1$_{0,}$$_{1}$) emission and (2$_{2,}$$_{0}$-2$_{2,}$$_{1}$) absorption line  transitions of c-C$_{3}$H$_{2}$ are also observed in other star-forming regions, for example, towards the TMC-1 (see Fig.\,2 of \citealt{1989AJ.....97.1403M}).

Figure\,\ref{FigB.2} shows the integrated intensity maps for the c-C$_{3}$H$_{2}$\,(1$_{1,}$$_{0}$-1$_{0,}$$_{1}$) and (2$_{2,}$$_{0}$-2$_{2,}$$_{1}$) transitions in the W33\,Main and W33\,Main1 regions. It is clear from the Figures that the c-C$_{3}$H$_{2}$\,(1$_{1,}$$_{0}$-1$_{0,}$$_{1}$) line shows an extended distribution, while the (2$_{2,}$$_{0}$-2$_{2,}$$_{1}$) transition can be found in the densest part.

\paragraph{SiS:} Silicon sulfide (SiS) is an important interstellar molecule with unique properties. \citet{1975ApJ...199L..47M} reported its first detection in the circumstellar environment of IRC+10216. In the carbon-rich circumstellar envelope of IRC\,+\,10216, even SiS masers have been detected, indicating that its radiation has high brightness and directionality, serving as an important probe for studying the physical conditions of the circumstellar envelopes of late type stars \citep{2006ApJ...646L.127F,2015A&A...574A..56Gb,2017ApJ...843...54G,2018ApJ...860..162F}. We may have detected the SiS $J$ = 1\,-\,0 line in the W33\,Main1 region (see Fig.\,\ref{FigA.12}). The observed line parameters are given in Table \ref{Tab:A.4}.

\paragraph{CCS:} Due to its abundance in low temperature environments, CCS is often used as a tracer of cold dark molecular clouds. The CCS emission was first observed in the Orion\,A Gaint Molecular cloud (GMC) by \citet{2010PASJ...62.1473T}. CCS, a carbon-chain molecule, is abundant in the early stages of chemical evolution, while NH$_{3}$ becomes more abundant in later stages \citep{2021SCPMA..6479511X}. Therefore, in cold dark clouds, the CCS to NH$_{3}$ abundance ratio is a good tracer of the chemical evolution \citep{2021SCPMA..6479511X}. In our survey, we detected one CCS line in the W33\,Main1 region (see Fig.\,\ref{FigA.12}). The observed line parameters can be found in Table \ref{Tab:A.4}.

\begin{figure}
\centering
 \includegraphics[width=0.45\textwidth]{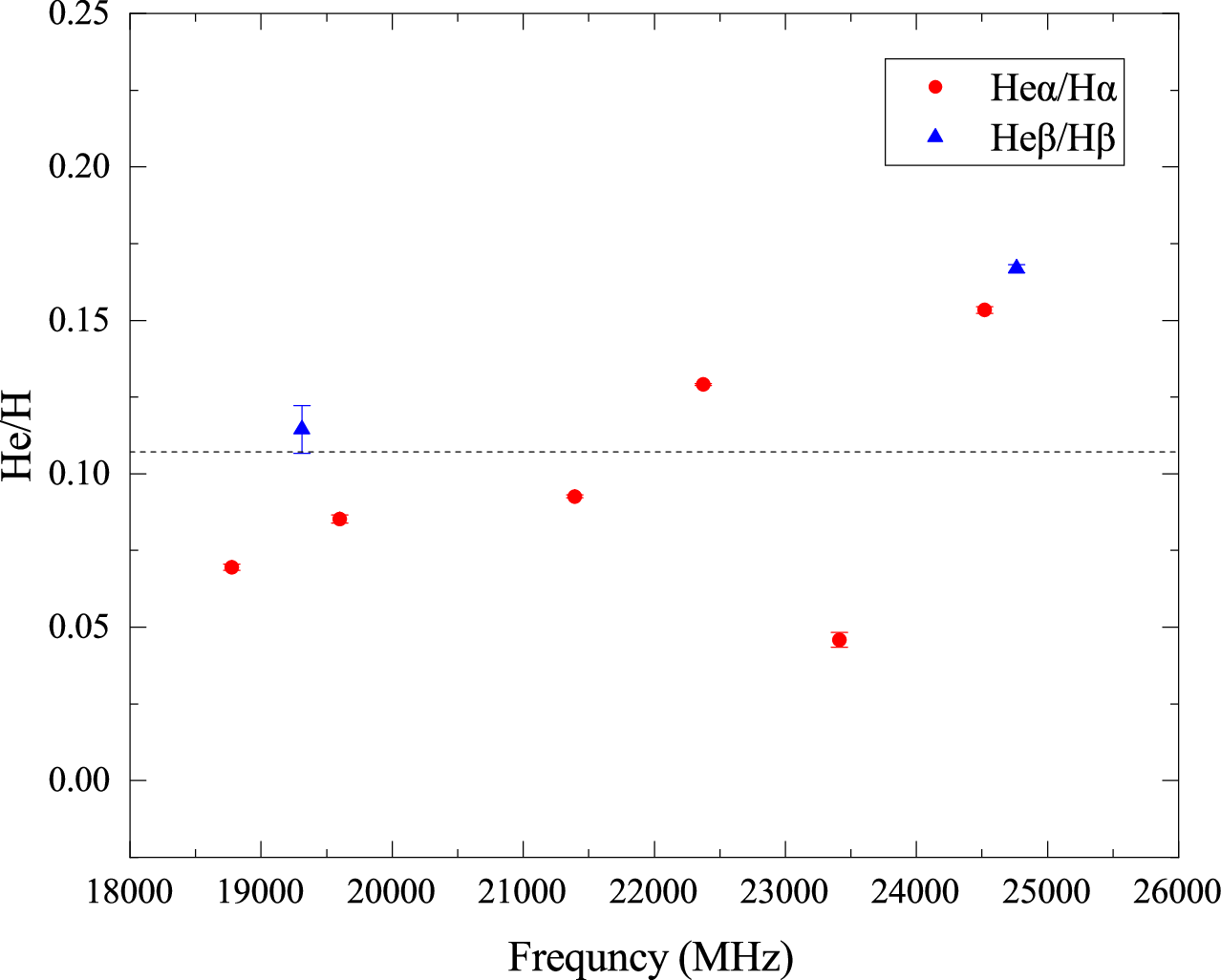}
 \caption{He/H abundance ratios derived from RRLs as a function of rest frequency. The abundance ratios derived from He$\alpha$/H$\alpha$ and He$\beta$/H$\beta$ are marked with red filled circles and blue filled triangles, respectively. The dashed line represents the sigma-weighted mean value of He/H abundance ratios. Each data point has a small error, ranging from 4.9 $\times10^{-4}$ to 1.3 $\times10^{-3}$, with an average of 1.6 $\times10^{-3}$.}
 \label{Fig:7}
 \end{figure}

\begin{figure*}[t]
\centering
\includegraphics[width=0.33\textwidth]{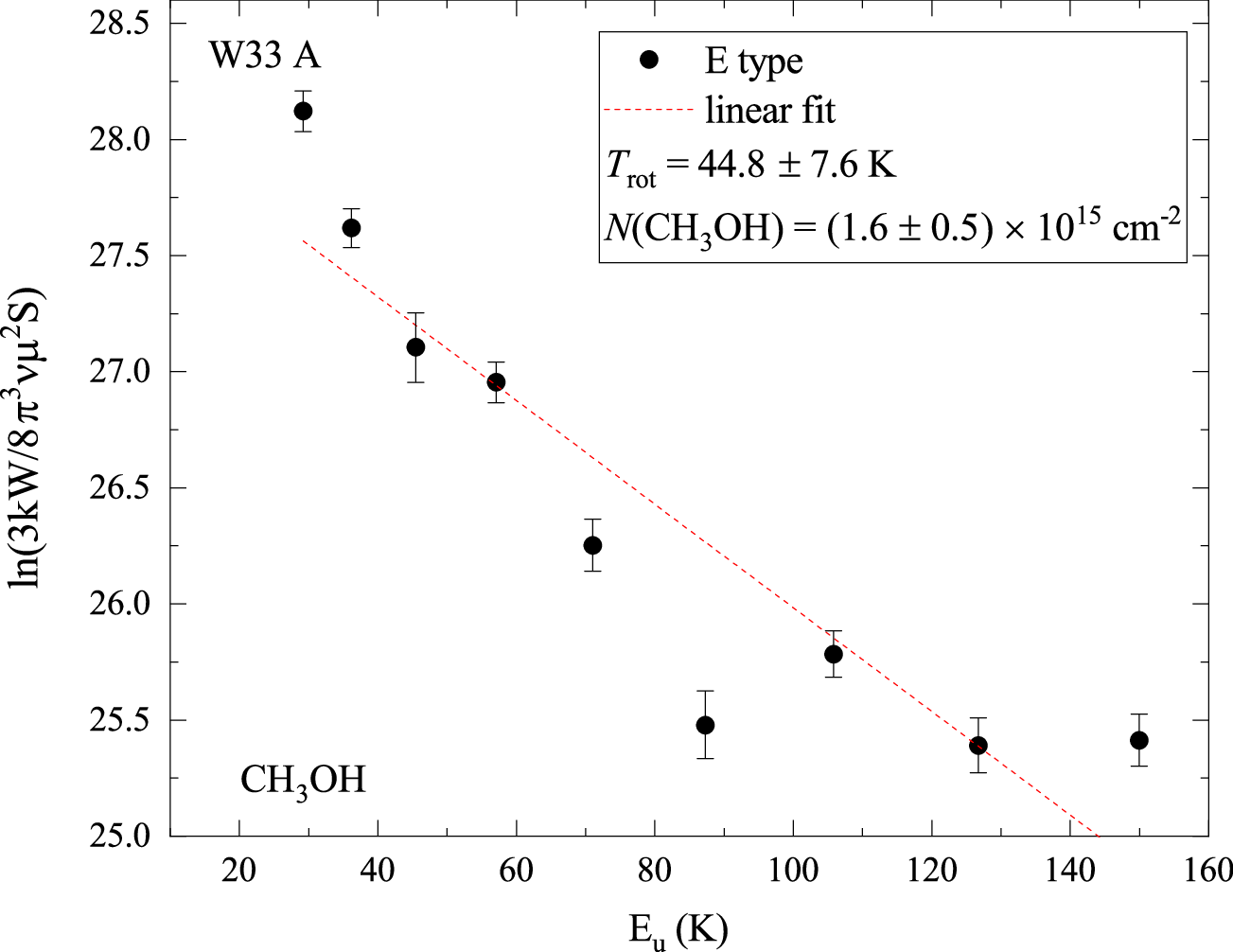}
\includegraphics[width=0.33\textwidth]{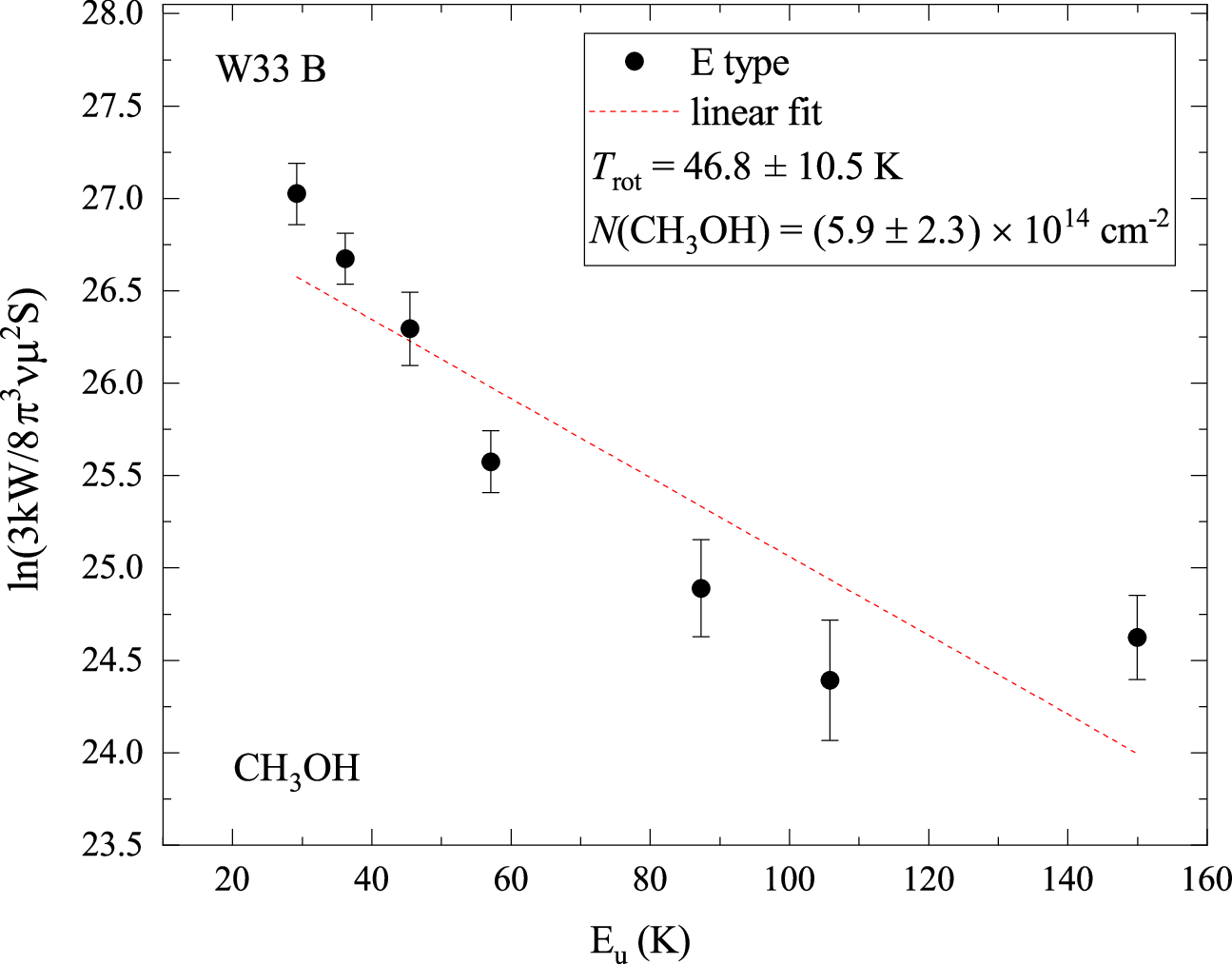}
\includegraphics[width=0.33\textwidth]{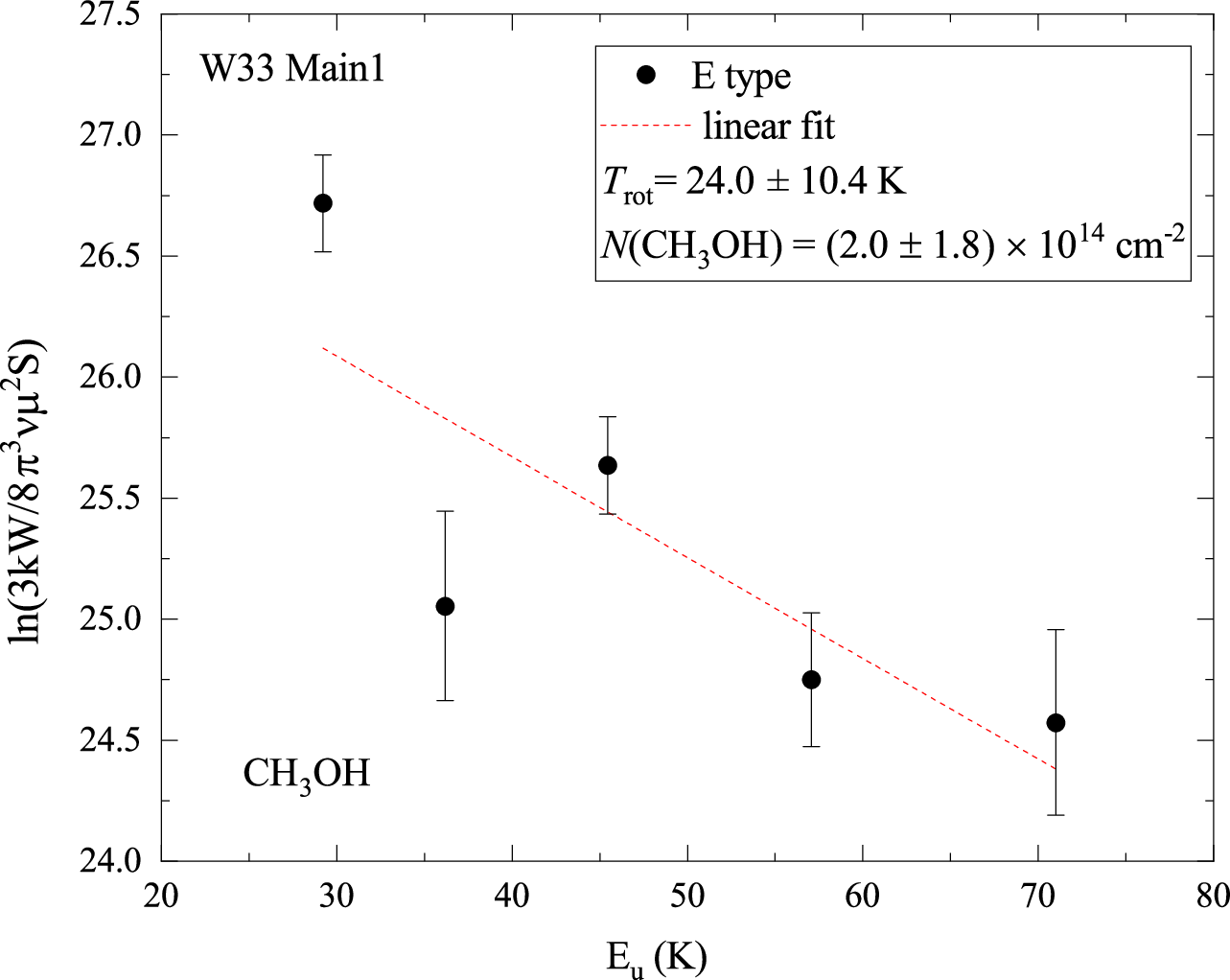}
\caption[]{Rotation diagrams for the weak inverted lines of CH$_{3}$OH. The Boltzmann plots display the W33\,A emission lines (\textit{left}), W33\,B emission lines (\textit{middle}), and W33\,Main1 emission lines (\textit{right}). The positions taken are those of Table\,\ref{table:1}. The dashed lines indicate the fitting for E-type transitions. The rotation temperatures were obtained from the corresponding slopes. The displayed errors were calculated via error propagation.}
\label{Fig:8}
\end{figure*}

\section{Discussion}
\label{sect:discussion}
\subsection{He/H abundance ratio in W\,33}
\label{4.1}
By utilizing the large number of RRLs, we could employ hydrogen and helium RRL pairs to determine the He$^{+}$/H$^{+}$ abundance ratio, which in turn allowed us to assess the He/H abundance ratio under LTE conditions. Adopting LTE conditions, the He/H abundance ratio can be determined by utilizing the integrated intensity from Gaussian fittings applied to the hydrogen and helium RRLs \citep[e.g.,][]{1974A&A....32..283C}:
\begin{equation}\label{f2}
y= \frac{N({\rm He})}{N({\rm H})} \simeq \frac{1}{R} \frac{\int_{\Omega_{\rm s}} {\rm d}\Omega \int S_{\nu}({\rm He^{+}}){\rm d}\upsilon}{\int_{\Omega_{\rm s}} {\rm d}\Omega \int S_{\nu}({\rm H^{+}}){\rm d}\upsilon} \;,
\end{equation}
where $R$ represents the volume ratio of the He$^{+}$ to H$^{+}$ Str\"omgren spheres, weighted by the square of the proton density. The term $\Omega_{\rm s}$ denotes the source’s solid angle, while $\int S_{\nu}({\rm He}^{+}){\rm d}\upsilon$ and $\int S_{\nu}({\rm H}^{+}){\rm d}\upsilon$ are the integrated intensities observed for the respective helium and hydrogen RRLs. If we consider that the H{\scriptsize II} region is equivalent in size to the He{\scriptsize II} region, then equation\,(\ref{f2}) simplifies to
\begin{equation}\label{f3}
y= \frac{N({\rm He})}{N({\rm H})}\simeq \frac{\int S_{\nu}({\rm He^{+}}){\rm d}\upsilon}{\int S_{\nu}({\rm H^{+}}){\rm d}\upsilon} \;.
\end{equation}
For this study we utilized eight distinct unblended pairs of lines originating from the H$\alpha$/He$\alpha$ and H$\beta$/He$\beta$ transitions. The rest frequencies of these pairs differ by approximately 10\,MHz, which implies that their ratios should be unaffected by uncertainties associated with pointing accuracy, calibration discrepancies, and deviations in beam width. Figure\,\ref{Fig:7} displays the calculated He/H abundance ratios based on equation\,(\ref{f3}). 

Our analysis revealed an increase in uncertainties from the H$\alpha$/He$\alpha$ to the H$\beta$/He$\beta$ ratios. Consequently, we adopted the sigma-weighted mean to determine the He/H abundance ratio, which is estimated to be (10.7\,$\pm$\,1.8)\%. This value is in good agreement with findings from earlier research \citep{1992ApJ...395..484P} and is marginally higher than the (9.1\,$\pm$\,0.5)\% previously reported for H70$\alpha$ and He70$\alpha$ measurements directed at the H{\scriptsize II} region’s peak \citep{1980A&A....87..269T}. Hence we validated, in spite of the differences encountered when analyzing turbulent velocity dispersions (Sect.\,\ref{sect-3-2} and Fig.\,\ref{Fig:1}), that the He/H abundance ratio aligns closely with the primordial He/H abundance ratio. The latter is approximately 8.3\%, as predicted by Big Bang nucleosynthesis (\citealt{2004ApJ...617...29O}), and has a solar abundance ratio of about 9.8\% \citep{1994ARA&A..32..191W}. This suggests that the He/H abundance ratio is predominantly set by Big Bang nucleosynthesis, with minimal influence from stellar nucleosynthesis processes. Additionally, the scatter observed between the ratios calculated from H$\alpha$/He$\alpha$ pairs and those from H$\beta$/He$\beta$ pairs is minimal (see Fig.\,\ref{Fig:7}), indicating that the departure coefficients, or the $b_{\rm n}$ factors, are largely consistent for both hydrogen and helium. This finding is in strong concordance with theoretical forecasts (e.g., \citealt{1995MNRAS.272...41S}).

\subsection{A comparison of turbulent velocity differences between He and H in H\,{\scriptsize II} regions}
\label{4.2}
As described in Sect.\,\ref{sect-3-2}, the turbulent line widths of helium are slightly greater than those of hydrogen.  This is not a common phenomenon in star-forming regions. For example, \citet{2015A&A...581A..48Ga} reported that the turbulent line widths of hydrogen are somewhat broader than those of helium in Orion\,A. They suggest that hydrogen RRLs trace a more extensive region of ionized gas compared to helium RRLs in Orion\,A. \citet{2006ApJS..165..338Q} demonstrate that the two H\,{\scriptsize II} regions, G10.315--0.15 and G81.681$+$0.54, exhibit turbulent line widths with  $\Delta \upsilon$\,(H)\,$\sim$\,2\,$\Delta \upsilon$\,(He). In W33\,Main, the difference in turbulent line widths between helium and hydrogen ranges from 5.1 to 12.9\,km\,s$^{-1}$ with an average of 9.3\,$\pm$\,3.5\,km\,s$^{-1}$. Studies often find that $\Delta \upsilon$\,(H)\,$>$\,$\Delta \upsilon$\,(He). However, in W33\,Main, we observe $\Delta \upsilon$\,(H)\,$<$\,$\Delta \upsilon$\,(He) for turbulent motions. This discrepancy likely arises from spatially unresolved fine structure that cannot be disentangled with the current data. Given that our measured He/H ratio, assuming $R$ = 1 (see eqs.\,\ref{f2} and \ref{f3}) is consistent with the normal abundance ratios (see Sect.\,\ref{4.1}), the inhomogeneities causing the different turbulent line widths do not appear to be dominant. However, a deeper study will require observations at a much higher angular resolution.

\begin{figure*}[t]
\centering
\includegraphics[width=0.32\textwidth]{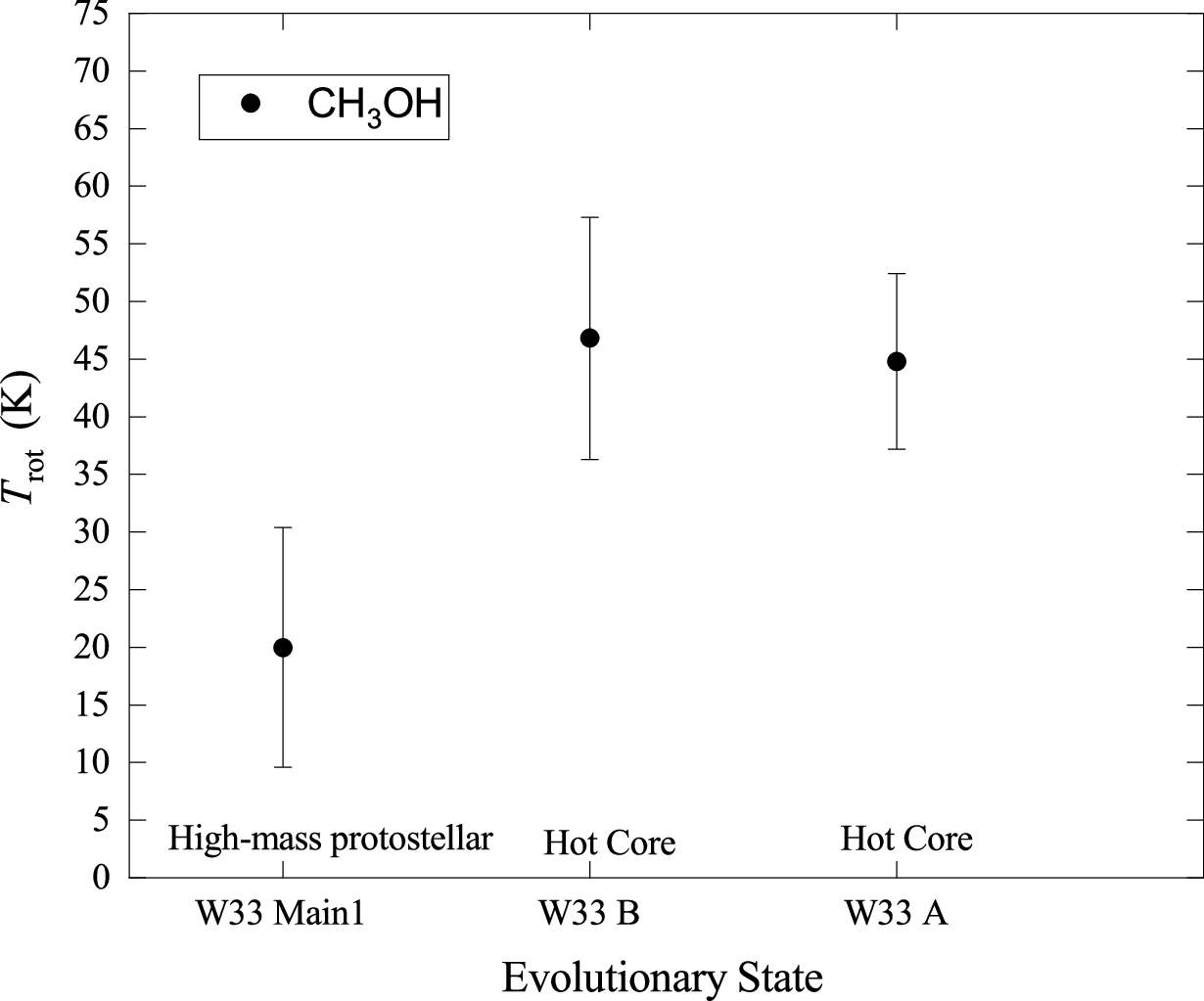}
\includegraphics[width=0.33\textwidth]{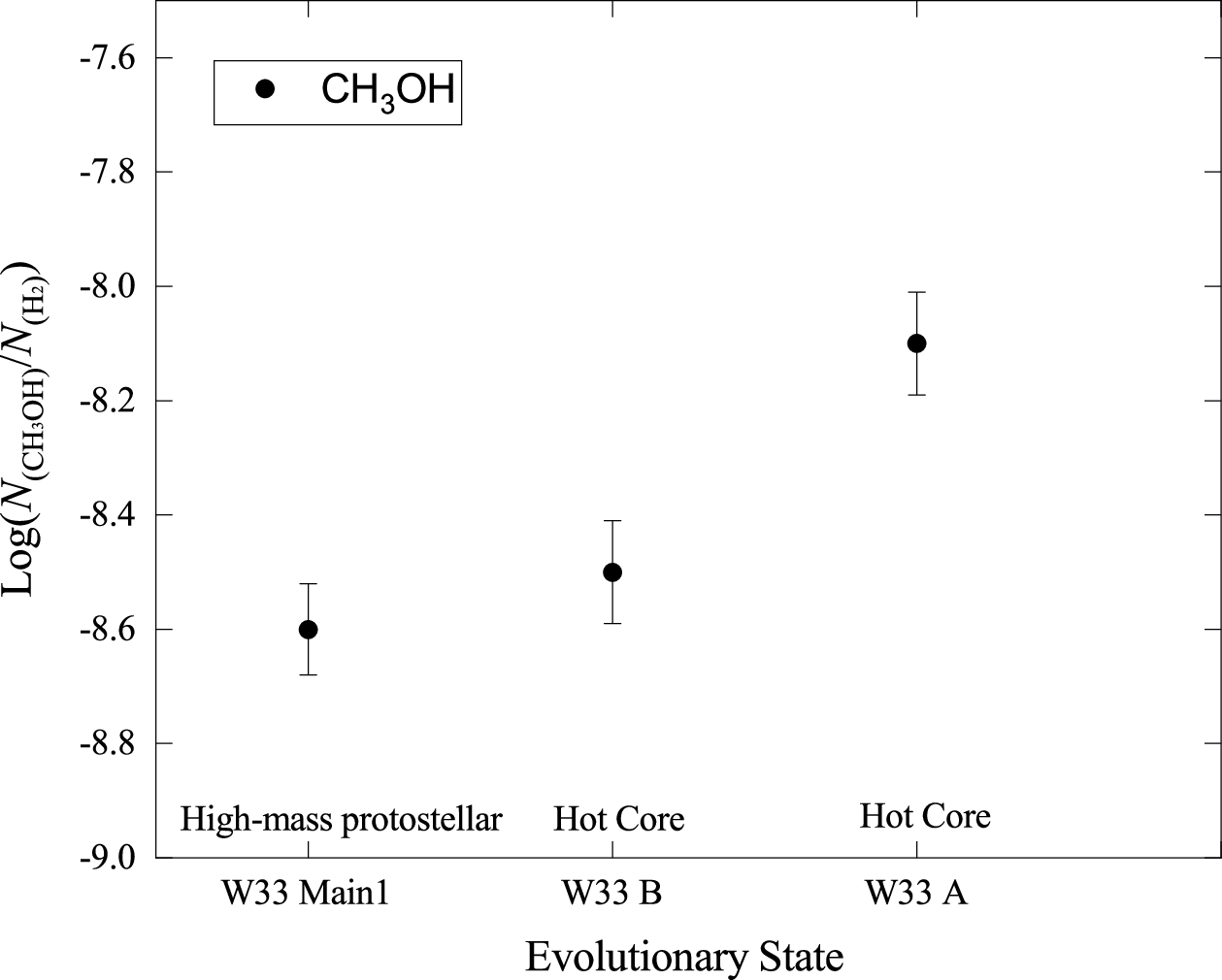}
\includegraphics[width=0.32\textwidth]{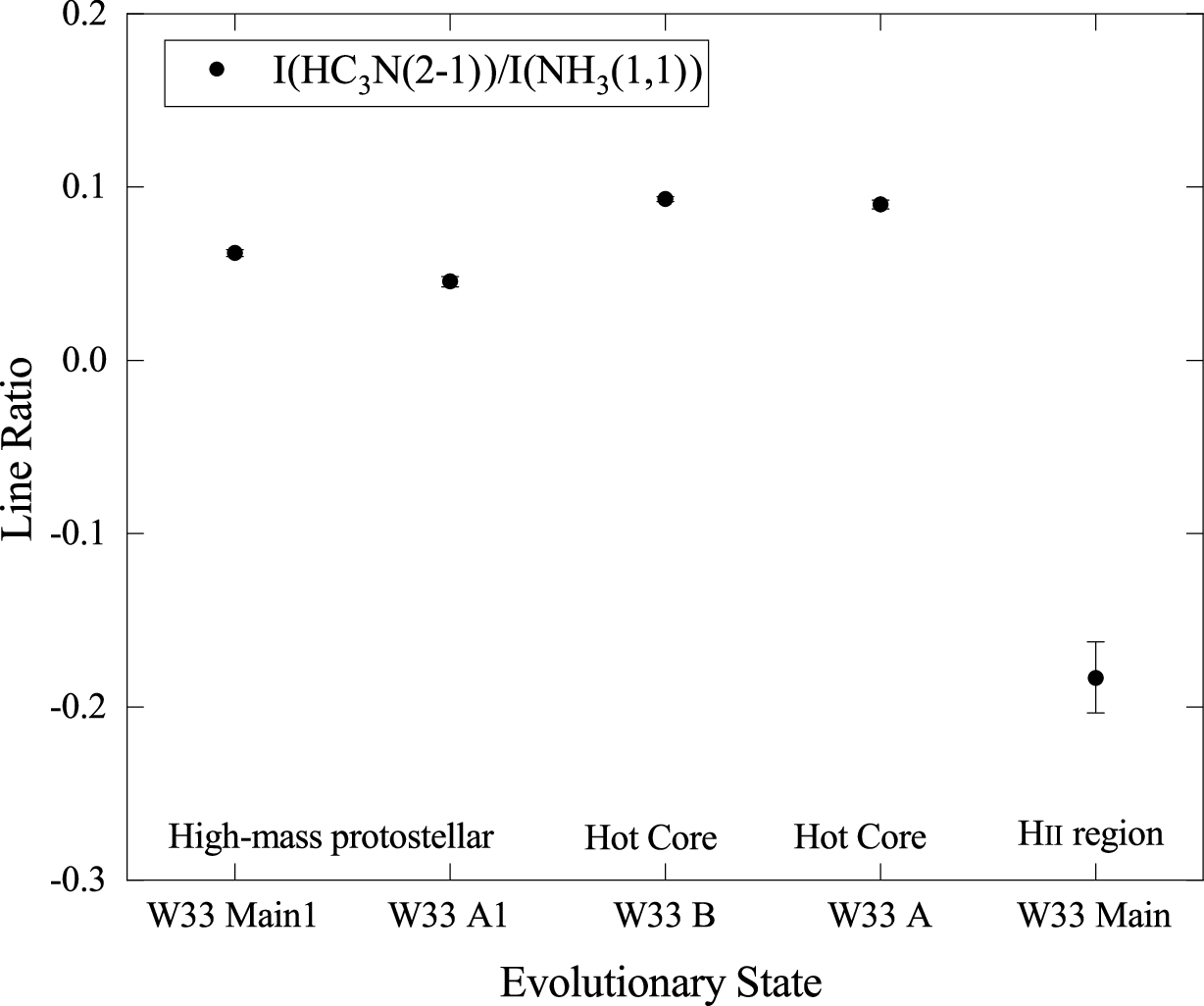}
\caption[]{Rotation temperatures derived from CH$_{3}$OH (\textit{left panel}); fractional CH$_{3}$OH abundances, $N$(CH$_{3}$OH)/$N$(H$_{2}$) (\textit{middle panel}); and integrated HC$_{3}$N\,(2-1)/NH$_{3}$\,(1,1) intensity ratios (\textit{right panel}) versus the evolutionary sequence of three to five W33 sources. For the chosen positions, see Table\,\ref{table:1}. The exact source classification, where known, is indicated above the horizontal axis (see Table\,\ref{table:1}, column\,6).}
\label{Fig:9}
\end{figure*}

\subsection{D/H abundance ratio in W33\,B1}                                                 
\label{4.3}
The D/H abundance is a key parameter for our understanding of cosmology and molecular fractionation processes \citep[e.g.,][]{1985ARA&A..23..319B,1994ARA&A..32..191W,2007ARNPS..57..463S}.
The deuterium fraction can be determined by analyzing the ratio of the column densities of para-NH$_{3}$ to para-NH$_{2}$D. We calculated the column density of para-NH$_{2}$D using the method described in \citet{1987A&A...172..311W}, which is
\begin{equation}\label{D}
N_l = 1.67\times10^{14} (2J + 1) \int T_B \, d\upsilon /(S \mu_c^2 \nu)\,\text{cm}^{-2},
\end{equation}
where the dipole moment along the $c$-axis $\mu_c$ is assumed to be 1.45 Debyes, the frequency $\nu$ is in GHz and the velocity $\upsilon$ is in km\,s$^{-1}$. The line strength, S, for the 3$_{1,3}$s--3$_{0,3}$a transition is 4.7. Realistically assuming that the line is optically thin, we determined the column density of para-NH$_{2}$D in W33\,B1 to be (0.69\,$\pm$\,0.20)\,$\times10^{12}$\,cm$^{-2}$. Since our line (Fig.\,\ref{FigA.12}) is only tentatively detected, this can be considered as an upper limit. The column density of para-NH$_3$ was calculated from our NH$_3$ observation data, using the calculation method in \citet{2022A&A...658A..34T}. The results indicate that the column density of para-NH$_3$ in the W33\,B1 region is 6.9\,($\pm$\,1.2)\,$\times10^{14}$\,cm$^{-2}$ (see Table\,6 of \citealt{2022A&A...658A..34T}). Then, we obtain for para-NH$_3$ a D/H ratio of (1.0\,$\pm$\,0.2)\,$\times10^{-3}$ toward W33\,B1. This value is lower than the (8.3\,$\pm$\,4.5)\,$\times10^{-3}$ measured in the star-forming region Orion\,A \citep{2015A&A...581A..48Ga} but is consistent with the ($\sim$\,2\,-\,8)\,$\times10^{-3}$ range found in a study of other deuterated molecules \citep{2013ApJ...770..142N}. This substantiates that the D/H ratio for NH$_3$ is likely influenced by fractionation being enhanced by up to two orders of magnitude relative to the abundance ratio found in the interstellar medium, D/H\,$\sim$\,1.5\,$\times10^{-5}$ (e.g., \citealt{1994ARA&A..32..191W,2003ApJ...587..235O,2007ARNPS..57..463S}). 

\citet{2005A&A...438..585R} measured the NH$_{2}$D/NH$_{3}$ abundance ratio using observations from the Caltech Submillimeter Observatory, the IRAM 30\,m telescope, and the Arecibo telescope. In the regions NGC\,1333, Barnard\,1, LDN\,1544C, LDN134\,N, and LDN\,1689N. They derived NH$_{2}$D/NH$_{3}$ ratios of  0.28\,$\pm$\,0.14, 0.23\,$\pm$\,0.05, 0.13\,$\pm$\,0.03, 0.10\,$\pm$\,0.03, and 0.19\,$\pm$\,0.05, respectively (see Table\,8 of \citealt{2005A&A...438..585R}). These NH$_{2}$D/NH$_{3}$ ratios are larger than our upper NH$_{2}$D/NH$_{3}$ abundance ratio limit we determined for W33 B1, likely due to the fact that these clouds are cooler than W33\,B1. The kinetic temperatures were derived from our NH$_3$ observation data using the method described in \citet{2022A&A...658A..34T}. The calculated values for NH$_3$\,(1,1) and (2,2) are 30\,$\pm$\,7, 16\,$\pm$\,1, 13\,$\pm$\,1, 13\,$\pm$\,1, 13\,$\pm$\,2, and 13\,$\pm$\,1\,K for W33\,Main, W33\,A, W33\,B, W33\,A1, W33\,B1, and W33\,Main1, respectively (see Table\.A.5 of \citealt{2022A&A...658A..34T}). Among these sources, W33\,B1 is the coldest, which explains why it may exhibit the highest NH$_{2}$D/NH$_{3}$ ratio. In addition, the 3$\sigma$ upper limits for the NH$_{2}$D/NH$_{3}$ ratio in the other W33 sources without a NH$_{2}$D detection are (5.0\,$\pm$\,0.4)\,$\times10^{-3}$.  

\subsection{Rotation temperatures, column densities, and fractional abundances relative to H$_{2}$}
\label{4.4}
The column densities and rotational temperatures for the observed lines were calculated using the following formula from \citet{1999ApJ...517..209G}:
\begin{equation}\label{f1}
\ln\biggl(\frac{3kW}{8\pi^3\nu\mu^2S}\biggr) = \ln\biggl(\frac{N_{\text{tot}}}{Q(T_{\text{rot}})}\biggr) - \frac{E_u}{kT_{\text{rot}}},
\end{equation}
where $k$ represents the Boltzmann constant; $W$ denotes the integrated intensity of the line; $\nu$ is the line’s rest frequency; $S$ refers to the transition’s intrinsic line strength; $\mu$ is the permanent dipole moment; $Q$ is the partition function, which is dependent on the rotational temperature $T_{\rm rot}$; and $E_{\rm u}$ denotes the excitation energy of the upper level. The quantities for $Q$, $E_{\rm u}$/k, and $S\mu^{2}$ are obtained from the Splatalogue spectral line catalogs (see footnote\,5). 

Figure\,\ref{Fig:8} displays the rotation diagrams for the CH$_{3}$OH transition lines in W33\,A, W33\,B, and W33\,Main\,1, respectively. We fit the E-type lines of CH$_{3}$OH.  
We apply the same source size to all molecules belonging to the same clump for computational simplicity. The adopted source sizes for each clump are taken from Table\,5 of \citet{2014A&A...572A..63I}. Using the adopted source sizes, we correct the observed intensities for beam dilution by dividing by the filling factor to derive source-averaged column densities. The column densities ($N_{\rm H_{2},\rm source}$) from Table\,5 of \citet{2014A&A...572A..63I} are then used to calculate the fractional abundances of the detected molecules. The beam sizes for the \citet{2014A&A...572A..63I} observations are 21.3$\arcsec$\,–\,22.2$\arcsec$ for the Atacama Pathfinder Experiment (APEX) data. 

\begin{figure*}[t]
\centering
\includegraphics[width=0.49\textwidth]{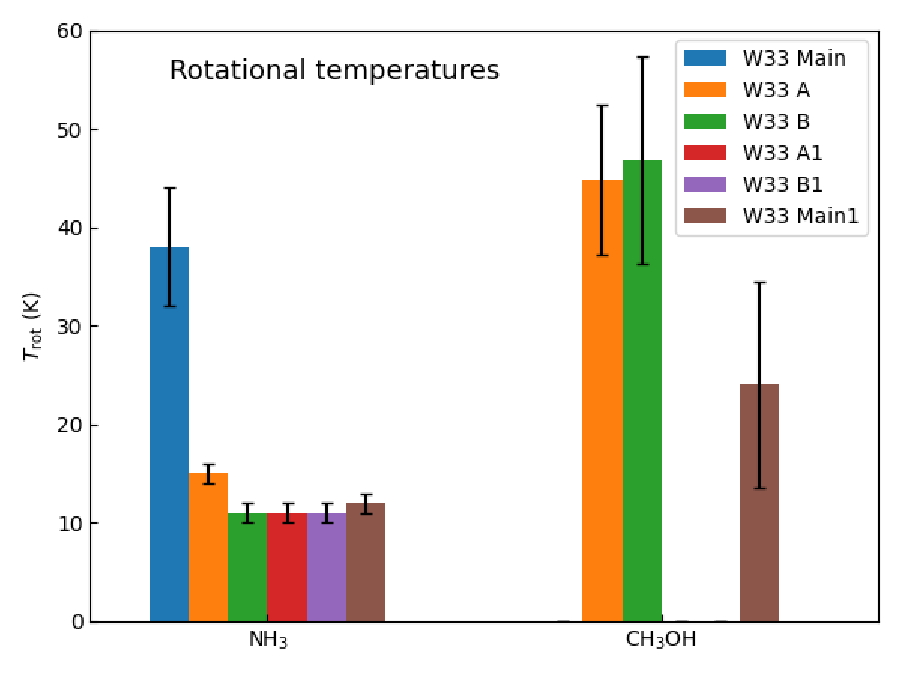}
\hspace{0.2cm}
\includegraphics[width=0.49\textwidth]{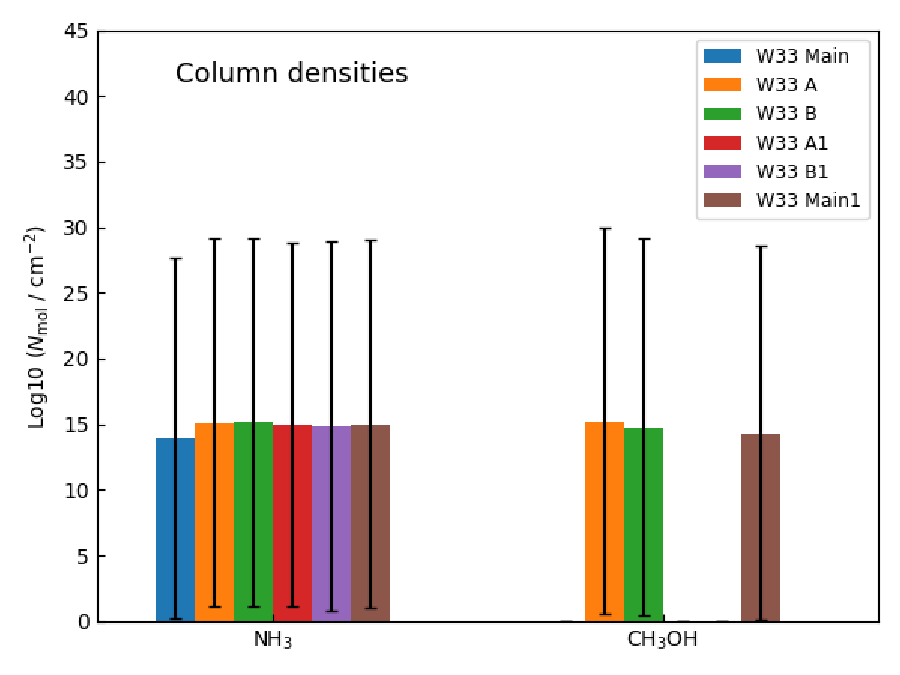}
\caption[]{Comparison of column densities and rotation temperatures among different molecules in our six sources (Table\,\ref{table:1}) belonging to the W33 complex. The source sizes are those derived in the study by \citet{2014A&A...572A..63I}.}
\label{Fig:10}
\end{figure*}

Due to the intense radio continuum radiation exhibited by W33\,Main \citep{1984ApJ...283..573S,2012PASP..124..939H,2022MNRAS.509.2234B}, we did not measure the methanol rotational temperatures and column densities toward this line of sight. The strong background emission contaminates the spectral features in the $J_{2}-J_{1}$ E-type transitions, making it impossible to derive reliable rotational temperatures and column densities from these lines. The rotation diagram analysis yields reliable results for the weakly inverted CH$_{3}$OH\,$J_{2}-J_{1}$ E-type transitions toward W33\,A, W33\,B, and W33\,Main1 (see Table\,\ref{Tab2} and Sect.\,\ref{sect-3-3-4}). This is justified by the radiative transfer equation, $T_{\rm L}$\,=\,($T_{\rm ex}$\,-\,$T_{\rm c}$)\,$\times$\,$\tau$. For inverted transitions, both $T_{\rm ex}$\,-\,$T_{\rm c}$ and $\tau$ are negative, yet the integrated intensity remains linearly dependent on $\tau$ in case of opacities well below unity, which is assumed as a consequence of the measured weak line emission. Thus, even for these transitions, a positive rotation temperature is obtained from a rotation temperature diagram, and the resulting value remains physically meaningful. We therefore applied the rotation temperature diagram method to derive rotation temperatures and column densities for these regions. The left panel of Fig.\,\ref{Fig:8} shows the Boltzmann plots of the W33\,A emission lines, where the least-square fit to the $J_{2}\,-\,J_{1}$ points with CH$_{3}$OH\,(2$_{2,1}$\,-\,2$_{1,2}$\,E) to (10$_{2,8}$\,-\,10$_{1,9}$\,E) gives the value $T_{\rm rot}$\,=\,44.8\,$\pm$\,7.6\,K and $N$\,=\,1.6\,($\pm$0.5)\,$\times$\,$10^{15}$\,cm$^{-2}$. For W33\,B, we get $T_{\rm rot}$\,=\,46.8\,$\pm$\,10.5\,K and $N$\,=\,5.9\,($\pm$2.3)\,$\times$\,$10^{14}$\,cm$^{-2}$ by fitting the seven E-type CH$_{3}$OH transitions (see Fig.\,\ref{Fig:8} middle panel). Toward the W33\,Main1 emission lines, we obtain $T_{\rm rot}$\,=\,24.0\,$\pm$\,10.4\,K and $N$\,=\,2.0\,($\pm$1.8)\,$\times$\,$10^{14}$\,cm$^{-2}$ by fitting the five $J_{2}-J_{1}$ lines (see Fig.\,\ref{Fig:8} right panel).

The rotation temperatures and column densities obtained in this study, along with those from other non-W33 sources found in the literature, are presented in Table\,\ref{table 3}. We find fitted CH$_{3}$OH rotation temperatures ranging from 24.0 to 46.8\,K and derive molecular column densities ranging from 2.0\,($\pm$1.8)\,$\times$\,$10^{14}$ (W33\,Main1) to 1.6\,($\pm$0.5)\,$\times$\,$10^{15}$\,cm$^{-2}$ (W33\,A). The other non-W33 sources listed in Table\,\ref{table 3} exhibit CH$_{3}$OH rotation temperatures ranging from 20 to 200\,K, with derived methanol column densities between 1.0\,$\times$\,$10^{14}$ and 1.1\,($\pm$0.4)\,$\times$\,$10^{18}$\,cm$^{-2}$. The CH$_{3}$OH rotation temperatures in our study are lower than those of other prominent non-W33 sources, while the calculated column densities fall within the elsewhere observed range. 

The rotation temperatures and total-NH$_3$(para+ortho) column densities of NH$_3$ were calculated from our NH$_3$ observational data, using the calculation method in \citet{2022A&A...658A..34T}. The results show rotation temperatures from NH$_3$\,(1,1) and (2,2) of 23\,$\pm$\,6, 15\,$\pm$\,1, 11\,$\pm$\,1, 11\,$\pm$\,1, 11\,$\pm$\,1, and 12\,$\pm$\,1\,K for W33\,Main, W33\,A, W33\,B, W33\,A1, W33\,B1, and W33\,Main1, respectively (see Sect.3.4 of \citealt{2022A&A...658A..34T}). These rotation temperatures are low. Since the (2,2) energy level is only 41.5\,K above the (1,1) level, this method cannot reliably measure $T_{\rm kin}$ above $\sim$50\,K. Even with the inclusion of higher metastable NH$_3$ lines, the derived temperatures, while increased, still do not reach the range typical of hot cores, which is 100\,K or even more. The corresponding total-NH$_3$ column densities ($N_{00}$+$N_{11}$+$N_{22}$+$N_{33}$+$N_{44}$+$N_{55}$+$N_{66}$) at the peak positions are 6.0\,($\pm$2.1)\,$\times$\,10$^{14}$, 3.5\,($\pm$0.1)\,$\times$\,10$^{15}$, 3.4\,($\pm$0.2)\,$\times$\,10$^{15}$, 2.8\,($\pm$0.2)\,$\times$\,10$^{15}$, 2.0\,($\pm$0.2)\,$\times$\,10$^{15}$, and 3.1\,($\pm$0.2)\,$\times$\,10$^{15}$\,cm$^{-2}$ for W33\,Main, W33\,A, W33\,B, W33\,A1, W33\,B1, and W33\,Main1, respectively (see Sect.3.6 and Eq.7 of \citealt{2022A&A...658A..34T}). Our main findings suggest that W33\,A and W33\,B are not hot cores. If they were hot cores, their ammonia rotation temperature and column density should be significantly higher than the currently observed values. For instance, typical hot core sources such as Orion, W51-IRS2, and Sagittarius\,B2 exhibit ammonia rotation temperatures of 160, $\sim$\,310, and $\sim$\,160\,K (e.g., \citealt{1988A&A...201..285H,2015A&A...581A..48Ga,1987A&A...173..352M,1993A&A...276..445H}), and column densities of 1.2\,$\times$\,10$^{17}$, $\sim$\,1.6\,$\times$\,10$^{19}$, and $\sim$\,1.5\,$\times$\,10$^{20}$\,cm$^{-2}$ (e.g., \citealt{2015A&A...581A..48Ga,1993A&A...276..445H}), respectively. These values of rotation temperature and column density in typical hot cores are significantly lager than the corresponding observed values in W33. This discrepancy might be attributed to potential beam dilution effects, which could reduce the beam averaged column densities and possibly also the derived rotation temperatures by including less excited lower temperature material from the surrounding environment. Since no high-resolution JVLA K-band spectra exist so far for this region, such observations are critical to resolve the small-scale gas structure and kinematics and to ascertain the presence of hot cores or corinos.

The column densities of CH$_{3}$OH, $N$(CH$_{3}$OH), are compared with the H$_{2}$ column densities derived from the 870\,$\mu$m continuum data obtained with the APEX telescope \citep{2009A&A...504..415S,2014A&A...572A..63I}. The APEX data are characterized by a full width at half maximum (FWHM)  beam size of  $\sim$\,23$\arcsec$. The H$_{2}$ peak column densities, as presented in Table\,5 of \citet{2014A&A...572A..63I}, are the most suitable for our comparison due to our beam size of $\sim$\,$40\arcsec$. The fractional abundances of CH$_{3}$OH, calculated as $\chi$\,(CH$_{3}$OH)\,=\,$N$(CH$_{3}$OH)\,/\,$N$(H$_{2}$) at the peak positions of the three W33 sources W33\,A, W33\,B, and W33\,Main1 and corrected for beam dilution with respect to 23$\arcsec$ are presented in Table\,\ref{table 3}. Therefore, the relative abundances of CH$_{3}$OH to molecular hydrogen at the peak positions of W33\,A, W33\,B, and W33\,Main1 are determined to be 6.6\,($\pm$0.6)\,$\times$\,10$^{-9}$, 2.8\,($\pm$0.5)\,$\times$\,10$^{-9}$, and 2.3\,($\pm$1.5)\,$\times$\,10$^{-9}$, respectively. The errors presented in parentheses are derived through the method of error propagation.

The left and middle panels of Figure\,\ref{Fig:9} show the rotation temperature and the fractional abundance of CH$_{3}$OH, respectively, as a function of the evolutionary sequence of the three sources W33\,A, W33\,B, and W33\,Main1. These rotation temperatures and fractional CH$_{3}$OH abundances are compatible with the stages of evolution outlined by \citet{2014A&A...572A..63I}. From our methanol $T_{\rm rot}$ and $N$(CH$_{3}$OH)\,/\,$N$(H$_{2}$) determinations, we may view a trend of increasing rotation temperature and fractional CH$_{3}$OH abundance with evolutionary stage (see Fig.\,\ref{Fig:9}). \citet{2007A&A...466..215L} demonstrated that methanol is a useful probe for measuring densities and temperatures in various regions of interstellar clouds, establishing it as an effective diagnostic tool for such environments. 

Figure\,\ref{Fig:10} compares two molecular species, CH$_{3}$OH and NH$_{3}$, across the six W33 sources in our sample. The rotation temperatures and column densities of para-NH$_{3}$\,(1,1) are taken from our NH$_{3}$ observational data (see Tables\,4 and 5 of \citealt{2022A&A...658A..34T}). The left panel of Fig.\,\ref{Fig:10} shows that the derived CH$_{3}$OH rotation temperatures are higher than those of NH$_{3}$ in W33\,A, W33\,B, and W33\,Main1. However, this conclusion is only valid when $T_{\rm rot}$ is derived from the (1,1) and (2,2) lines alone. The NH$_{3}$ rotation temperatures in W33\,Main are notably higher, as also seen in the left panel of Fig.\,\ref{Fig:10}. Meanwhile, in W33\,A, the column densities of NH$_{3}$ and CH$_{3}$OH show good agreement. However, in W33\,B and W33\,Main1, the column density of CH$_{3}$OH is lower than that of NH$_{3}$ (see the right panel of Fig.\,\ref{Fig:10}).

Molecular lines serve as effective diagnostic probes for analyzing the physical conditions and chemical evolution in star-forming regions \citep{2012A&ARv..20...56C}. The combination of carbon-chain molecules (CCMs) and NH$_{3}$, in particular, is commonly used to investigate star-forming regions \citep{2014ApJ...788..108S}. During the early phases of star formation, initially abundant CCMs often deplete, whereas nitrogen-bearing species such as NH$_{3}$ show increasing abundances \citep{2017ApJS..228...12T}. Thus, the column density ratios of N-bearing molecules to CCMs serve as reliable tracers of chemical evolution in star-forming regions \citep{1992ApJ...392..551S,2017ApJS..228...12T}. Previous studies by \citet{2021SCPMA..6479511X} report that the similar line widths of CCS, HC$_{3}$N, and NH$_{3}$ suggest these molecules trace similar regions. They also show that in Infrared Dark Clouds (IRDCs), the $N$(CCS)/$N{(\rm NH_3)}$ ratios range from 3.7$\times 10^{-4}$ to 1.5$\times 10^{-2}$, while $N{(\rm HC_3N)}$/$N{(\rm NH_3)}$ range from 5.5$\times 10^{-4}$ to 3.2$\times 10^{-2}$. In contrast, in low-mass star-forming regions, $N$(CCS)/$N{(\rm NH_3)}$ and $N{(\rm HC_3N)}$/ $N{(\rm NH_3)}$ span broader ranges, from 2.1$\times 10^{-3}$ to 6.9$\times 10^{-1}$ and 1.3$\times 10^{-3}$ to 9.0$\times 10^{-1}$, respectively. In our survey, we observe that cyanopolyynes are predominantly enriched in sources, such as W33\,A, W33\,B, and W33\,Main1 (see Sect.\,\ref{sect-3-3-3}). In W33\,Main, only HC$_{3}$N was detected. Additionally, the carbon-chain molecule CCS is detected only in the youngest and earliest-stage high-mass protostellar source studied by us, i.e., W33\,Main1. Meanwhile, W33\,Main has a total-NH$_3$ fractional abundance ($\chi$\,(total-NH$_3$)\,=\,(total-$N$(NH$_3$))/$N$(H$_2$)) of 1.3\,($\pm$0.1)\,$\times$\,10$^{-9}$ at the peak position. Higher values of 1.4\,($\pm$0.3)\,$\times$\,10$^{-8}$, 1.6\,($\pm$0.3)\,$\times$\,10$^{-8}$, 1.6\,($\pm$0.5)\,$\times$\,10$^{-8}$, 4.0\,($\pm$1.2)\,$\times$\,10$^{-8}$, and 3.4\,($\pm$0.5)\,$\times$\,10$^{-8}$, are obtained at the central positions of W33\,A, W33\,B, W33\,A1, W33\,B1, and W33\,Main1 (see Table\,6 of \citealt{2022A&A...658A..34T}). \citet{2014A&A...572A..63I} pointed out that the integrated intensity ratios of N$_{2}$H$^{+}$\,(3--2)/CS\,(6--5) and N$_{2}$H$^{+}$\,(3--2)/H$_{2}$CO\,(4$_{2,2}$--3$_{2,1}$) show clear trends as a function of evolutionary stages of the W33 clumps. Accordingly, we also calculated the HC$_{3}$N\,(2--1)/NH$_{3}$\,(1,1) integrated intensity ratio in our study. The line intensity ratios are found to decrease in the more evolved W33\,Main region (see the right panel of Fig.\,\ref{Fig:9}), which is consistent with standard astrochemical models \citep{2012MNRAS.419..238B,2015ASPC..499..191O,2021SCPMA..6479511X}.

Note that we have identified only two or three HC$_{5}$N transitions in the W33 region. Energy differences between their upper levels are quite small ($\sim$2\,K). Therefore, rotation temperatures and column densities could not be reliably established for HC$_{5}$N.

To further investigate the physical properties of the gas, we combined observations of the ortho c-C$_{3}$H$_{2}$\,(1$_{1,}$$_{0}$\,-\,1$_{0,}$$_{1}$) and para c-C$_{3}$H$_{2}$\,(2$_{2,}$$_{0}$\,-\,2$_{2,}$$_{1}$) transitions at the 18.3 and 21.6\,GHz toward W33\,Main. We used the RADEX radiative transfer code \citep{2007A&A...468..627V} with fixed physical parameters: a kinetic temperature of $T_{\rm kin}$\,=\,30\,K and an H$_{2}$ volume density of $n_{\rm H_{2}}$\,=\,1.2\,$\times$\, 10$^{4}$\,cm$^{-3}$, based on NH$_3$ results from \citet{2022A&A...658A..34T}. The RADEX  reveals c-C$_{3}$H$_{2}$ excitation temperature ($T_{\rm ex}$) of 4.5\,K and 3.7\,K, and column densities of 5.7\,$\times$\,10$^{13}$\,cm$^{-2}$ and 4.3\,$\times$\,10$^{16}$\,cm$^{-2}$ for the ortho c-C$_{3}$H$_{2}$\,(1$_{1,}$$_{0}$ -1$_{0,}$$_{1}$) and para c-C$_{3}$H$_{2}$\,(2$_{2,}$$_{0}$ - 2$_{2,}$$_{1}$) transitions, respectively. These values were derived using line widths of $\Delta v$\,=\,3.0\,km\,s$^{-1}$ and 2.5\,km\,s$^{-1}$, and mean beam brightness temperatures of $T_{\rm mb}$\,=\,0.89\,K and -0.97\,K (see Table\,\ref{Tab:A.5}). The discrepancy in column densities—differing by approximately 2.5 orders of magnitude—arises because most of the absorption originates from in front of the continuum source, so the absorption line arises from a shorter line of sight.                                                                 
             
\section{Summary}
\label{sect:summary}
Using the 100\,m telescope at Effelsberg, we carried out a 1.3\,cm spectral line survey in the prominent massive star-forming regions of the W33 complex. The main results are summarized as follows:

\begin{enumerate}
\item
In our observations, we identified a total of 68 spectral lines. The identified lines are assigned to 44 RRLs and 24 molecular lines. The RRLs, from hydrogen and helium, stem from the ionized material of W33\,Main. The molecular lines were assigned to nine different molecular species: CH$_{3}$OH, HC$_{n}$N, SiS, c-C$_{3}$H$_{2}$, CH$_{3}$CN, NH$_{2}$D, HNCO, H$_{2}$O, and CCS. 

\item
The ratios of intensity between RRL pairs with varying $\Delta n$ from the same element at adjacent frequencies align with the LTE ratios, indicating that deviations from LTE are negligible within this frequency range.

\item
The ratios of elemental abundances were computed as follows: The helium to hydrogen ratio (He/H) is determined to be (10.7\,$\pm$\,1.8)\% based on the comparison of helium and hydrogen RRLs. The difference in the He/H ratios, as inferred from H$\alpha$/He$\alpha$ and H$\beta$/He$\beta$ pairs, is minimal and consistent with theoretical predictions that suggest the departure coefficient ($b_{\rm n}$) values for hydrogen and helium are almost the same. 

\item
The turbulent line widths of helium are slightly greater than those of hydrogen. The difference between the two ranges from 5.1 to 12.9\,km\,s$^{-1}$, with an average of 9.3\,$\pm$\,3.5\,km\,s$^{-1}$. We argue this difference likely arises from a spatially unresolved fine structure that cannot be disentangled with current data. However, our derived He/H ratio is consistent with the normal abundance, indicating that these inhomogeneities are not a dominant effect. Ultimately, verifying this interpretation will require much higher angular resolution observations.

\item
Maser emissions from water and methanol have been observed in W33\,Main, W33\,A, and W33\,B. Our survey has led to the discovery of a (10$_{2,8}$--10$_{1,9}$\,E) maser in W33\,Main. 
A rotational diagram analysis of CH$_{3}$OH transitions in W33\,A, W33\,B, and W33\,Main1 yielded rotation temperatures and column densities, that indicate an increase in rotation temperature with evolutionary stage. Additionally, we calculated the HC$_{3}$N\,(2--1)/NH$_{3}$\,(1,1) integrated intensity ratio in our study. The line intensity ratios were found to decrease in the more evolved W33\,Main region, which is consistent with standard astrochemical models \citep{2012MNRAS.419..238B,2015ASPC..499..191O,2021SCPMA..6479511X}.

\item
The deuterium-to-hydrogen ratio (D/H) toward W33\,B1 is estimated to be (1.0\,$\pm$\,0.2)\,$\times10^{-3}$ from NH$_{2}$D/NH$_{3}$. Given that the NH$_{2}$D line is only tentatively detected, this value should be considered an upper limit. For the other W33 sources with non-detections, the 3$\sigma$ upper limits on the NH$_{2}$D/NH${3}$ ratio are (5.0\,$\pm$\,0.4)\,$\times10^{-3}$.

\item
At the linear scales considered here (0.5\,pc), fractional abundances and excitation temperatures do not reach values close to those in well-established hot cores, but higher-resolution measurements may alter this picture.

\end{enumerate}

\begin{acknowledgements}
We like to thank the anonymous referee for the useful suggestions that improved this study. This work is based on observations made with the 100\,m Effelsberg telescope, which is operated by the Max-Planck-Institut f\"ur Radioastronomie. We thank the staff of the Effelsberg 100\,m radio telescope for their assistance during the observations. This work was mainly supported by the National Key R\&D Program of China under grant Nos.\,(2023YFA1608002 and 2022YFA1603103), the Natural Science Foundation of Xinjiang Uygur Autonomous Region of China (No.\,2025D01B171), the Tianshan Talent Training Program 2024TSYCTD0013, the Tianchi Talent Project of Xinjiang Uyghur Autonomous Region. It was also partially funded by the NSFC under grants 12373029, 12103082, and 12403033, the Chinese Academy of Sciences (CAS) “Light of West China” Program under grant No. xbzg-zdsys-202212, G.W was supported by the Xinjiang Tianchi Talents Program, the Youth Innovation Promotion Association CAS. The Central Guidance for Local Science and Technology Development Fund ZYYD2025ZY23. C. Henkel has been funded by Chinese Academy of Sciences President’s International Fellowship Initiative grant No. 2025PVA0048. Moreover, this research has been funded by the Science Committee of the Ministry of Science and Higher Education of the Republic of Kazakhstan (Grant No. AP26102915). 

\end{acknowledgements}

\onecolumn

\clearpage

\renewcommand{\tabcolsep}{0.1 cm}
\begin{table}[t]
\centering
\caption{Methanol excitation temperatures and optical depths (see Sect.\,\ref{sect-3-3-4}).}
\label{Tab2}
\begin{tabular}{cccccccccc}
\hline
  &      & \multicolumn{2}{c}{W33\,Main}                  &  \multicolumn{2}{c}{W33\,A} &   \multicolumn{2}{c}{W33\,B}     & \multicolumn{2}{c}{W33\,Main1}     \\
\cline{3-10}
Transition  &  Freq.    & $T_{\rm ex}$   &   $\tau$   &  $T_{\rm ex}$  &   $\tau$  &    $T_{\rm ex}$ & $\tau$ & $T_{\rm ex}$  &  $\tau$      \\
            & (MHz)     & (K)            &            &  (K)           &           &    (k)          &        &    (K)        &               \\
\hline
\hline 
$3_{2}-3_{1}E$    & 24928.7  & -3.6   & -2.5\,$\times$\,10$^{-4}$ & -3.6 & -3.6\,$\times$\,10$^{-3}$ & -3.3 & -3.1\,$\times$\,10$^{-5}$  & -2.2 & -2.5\,$\times$\,10$^{-4}$ \\   
$4_{2}-4_{1}E$    & 24933.5  & -1.8   & -2.0\,$\times$\,10$^{-4}$ & -2.0 & -1.9\,$\times$\,10$^{-3}$ & -2.7 & -3.1\,$\times$\,10$^{-4}$  & -2.3 & -2.5\,$\times$\,10$^{-4}$ \\
$2_{2}-2_{1}E$    & 24934.4  & -2.9   & -7.3\,$\times$\,10$^{-4}$ & -2.7 & -1.3\,$\times$\,10$^{-2}$ & -7.1 & -4.1\,$\times$\,10$^{-3}$  & -3.5 & -2.5\,$\times$\,10$^{-4}$  \\
$5_{2}-5_{1}E$    & 24959.1  & -1.4   & -8.2\,$\times$\,10$^{-5}$ & -1.6 & -3.2\,$\times$\,10$^{-4}$ & -1.8 & -1.9\,$\times$\,10$^{-4}$  & -1.8 & -2.5\,$\times$\,10$^{-4}$  \\
$6_{2}-6_{1}E$    & 25018.1  & -2.0   & -1.8\,$\times$\,10$^{-5}$ & -2.1 & -9.7\,$\times$\,10$^{-5}$ &      &                            & -2.2 & -2.5\,$\times$\,10$^{-4}$  \\
$7_{2}-7_{1}E$    & 25124.9  & -2.1   & -5.4\,$\times$\,10$^{-6}$ & -2.1 & -3.0\,$\times$\,10$^{-5}$ & -2.4 & -4.4\,$\times$\,10$^{-6}$  &      &                              \\
$8_{2}-8_{1}E$    & 25294.4  & -2.1   & -2.1\,$\times$\,10$^{-6}$ & -1.9 & -4.2\,$\times$\,10$^{-6}$ & -2.3 & -6.1\,$\times$\,10$^{-7}$  &      &                              \\
$9_{2}-9_{1}E$    & 25541.4  & -1.4   & -9.1\,$\times$\,10$^{-7}$ & -1.3 & -9.9\,$\times$\,10$^{-7}$ & -1.2 & -7.9\,$\times$\,10$^{-9}$  &      &                              \\
$10_{2}-10_{1}E$  & 25878.3  & -1.3   & -2.9\,$\times$\,10$^{-7}$ & -1.2 & -1.1\,$\times$\,10$^{-7}$ &      &                            &      &                               \\
\hline  \\
\end{tabular}
\normalsize
\normalsize 
\end{table}

\renewcommand{\tabcolsep}{0.1 cm}
\begin{table}[t]
\centering
\caption{Rotation temperatures and  column densities of CH$_{3}$OH.}\label{table 3}
\begin{tabular}{ccccccccccc}
\hline
  &      & \multicolumn{4}{c}{This Work (W33)}                  &                    & \multicolumn{3}{c}{Other Sources (non-W33)} &              \\
\cline{2-5} \cline{7-10}
  Species     & $T_{\rm rot}$   &   $N$(CH$_{3}$OH)   &  $\chi$($N/N_{\rm H_{2}}$)  &   Detected sources   &   & $T_{\rm rot}$ & $N$(CH$_{3}$OH) &  Detected sources & Ref.      \\
   &            K          &  cm$^{-2}$      &  &          &    &  K   & cm$^{-2}$ & &    \\
\hline
\hline  
CH$_{3}$OH & 44.8 $\pm$ 7.6  & (1.6 $\pm$ 0.5) $\times10^{15}$ &  (6.6 $\pm$ 0.6) $\times10^{-9}$ & W33\,A    &  & 115 $\pm$ 8.0  & (1.1 $\pm$ 0.4) $\times10^{18}$ & Orion\,KL & 1 &  \\
CH$_{3}$OH & 46.8 $\pm$ 10.5 & (5.9 $\pm$ 2.3) $\times10^{14}$ & (2.8 $\pm$ 0.5) $\times10^{-9}$  & W33\,B   &  & 208 $\pm$ 70  & (2.6 $\pm$ 1.8) $\times10^{14}$ & IRAS\,16293-2411 & 3 &  \\
CH$_{3}$OH & 24.0 $\pm$ 10.4  & (2.0 $\pm$ 1.8) $\times10^{14}$ &  (2.3 $\pm$ 1.5) $\times10^{-9}$ & W33\,Main1   &  & 150   & (293.5 $\pm$ 19.4 $\times10^{15}$ & W51e1 & 4 &    \\
           &   &  &    &    & & 20             & 1.0 $\times10^{14}$ & G79.3p1 & 2 &   \\
           &   &  &    &    &  & 150   & (938 $\pm$ 29.1) $\times10^{15}$ & W51e2  & 4 &    \\
           &   &  &    &    &  & 150   & (251 $\pm$ 19.4) $\times10^{15}$ & W51IRS2 & 4 &    \\
           &   &  &    &    &  & 100   &    & NGC253 & 5 &    \\
\hline 
\end{tabular}
\normalsize
\flushleft \textbf{References.} References for rotation temperatures and column densities from the literature.-- (1) \citet{2015A&A...581A..48Ga}; (2) \citet{2004A&A...422..573L}; (3) \citet{2004A&A...416..159P}; (4) \citet{2004A...34..691}; (5) \citet{2020NatAs...4.1170C}.
\normalsize 
\end{table} 
                                           
\begin{appendix}
\onecolumn
\section{Zoomed-in plots of observed spectra}
\label{Appendix A}

\renewcommand{\tabcolsep}{0.15 cm}
\begin{longtable}{clll}
\caption{Lines detected in the spectroscopic survey of W33.}\label{Tab:A.1}\\
\hline
Rest frequency      & Transition  &   $E_{\rm u}$/k         &        \\
(MHz)	       &	     &   (K)                 & Detected sources  \\
\hline
\endfirsthead
\caption{continued.} \\
\hline
Rest frequency  & Transition   &  $E_{\rm u}$/k            &               \\
(MHz)	   &		  &  (K)	              & Detected sources       \\ 
\hline
\endhead
\hline  \\
\endfoot
\hline
\endlastfoot

18045.9     &	 H89$\beta$			                     &      & W33\,Main            \\ 
18154.9     &    SiS(1--0)                                     &      & W33\,Main1            \\  
18196.2     &	 HC$_{3}$N (2--1)		                     & 1    & W33\,Main, A, B, A1, Main1           \\ 
18241.7     &	 H111$\delta$			                     &       & W33\,Main             \\ 
18327.5     &	 H101$\gamma$			                     &       & W33\,Main             \\ 
18343.1     &    c-C$_{3}$H$_{2}$\,(1$_{1,}$$_{0}$ -1$_{0,}$$_{1}$)                            &    & W33\,Main, A, B, A1, B1, Main1        \\
18398.0     &	 CH$_{3}$CN\,(1$_{0}$--0$_{0}$)		             & 1     & W33\,Main, A, B, Main1          \\ 
18638.6      &   HC$_{5}$N                        &       & W33\,A, B, Main1            \\ 
18661.1     &	 H88$\beta$			                     &        &W33\,Main             \\ 
18734.9     &	 H110$\delta$			                     &       & W33\,Main, A             \\ 
18769.2     &	 H70$\alpha$			                     &      & W33\,Main             \\ 
18776.8     &	 He70$\alpha$			                     &      &W33\,Main          \\ 
18807.9     &	 NH$_{2}$D\,(3$_{1,3}$s--3$_{0,3}$a)                     & 94    &W33\,B1          \\ 
18874.8     &	 H100$\gamma$			                     &          & W33\,Main, B         \\ 
19246.1     &	 H109$\delta$			                     &         &  W33\,Main           \\ 
19304.7     &	 H87$\beta$			                     &       & W33\,Main           \\ 
19312.6     &	 He87$\beta$			                     &       & W33\,Main            \\ 
19444.0     &	 H99$\gamma$			                     &        & W33\,Main           \\ 
19591.1     &	 H69$\alpha$			                     &        & W33\,Main, Main1           \\ 
19599.1     &	 He69$\alpha$			                     &       & W33\,Main         \\ 
19776.0     &	 H108$\delta$			                     &       & W33\,Main         \\ 
19978.2     &	 H86$\beta$			                     &       & W33\,Main            \\ 
20653.0     &	 H97$\gamma$			                     &        & W33\,Main         \\ 
20683.3     &	 H85$\beta$			                     &         &  W33\,Main        \\ 
20908.9     &	 CH$_{3}$OH (16$_{-4}$--15$_{-5}$ E)                    & 403    & W33\,Main1         \\ 
21295.2     &	 H96$\gamma$			                     &        & W33\,Main           \\ 
21301.3     &    HC$_{5}$N ($8-7$)                                    & 2     & W33\,A, B           \\
21384.8     &	 H67$\alpha$			                     &        & W33\,Main        \\ 
21393.5     &	 He67$\alpha$			                     &       & W33\,Main         \\ 
21422.1     &	 H84$\beta$			                     &        & W33\,Main       \\ 
21431.9     &	 HC$_{7}$N 			                     &      & W33\,B          \\ 
21487.2     &	 H105$\delta$			                     &        & W33\,Main          \\ 
21587.4      &	 c-C$_{3}$H$_{2}$\,(2$_{2,}$$_{0}$ -2$_{2,}$$_{1}$)			                     &         & W33\,Main         \\ 
21964.3     &	 H95$\gamma$			                     &       & W33\,Main          \\ 
21981.6     &	 HNCO (1$_{0,1}$--0$_{0,0}$)		             & 1     & W33\,Main, A, B             \\ 
22101.4     &	 H104$\delta$			                     &          & W33\,Main          \\ 
22196.5     &	 H83$\beta$			                     &          & W33\,Main        \\ 
22235.1     &	 H$_{2}$O (6$_{1,6}$--5$_{2,3}$)		             & 643    & W33\,Main, A, B        \\ 
22344.0     &	 CCS			                     &        &  W33\,A1, Main1         \\
22364.2     &	 H66$\alpha$			                     &        &  W33\,Main          \\ 
22373.3     &	 He66$\alpha$			                     &        &  W33\,Main        \\ 
22661.8     &	 H94$\gamma$			                     &         & W33\,Main          \\ 
22739.2     &	 H103$\delta$			                     &         & W33\,Main        \\ 
23008.6     &	 H82$\beta$			                     &        & W33\,Main        \\ 
23389.1     &	 H93$\gamma$			                     &          & W33\,Main     \\ 
23404.3     &	 H65$\alpha$			                     &        & W33\,Main      \\ 
23413.8     &	 He65$\alpha$			                     &       & W33\,Main \\ 
23444.8     &	 CH$_{3}$OH (10$_{1}$--9$_{2}$ A$^{-}$)                 & 143  & W33\,Main       \\ 
23860.9     &	 H81$\beta$			                     &           & W33\,Main       \\ 
23963.9     &	 HC$_{5}$N (9--8)			             & 6       &  W33\,B,  Main1      \\ 
24090.4     &	 H101$\delta$			                     &        & W33\,Main         \\ 
24147.9     &	 H92$\gamma$			                     &           & W33\,Main       \\ 
24509.9     &	 H64$\alpha$			                     &         & W33\,Main        \\ 
24519.9     &	 He64$\alpha$			                     &         & W33\,Main      \\ 
24755.7     &	 H80$\beta$			                     &          & W33\,Main        \\ 
24765.8     &	 He80$\beta$			                     &        & W33\,Main          \\ 
24928.7     &	 CH$_{3}$OH (3$_{2}$--3$_{1}$ E)                       & 36     & W33\,Main, A, B, Main1          \\ 
24933.5     &	 CH$_{3}$OH (4$_{2}$--4$_{1}$ E)                       & 45       &  W33\,Main, A, B, Main1       \\ 
24934.4     &	 CH$_{3}$OH (2$_{2}$--2$_{1}$ E)                       & 29        &  W33\,Main, A, B, Main1     \\ 
24959.1     &	 CH$_{3}$OH (5$_{2}$--5$_{1}$ E)                       & 57      &  W33\,Main, A, B, Main1        \\ 
25018.1     &	 CH$_{3}$OH (6$_{2}$--6$_{1}$ E)                       & 71       & W33\,Main, A, Main1        \\ 
25124.9     &	 CH$_{3}$OH (7$_{2}$--7$_{1}$ E)                       & 87     & W33\,Main, A, B           \\ 
25294.4     &	 CH$_{3}$OH (8$_{2}$--8$_{1}$ E)                       & 106     & W33\,Main, A, B          \\ 
25541.4     &	 CH$_{3}$OH (9$_{2,7}$--9$_{1,8}$ E)	             & 127    & W33\,Main, A            \\ 
25550.8     &	 H99$\delta$			                     &      & W33\,Main          \\
25686.3     &	 H63$\alpha$			                     &        & W33\,Main      \\ 
25695.9     &	 H79$\beta$			                     &        & W33\,Main      \\ 
25878.3     &	 CH$_{3}$OH (10$_{2}$--10$_{1}$ E)	             & 150   &  W33\,Main, A, B         \\ 

\end{longtable}   

\clearpage

\begin{longtable}{ccccccc}
\caption{Gaussian fits to the recombination lines.}\label{Tab:A.2} \\
\hline
Transition   & Rest frequency      &  $\int$$T_{\rm MB}$d$v$   & $V_{\rm LSR}$  & $\Delta v$ & $T_{\rm MB}$     &  \\
             & (MHz)      & (K\,km\,s$^{-1}$)        & (km\,s$^{-1}$)      & (km\,s$^{-1}$)     & (K)               & Detected source  \\
\hline
\endfirsthead
\caption{continued.} \\
\hline
Transition   & Rest frequency      &  $\int$$T_{\rm MB}$d$v$   & $V_{\rm LSR}$   & $\Delta v$ & $T_{\rm MB}$     &  \\
             & (MHz)      & (K\,km\,s$^{-1}$)        & (km\,s$^{-1}$)      & (km\,s$^{-1}$s)     & (K)              & Detected source \\
\hline
\endhead
\hline  \\
\endfoot
\hline
\endlastfoot

H$\alpha$      &\multicolumn{6}{c}{}\\
\hline
H70$\alpha$    & 18769.2 & 25.72 $\pm$ 0.48   & 35.2 $\pm$ 0.2 & 18.6 $\pm$ 0.4 & 1.29 $\pm$ 0.02 &  W33\,Main    \\
H69$\alpha$    & 19591.1 & 34.23 $\pm$ 0.42   & 34.1 $\pm$ 0.2 & 19.9 $\pm$ 0.3 & 1.62 $\pm$ 0.01  &  W33\,Main   \\
               &         & 1.05  $\pm$ 0.06   & 30.4 $\pm$ 0.4 & 23.3 $\pm$ 1.6 & 0.04 $\pm$ 0.09  &  W33\,Main1   \\
H67$\alpha$    & 21384.8 & 30.12 $\pm$ 0.81   & 36.6 $\pm$ 0.2 & 17.4 $\pm$ 0.5 & 1.62 $\pm$ 0.01  &  W33\,Main   \\
H66$\alpha$    & 22364.2 & 27.89 $\pm$ 0.64   & 36.8 $\pm$ 0.2 & 17.7 $\pm$ 0.4 & 1.47 $\pm$ 0.03  & W33\,Main    \\
H65$\alpha$    & 23404.3 & 77.09 $\pm$ 1.15   & 34.2 $\pm$ 0.2 & 20.5 $\pm$ 0.3 & 3.53 $\pm$ 0.02  & W33\,Main    \\
H64$\alpha$    & 24509.9 & 60.53 $\pm$ 1.22   & 35.9 $\pm$ 0.2 & 18.2 $\pm$ 0.4 & 3.12 $\pm$ 0.01  & W33\,Main   \\
H63$\alpha$    & 25686.3 & 54.95 $\pm$ 1.27   & 36.5 $\pm$ 0.2 & 17.6 $\pm$ 0.4 & 2.93 $\pm$ 0.03  & W33\,Main      \\
\hline
He$\alpha$     &\multicolumn{6}{c}{}\\
\hline

He70$\alpha$   & 18776.8 & 1.79 $\pm$ 0.06  & 35.8 $\pm$ 0.3 & 15.2 $\pm$ 0.5 & 0.11 $\pm$ 0.02  &  W33\,Main    \\
He69$\alpha$   & 19599.1 & 2.92 $\pm$ 0.08  & 34.6 $\pm$ 0.3 & 18.7 $\pm$ 0.6 & 0.14 $\pm$ 0.01  & W33\,Main    \\
He67$\alpha$   & 21393.5 & 2.79 $\pm$ 0.09  & 37.0 $\pm$ 0.3 & 15.6 $\pm$ 0.5 & 0.16 $\pm$ 0.01  & W33\,Main    \\
He66$\alpha$   & 22373.3 & 3.60 $\pm$ 0.09  & 35.9 $\pm$ 0.3 & 19.7 $\pm$ 0.5 & 0.17 $\pm$ 0.01  & W33\,Main    \\
He65$\alpha$   & 23413.8 & 3.54 $\pm$ 0.24  & 37.6 $\pm$ 0.4 & 14.2 $\pm$ 1.4 & 0.23 $\pm$ 0.03  & W33\,Main    \\
He64$\alpha$   & 24519.9 & 9.28 $\pm$ 0.12  & 33.5 $\pm$ 0.1 & 20.8 $\pm$ 1.3 & 0.42 $\pm$ 0.01  & W33\,Main    \\

\hline
H$\beta$       &\multicolumn{6}{c}{}\\
\hline
H89$\beta$     & 18045.9 & 2.34 $\pm$ 0.09 & 36.4$\pm$0.4 & 17.7 $\pm$ 0.7 & 0.12 $\pm$ 0.01  & W33\,Main    \\
H88$\beta$     & 18661.1 & 6.54 $\pm$ 0.09 & 34.2 $\pm$ 0.1 & 20.5 $\pm$ 0.3 & 0.29 $\pm$ 0.02  & W33\,Main    \\
H87$\beta$     & 19304.7 & 5.94 $\pm$ 0.12 & 35.5 $\pm$ 0.2 & 19.3 $\pm$ 0.4 & 0.28 $\pm$ 0.01  & W33\,Main    \\
H86$\beta$     & 19978.2 & 4.66 $\pm$ 0.23 & 37.3 $\pm$ 0.4 & 16.7 $\pm$ 0.8 & 0.26 $\pm$ 0.03  & W33\,Main    \\
H85$\beta$     & 20683.3 & 6.29 $\pm$ 0.22 & 36.9 $\pm$ 0.3 & 17.3 $\pm$ 0.6 & 0.34 $\pm$ 0.03  & W33\,Main    \\
H84$\beta$     & 21422.1 & 6.96 $\pm$ 0.20 & 37.2 $\pm$ 0.3 & 17.3 $\pm$ 0.5 & 0.37 $\pm$ 0.02  & W33\,Main    \\
H83$\beta$     & 22196.5 & 6.50 $\pm$ 0.30 & 38.4 $\pm$ 0.4 & 15.9 $\pm$ 0.7 & 0.38 $\pm$ 0.01  & W33\,Main    \\
H82$\beta$     & 23008.6 & 15.03 $\pm$ 0.46 & 35.4 $\pm$ 0.3 & 16.9 $\pm$ 0.5 & 0.83 $\pm$ 0.01  & W33\,Main    \\
H81$\beta$     & 23860.9 & 43.12 $\pm$ 0.89 & 37.2 $\pm$ 0.3 & 27.8 $\pm$ 0.6 & 1.45 $\pm$ 0.03  & W33\,Main    \\
H80$\beta$     & 24755.7 & 10.82 $\pm$ 0.41 & 36.8 $\pm$ 0.3 & 16.6 $\pm$ 0.6 & 0.61 $\pm$ 0.02  & W33\,Main    \\
H79$\beta$     & 25695.9 & 22.33 $\pm$ 0.34 & 33.9 $\pm$ 0.2 & 20.5 $\pm$ 0.3 & 1.02 $\pm$ 0.03  & W33\,Main    \\

\hline
He$\beta$      &\multicolumn{6}{c}{}\\

\hline
He87$\beta$    & 19312.6 & 0.68 $\pm$ 0.06 & 40.4 $\pm$ 1.0 & 20.9 $\pm$ 1.6 & 0.03 $\pm$ 0.02  & W33\,Main     \\ 
He80$\beta$    & 24765.8 & 3.32 $\pm$ 0.14 & 39.4 $\pm$ 0.5 & 22.2 $\pm$ 0.9 & 0.14 $\pm$ 0.02  & W33\,Main     \\

\hline
H$\gamma$      &\multicolumn{6}{c}{}\\
\hline
H101$\gamma$   & 18327.5 & 1.85 $\pm$ 0.05  & 35.6 $\pm$ 0.3 & 18.7 $\pm$ 0.6 & 0.09 $\pm$ 0.01  & W33\,Main    \\ 
H100$\gamma$   & 18874.8 & 1.49 $\pm$ 0.07  & 37.5 $\pm$ 0.4 & 16.7 $\pm$ 0.8 & 0.08 $\pm$ 0.01  &  W33\,Main   \\ 
H99$\gamma$    & 19444.0 & 1.86 $\pm$ 0.09  & 36.8 $\pm$ 0.5 & 17.2 $\pm$ 0.9 & 0.10 $\pm$ 0.02  & W33\,Main     \\  
H97$\gamma$    & 20653.0 & 2.00 $\pm$ 0.08 &  38.9 $\pm$ 0.4 & 16.5 $\pm$ 0.7 & 0.10 $\pm$ 0.01  & W33\,Main    \\  
H96$\gamma$    & 21295.2 & 1.62 $\pm$ 0.08 & 36.5 $\pm$ 0.4 & 15.2 $\pm$ 0.8 & 0.10 $\pm$ 0.01  & W33\,Main    \\  
H95$\gamma$    & 21964.3 & 11.74 $\pm$ 0.20 & 34.8 $\pm$ 0.2 & 26.0 $\pm$ 0.3 & 0.42 $\pm$ 0.01  & W33\,Main    \\ 
H94$\gamma$    & 22661.8 & 10.07 $\pm$ 0.35 & 38.0 $\pm$ 0.4 & 21.2 $\pm$ 0.8 & 0.45 $\pm$ 0.01  & W33\,Main    \\  
H93$\gamma$    & 23389.1 & 11.81 $\pm$ 0.27 & 35.3 $\pm$ 0.3 & 22.6 $\pm$ 0.6 & 0.47 $\pm$ 0.04  & W33\,Main  \\  
H92$\gamma$    & 24147.9 & 4.33 $\pm$ 0.14 & 38.1 $\pm$ 0.3 & 17.5 $\pm$ 0.6 & 0.23 $\pm$ 0.03  & W33\,Main    \\  

\hline
H$\delta$      &\multicolumn{6}{c}{}\\
\hline
H111$\delta$   & 18241.7 & 0.76 $\pm$ 0.05 & 34.8 $\pm$ 0.6 & 16.6 $\pm$ 1.1 & 0.04 $\pm$ 0.01  & W33\,Main    \\     
H110$\delta$   & 18734.9 & 0.79 $\pm$ 0.04 & 36.8 $\pm$ 0.5 & 16.6 $\pm$ 0.9 & 0.04 $\pm$ 0.01  & W33\,Main    \\     
H109$\delta$   & 19246.1 & 0.48 $\pm$ 0.06 & 40.9 $\pm$ 0.9 & 13.7 $\pm$ 1.7 & 0.03 $\pm$ 0.01  & W33\,Main    \\     
H108$\delta$   & 19776.0 & 1.07 $\pm$ 0.11 & 39.5 $\pm$ 0.9 & 16.2 $\pm$ 1.7 & 0.06 $\pm$ 0.03  & W33\,Main    \\    
H105$\delta$   & 21487.2 & 0.83 $\pm$ 0.06 & 39.5 $\pm$ 0.5 & 14.4 $\pm$ 1.0 & 0.05 $\pm$ 0.02  & W33\,Main   \\     
H104$\delta$   & 22101.4 & 0.57 $\pm$ 0.11 & 39.3 $\pm$ 1.2 & 11.3 $\pm$ 2.5 & 0.05 $\pm$ 0.02  & W33\,Main    \\     
H103$\delta$   & 22739.2 & 3.21 $\pm$ 0.19 & 37.6 $\pm$ 0.7 & 17.1 $\pm$ 1.1 & 0.17 $\pm$ 0.01  & W33\,Main    \\
H101$\delta$   & 24090.4 & 1.72 $\pm$ 0.09 & 36.9 $\pm$ 0.4 & 14.8 $\pm$ 0.9 & 0.11 $\pm$ 0.01  & W33\,Main    \\
H99$\delta$   & 25550.8  & 1.66 $\pm$ 0.09 & 36.6 $\pm$ 0.4 & 14.5 $\pm$ 0.9 & 0.11 $\pm$ 0.01  & W33\,Main    \\  
\hline
\end{longtable} 

\clearpage

\renewcommand{\tabcolsep}{0.15 cm}
\begin{longtable}{ccccccc}
\caption{Gaussian CH$_{3}$OH line parameters.}\label{Tab:A.3}                                                     \\
\hline
Rest frequency    &               & $\int$$T_{\rm MB}$d$v$  & $V_{\rm LSR}$   & $\Delta v$          & $T_{\rm MB}$      &                       \\
 (MHz)    & $J_{k_{\rm a}}$  & (K\,km\,s$^{-1}$)         & (km\,s$^{-1}$)       & (km\,s$^{-1}$)      & (K)                &  Detected source                 \\
\hline
\endfirsthead
\caption{continued.}                                                                                                             \\
\hline
Rest frequency    &               & $\int$$T_{\rm MB}$d$v$  & $V_{\rm LSR}$     & $\Delta v$        & $T_{\rm MB}$    &                       \\
 (MHz)    & $J_{k_{\rm a}}$  & (K\,km\,s$^{-1}$)         & (km\,s$^{-1}$)       & (km\,s$^{-1}$)      & (K)               &  Detected source                \\
\hline
\endhead
\hline                                                                                                                           \\
\endfoot
\hline
\endlastfoot
CH$_{3}$OH     &\multicolumn{6}{c}{}                                                                                              \\
\hline                                      
20908.9 & $16_{-4} -15_{-5}E$	   &0.41$\pm$0.16   & 40.6$\pm$0.9 & 4.9$\pm$2.7 & 0.08$\pm$0.02   & W33\,Main1   \\
23444.8 & $10_{1}-9_{2}A^{-}$	   &5.72$\pm$0.48   & 37.4$\pm$0.9 & 2.7$\pm$0.2 & 0.18$\pm$0.06   & W33\,Main \\
24928.7 & $3_{2}-3_{1}E$	       &9.52$\pm$0.59   & 37.3$\pm$0.2 & 5.5$\pm$0.4 & 1.62$\pm$0.12   & W33\,Main  \\
        & 	                       &1.65$\pm$0.14   & 38.3$\pm$0.2 & 4.3$\pm$0.4 & 0.36$\pm$0.02   & W33\,A  \\
        & 	                       &0.64$\pm$0.08   &56.1$\pm$0.2 & 2.4$\pm$0.4 & 0.25$\pm$0.03    & W33\,B  \\
        & 	                       &0.13$\pm$0.04   & 36.6$\pm$0.2 & 1.1$\pm$0.5 & 0.11$\pm$0.01   & W33\,Main1  \\        
        
24933.5 & $4_{2}-4_{1}E$	       & 9.87$\pm$0.12 & 25.74$\pm$0.1 & 6.9$\pm$0.1 & 1.48 $\pm$ 0.32  & W33\,Main   \\
        & 	                       &1.38$\pm$0.21   &37.7$\pm$0.3 & 4.4$\pm$0.8 & 0.29$\pm$0.04   & W33\,A  \\
        & 	                       &0.62$\pm$0.12   &57.5$\pm$0.5 & 5.1$\pm$1.3 & 0.11$\pm$0.03    & W33\,B  \\
        & 	                       &0.32$\pm$0.06   &37.4$\pm$0.3 & 3.1$\pm$0.6 & 0.09$\pm$0.02  & W33\,Main1  \\

24934.4 & $2_{2}-2_{1}E$		   & 6.00$\pm$0.29 & 36.8$\pm$0.1 & 5.3$\pm$0.3 & 1.06$\pm$0.07   & W33\,Main	 \\
        & 	                       &1.55$\pm$0.14   &37.7$\pm$0.2 & 3.8$\pm$0.4 & 0.38$\pm$0.03   & W33\,A  \\
        & 	                       &0.52$\pm$0.08   &56.0$\pm$0.2 & 2.7$\pm$0.6 & 0.18$\pm$0.03   & W33\,B  \\
        & 	                       &0.38$\pm$0.07   &37.2$\pm$0.3 & 3.4$\pm$0.9 & 0.11$\pm$0.02   & W33\,Main1  \\    
        
24959.1 & $5_{2}-5_{1}E$		   &8.67$\pm$0.09 & 37.6$\pm$0.1 & 5.5$\pm$0.1 & 1.49$\pm$0.07  & W33\,Main	 \\
        & 	                       &1.52$\pm$0.13   & 38.7$\pm$0.3 & 7.3$\pm$0.9 & 0.19$\pm$0.05  & W33\,A  \\
        & 	                       &0.38$\pm$0.06   & 55.9$\pm$0.2 & 2.3$\pm$0.4 & 0.15$\pm$0.01  & W33\,B  \\
        & 	                       &0.17$\pm$0.05   &35.9$\pm$0.3 & 2.2$\pm$0.6 & 0.07$\pm$0.01  & W33\,Main1  \\        
        
25018.1 & $6_{2}-6_{1}E$		    & 7.08$\pm$0.14 & 37.6$\pm$0.1 & 5.6$\pm$0.1 & 1.18$\pm$0.05 & W33\,Main	 \\
        & 	                       &0.92$\pm$0.10   & 38.6$\pm$0.3 & 4.1$\pm$0.6 & 0.21$\pm$0.05  & W33\,A  \\
        & 	                       &0.17$\pm$0.06   & 35.5$\pm$0.6 & 3.0$\pm$1.2 & 0.05$\pm$0.01  & W33\,Main1  \\

25124.9 & $7_{2}-7_{1}E$		    & 5.57$\pm$0.08 & 37.6$\pm$0.1 & 5.6$\pm$0.1 & 0.93$\pm$0.03  & W33\,Main	 \\
        & 	                       &0.51$\pm$0.07   & 38.8$\pm$0.2 & 2.6$\pm$0.4 & 0.18$\pm$0.02   & W33\,A  \\
        & 	                        &0.28$\pm$0.07   &57.1$\pm$0.2 & 1.8$\pm$0.7 & 0.14$\pm$0.02   & W33\,B  \\
        
25294.4 & $8_{2}-8_{1}E$		    & 3.77$\pm$0.05 & 37.9$\pm$0.1 & 5.5$\pm$0.1 & 0.61$\pm$0.03  & W33\,Main \\
        & 	                       & 0.80$\pm$0.07   & 38.9$\pm$0.2 & 3.5$\pm$0.4 & 0.22$\pm$0.03    & W33\,A  \\
        & 	                       & 0.19$\pm$0.06   &56.1$\pm$0.3 & 1.9$\pm$0.6 & 0.09$\pm$0.01    & W33\,B  \\
        
25541.4 & $9_{2}-9_{1}E$           & 2.09$\pm$0.09  & 37.7$\pm$0.1 & 5.4$\pm$0.3 & 0.36$\pm$0.02  & W33\,Main	 \\
        & 	                       &0.62$\pm$0.07   & 39.0$\pm$0.2 & 3.6$\pm$0.4 & 0.16$\pm$0.02  & W33\,A  \\
       
25878.3 & $10_{2}-10_{1}E$	       & 2.04$\pm$0.12 & 37.6$\pm$0.2 & 6.8$\pm$0.6 & 0.27$\pm$0.02  & W33\,Main	 \\
        & 	                       & 0.72$\pm$0.08   &38.1$\pm$0.3 & 4.8$\pm$0.5 & 0.14$\pm$0.04   & W33\,A  \\
        & 	                       &0.33$\pm$0.07   & 55.4$\pm$0.4 & 3.1$\pm$0.7 & 0.09$\pm$0.03   & W33\,B  \\
\end{longtable} 
        
\clearpage

\renewcommand{\tabcolsep}{0.15 cm}
\begin{longtable}{ccccccc}
\caption{Gaussian cyanopolyyne line parameters in W33.}\label{Tab:A.4}                                                     \\
\hline
Transitions    &   Rest frequency       & $\int$$T_{\rm MB}$d$v$  & $V_{\rm LSR}$   & $\Delta v$          & $T_{\rm MB}$      &                       \\
      & (MHz)  & (K\,km\,s$^{-1}$)         & (km\,s$^{-1}$)       & (km\,s$^{-1}$)      & (K)                &  Detected source                 \\
\hline
\endfirsthead
\caption{continued.}                                                                                                             \\
\hline
Transitions    &  Rest frequency     & $\int$$T_{\rm MB}$d$v$  & $V_{\rm LSR}$     & $\Delta v$        & $T_{\rm MB}$    &                       \\
      & (MHz)  & (K\,km\,s$^{-1}$)         & (km\,s$^{-1}$)       & (km\,s$^{-1}$)      & (K)               &  Detected source                \\
\hline
\endhead
\hline                                                                                                                           \\
\endfoot
\hline
\endlastfoot
Cyanopolyynes     &\multicolumn{6}{c}{}                                                                                              \\
\hline                                      
HC$_{3}$N\,(2-1) & 18196.2	   & 0.39$\pm$0.04   & 32.1$\pm$0.2 & 4.0$\pm$0.4 & 0.09$\pm$0.01   & W33\,Main   \\
                 &             & 0.99$\pm$0.04   & 34.2$\pm$0.1 & 4.4$\pm$0.3 & 0.21$\pm$0.02   & W33\,A  \\
                 & 	          & 0.94$\pm$0.04   & 53.2$\pm$0.1 & 4.2$\pm$0.2 & 0.21$\pm$0.01    & W33\,B  \\
                 & 	          & 0.33$\pm$0.03   & 33.7$\pm$0.2 & 4.2$\pm$0.4 & 0.07$\pm$0.01   & W33\,A1  \\
                 & 	          & 0.68$\pm$0.03   & 33.2$\pm$0.1 & 3.6$\pm$0.2 & 0.18$\pm$0.01   & W33\,Main1 \\
        
HC$_{5}$N\,(7-6) & 18638.6	  &0.17$\pm$0.02   &35.7$\pm$0.2 & 2.5$\pm$0.3 & 0.06$\pm$0.01   & W33\,A  \\  
 	        & 	           &0.21$\pm$0.04   &54.3$\pm$0.3 & 3.6$\pm$0.7 & 0.05$\pm$0.01    & W33\,B  \\
                &              &0.07$\pm$0.01 & 34.9$\pm$0.2 & 1.8$\pm$0.3 & 0.04$\pm$0.01  & W33\,Main1   \\
        
HC$_{5}$N\,(8-7) & 21301.3		    &0.32$\pm$0.05   &38.2$\pm$0.3 & 3.7$\pm$0.6 & 0.08$\pm$0.02   & W33\,A	 \\
                 & 	                       &0.21$\pm$0.04   &56.7$\pm$0.2 & 1.8$\pm$0.4 & 0.11$\pm$0.01  & W33\,B \\
                 
HC$_{7}$N\,(19-18)        & 21431.9		    & 0.16$\pm$0.03 & 56.1$\pm$0.2 & 1.7$\pm$0.4 & 0.09$\pm$0.01 & W33\,B	 \\

HC$_{5}$N\,(9-8) & 23963.9		&0.57$\pm$0.14   &54.8$\pm$0.5 & 4.4$\pm$1.5 & 0.12$\pm$0.05    & W33\,B  \\              
                 & 	            & 0.51$\pm$0.07 & 35.2$\pm$0.2 & 3.5$\pm$0.6 & 0.14$\pm$0.02 & W33\,Main1 \\    

\end{longtable}

\renewcommand{\tabcolsep}{0.15 cm}
\begin{longtable}{ccccccc}
\caption{Gaussian SiS, c-C$_{3}$H$_{2}$, CH$_{3}$CN, CH$_{2}$CHCN, NH$_{2}$D, CH$_{3}$CHO, HNCO, H$_{2}$O, CCS line parameters.}
\label{Tab:A.5}                                                     \\
\hline
Transitions    &   Rest frequency       & $\int$$T_{\rm MB}$d$v$  & $V_{\rm LSR}$   & $\Delta v$          & $T_{\rm MB}$      &                       \\
      & (MHz)  & (K\,km\,s$^{-1}$)         & (km\,s$^{-1}$)       & (km\,s$^{-1}$)      & (K)                &  Detected source                 \\
\hline
\endfirsthead
\caption{continued.}                                                                                                             \\
\hline
Transitions    &  Rest frequency     & $\int$$T_{\rm MB}$d$v$  & $V_{\rm LSR}$     & $\Delta v$        & $T_{\rm MB}$    &                       \\
      & (MHz)  & (K\,km\,s$^{-1}$)         & (km\,s$^{-1}$)       & (km\,s$^{-1}$)      & (K)               &  Detected source                \\
\hline
\endhead
\hline                                                                                                                           \\
\endfoot
\hline
\endlastfoot
SiS(1-0)              & 18154.9	   & 0.11$\pm$0.02  & 37.9$\pm$0.7 & 6.3$\pm$1.8 & 0.02$\pm$0.01   & W33\,Main1   \\
c-C$_{3}$H$_{2}$\,(1$_{1,}$$_{0}$ -1$_{0,}$$_{1}$) & 18343.1	   & 2.89$\pm$0.08  & 34.9$\pm$0.1 & 3.0$\pm$0.1 & 0.89$\pm$0.10   & W33\,Main   \\  
                 &             & 1.21$\pm$0.04  & 35.1$\pm$0.1 & 5.2$\pm$0.2 & 0.22$\pm$0.01   & W33\,A  \\
                 & 	          & 0.96$\pm$0.06   & 54.1$\pm$0.2 & 7.5$\pm$0.6 & 0.12$\pm$0.01    & W33\,B  \\
                 & 	          & 0.94$\pm$0.04   & 34.9$\pm$0.1 & 4.7$\pm$0.2 & 0.19$\pm$0.01   & W33\,A1  \\
                 & 	          & 0.54$\pm$0.03   & 33.9$\pm$0.1 & 4.2$\pm$0.3 & 0.12$\pm$0.01    & W33\,B1  \\  
                 & 	          & 1.56$\pm$0.03   & 34.8$\pm$0.1 & 4.4$\pm$0.1 & 0.34$\pm$0.02   & W33\,Main1 \\
        
CH$_{3}$CN\,(1$_{0}$--0$_{0}$)    & 18398.0	    & 0.29$\pm$0.03 & 32.1$\pm$0.2 & 4.9$\pm$0.5 & 0.05$\pm$0.01  & W33\,Main   \\    
                 & 	           &0.28$\pm$0.04   &35.2$\pm$0.4 & 4.9$\pm$0.6 & 0.05$\pm$0.01   & W33\,A  \\
                 & 	           &0.29$\pm$0.04   &54.6$\pm$0.4 & 5.8$\pm$0.9 & 0.05$\pm$0.01    & W33\,B  \\
                 & 	            & 0.18$\pm$0.02  & 35.4$\pm$0.3 & 4.4$\pm$0.6 & 0.04$\pm$0.01  & W33\,Main1   \\
        
NH$_{2}$D\,(3$_{1,3}$s--3$_{0,3}$a)        & 18807.9		   &0.11$\pm$0.02 & 31.3$\pm$0.2 & 2.4$\pm$0.4 & 0.04$\pm$0.01  & W33\,B1	 \\
                 
c-C$_{3}$H$_{2}$\,(2$_{2,}$$_{0}$ - 2$_{2,}$$_{1}$)       & 21587.4		    & -2.58$\pm$0.04 & 33.5$\pm$0.1 & 2.5$\pm$0.1 & -0.97$\pm$0.15  & W33\,Main \\

HNCO\,(1$_{0,}$$_{1}$ -0$_{0,}$$_{0}$)       & 21981.6	    & 0.36$\pm$0.06 & 39.3$\pm$0.7 & 9.1$\pm$1.8 & 0.04$\pm$0.01  & W33\,Main \\
                                             &              & 0.49$\pm$0.10 & 37.5$\pm$0.6 & 6.3$\pm$1.6 & 0.07$\pm$0.02  & W33\,A \\
                                             &             & 1.12$\pm$0.13 & 56.4$\pm$0.4 & 6.7$\pm$0.9 & 0.16$\pm$0.03  & W33\,B \\                                                            
                                                            
H$_{2}$O\,(6$_{1,}$$_{6}$ -5$_{2,}$$_{3}$)        & 22235.1		    & 27.81$\pm$0.35 & 36.5$\pm$0.1 & 4.8$\pm$0.1 & 5.37$\pm$0.96  & W33\,Main \\
                                                  & 	           &65.29$\pm$1.25   &37.6$\pm$0.1 & 1.3$\pm$0.1 & 46.40$\pm$1.06  & W33\,A  \\
                                                  &     & 469.5$\pm$8.99 & 62.8$\pm$0.1 & 1.8$\pm$0.1 & 248.00$\pm$16.6  & W33\,B \\

CCS      & 22344.0	   &0.73$\pm$0.10   &35.9$\pm$0.2 & 2.9$\pm$0.4 & 0.23$\pm$0.01   & W33\,A1  \\                  
         & 	       & 0.59$\pm$0.07 & 34.1$\pm$0.1 & 2.4$\pm$0.4 & 0.23$\pm$0.01  & W33\,Main1 \\    

\end{longtable} 
           
\clearpage 

\begin{figure*}[t]
\centering
\includegraphics[width=0.79\textwidth]{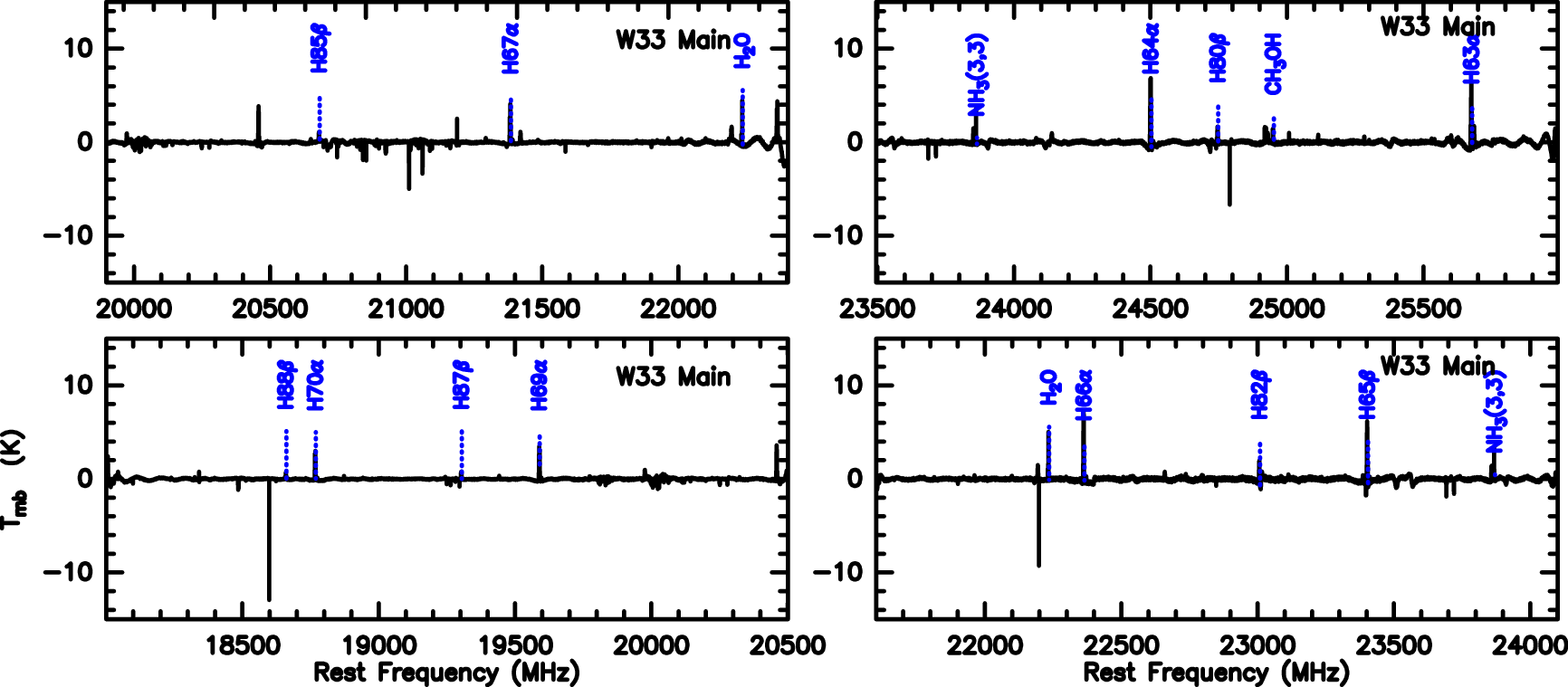}
\caption{Overview of the 1.3\,cm line survey toward W33\,Main with strong lines marked. The displayed frequency scale is based on the LSR velocity with 0\,km\,s$^{-1}$. The absorption features are due to bad channels.}
\label{FigA.1}
\end{figure*}

\begin{figure*}[h]
\centering
\includegraphics[width=0.79\textwidth]{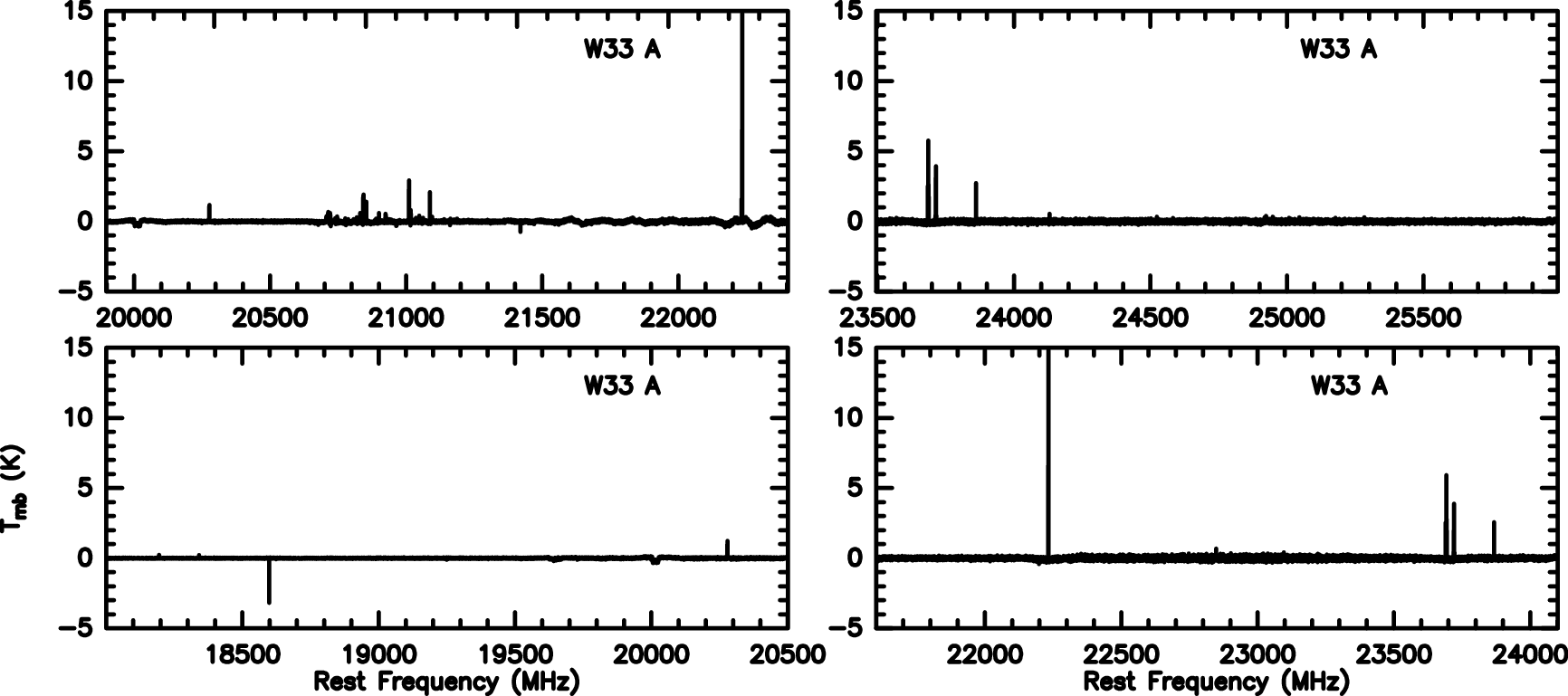}
\caption{Overview of the 1.3\,cm line survey toward W33\,A with strong lines marked. For more details, see Fig.\,\ref{FigA.1}.}
\label{FigA.2}
\end{figure*}

\begin{figure*}[h]
\centering
\includegraphics[width=0.79\textwidth]{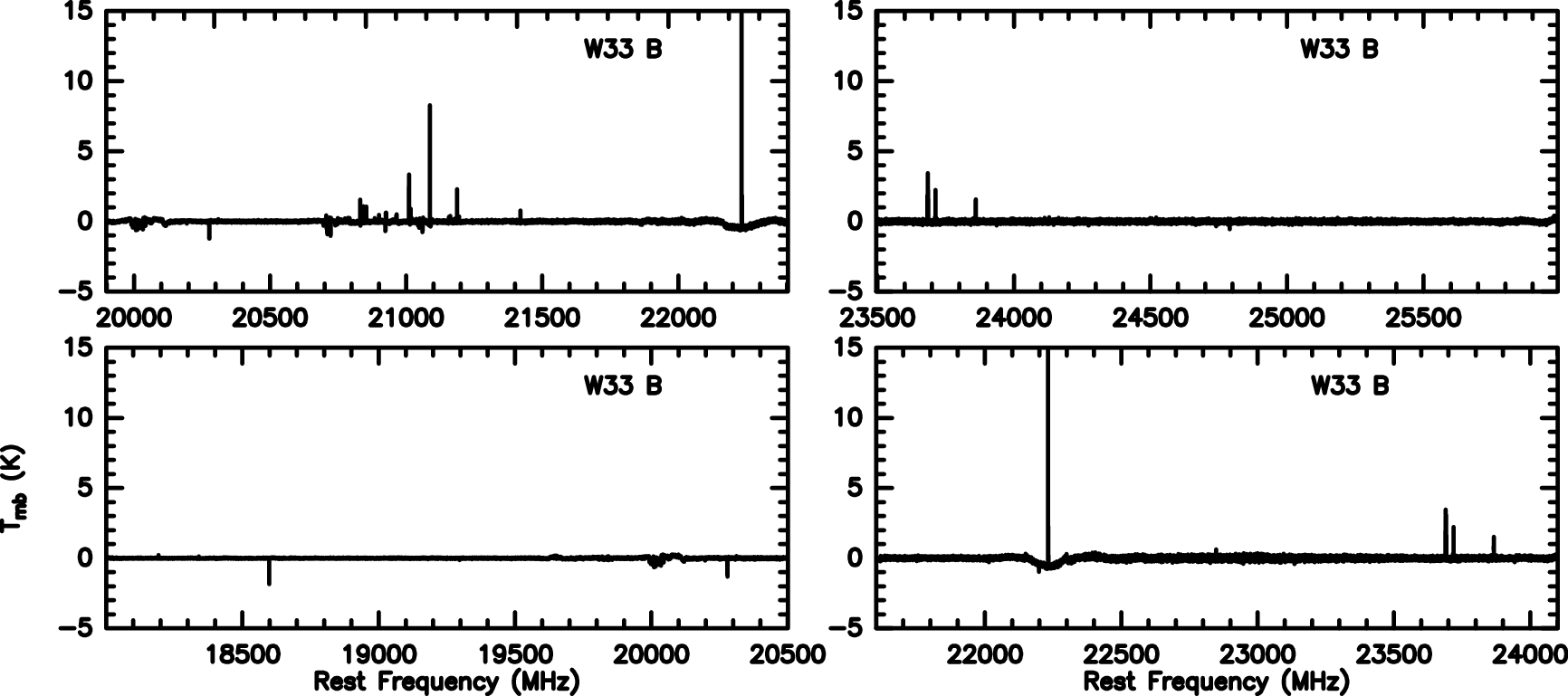}
\caption{Overview of the 1.3\,cm line survey toward W33\,B with strong lines marked. For more details, see Fig.\,\ref{FigA.1}.}
\label{FigA.3}
\end{figure*}

\begin{figure*}[t]
\centering
\includegraphics[width=0.80\textwidth]{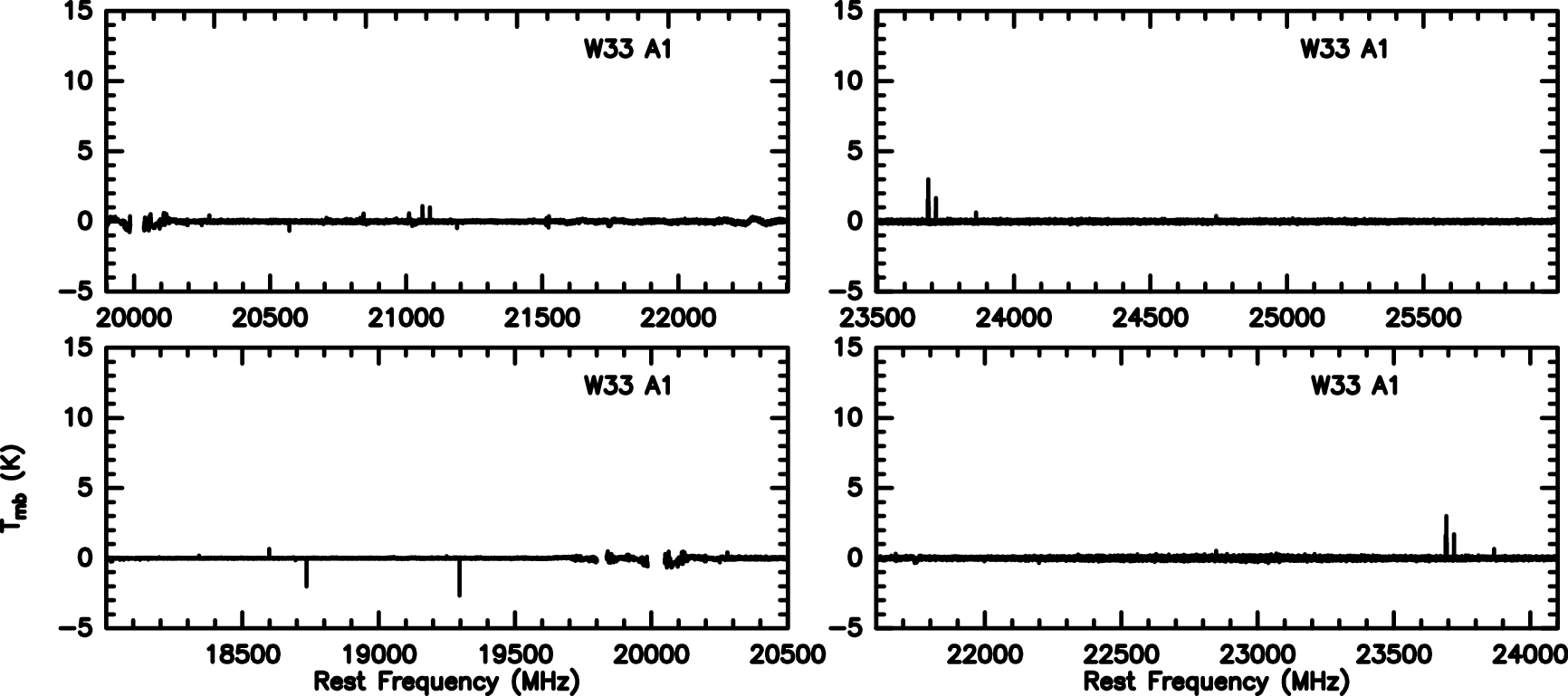}
\caption{Overview of the 1.3\,cm line survey toward W33\,A1 with strong lines marked. For more details, see Fig.\,\ref{FigA.1}. The frequencies near 20.0\,GHz are interference signals.}
\label{FigA.4}
\end{figure*}

\begin{figure*}[h]
\centering
\includegraphics[width=0.80\textwidth]{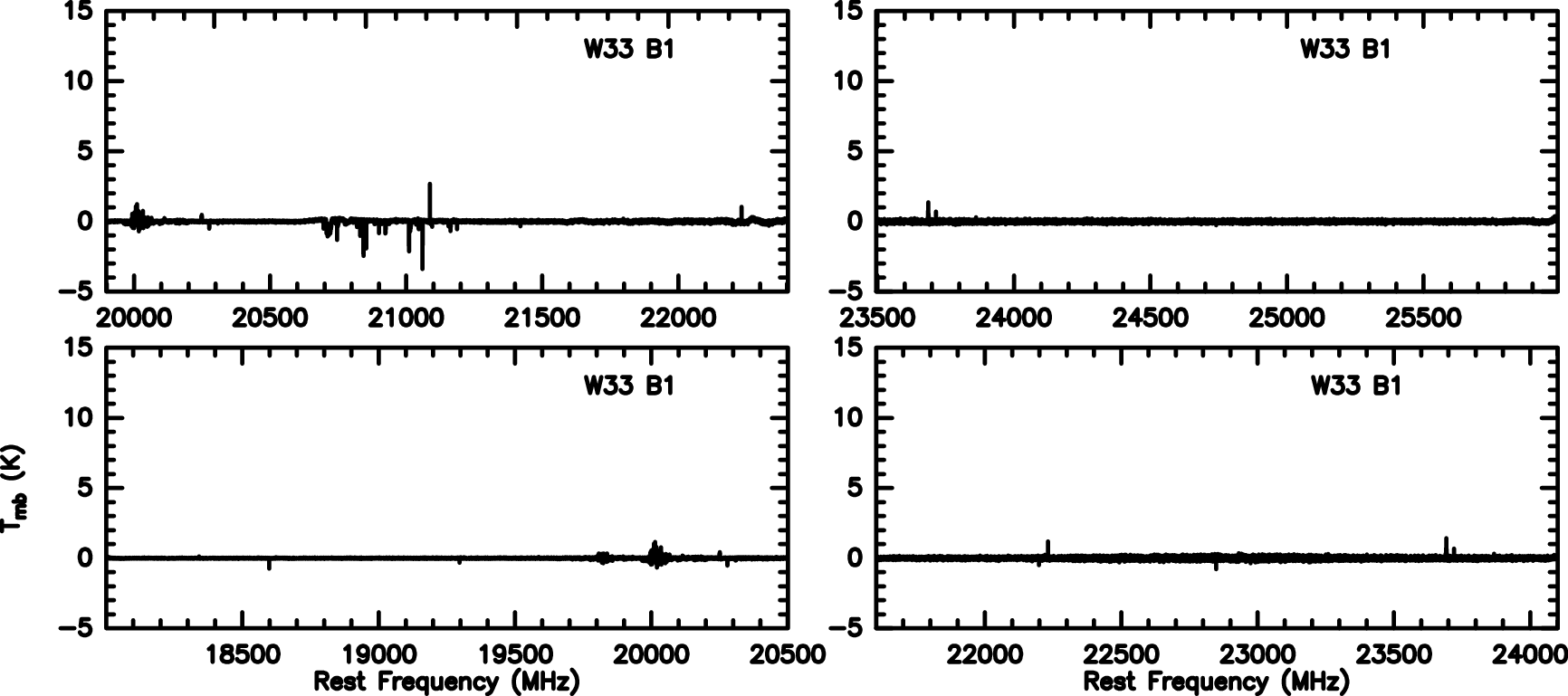}
\caption{Overview of the 1.3\,cm line survey toward W33\,B1 with strong lines marked. For more details, see Figs.\,\ref{FigA.1} and \ref{FigA.4}.}
\label{FigA.5}
\end{figure*}

\begin{figure*}[h]
\centering
\includegraphics[width=0.80\textwidth]{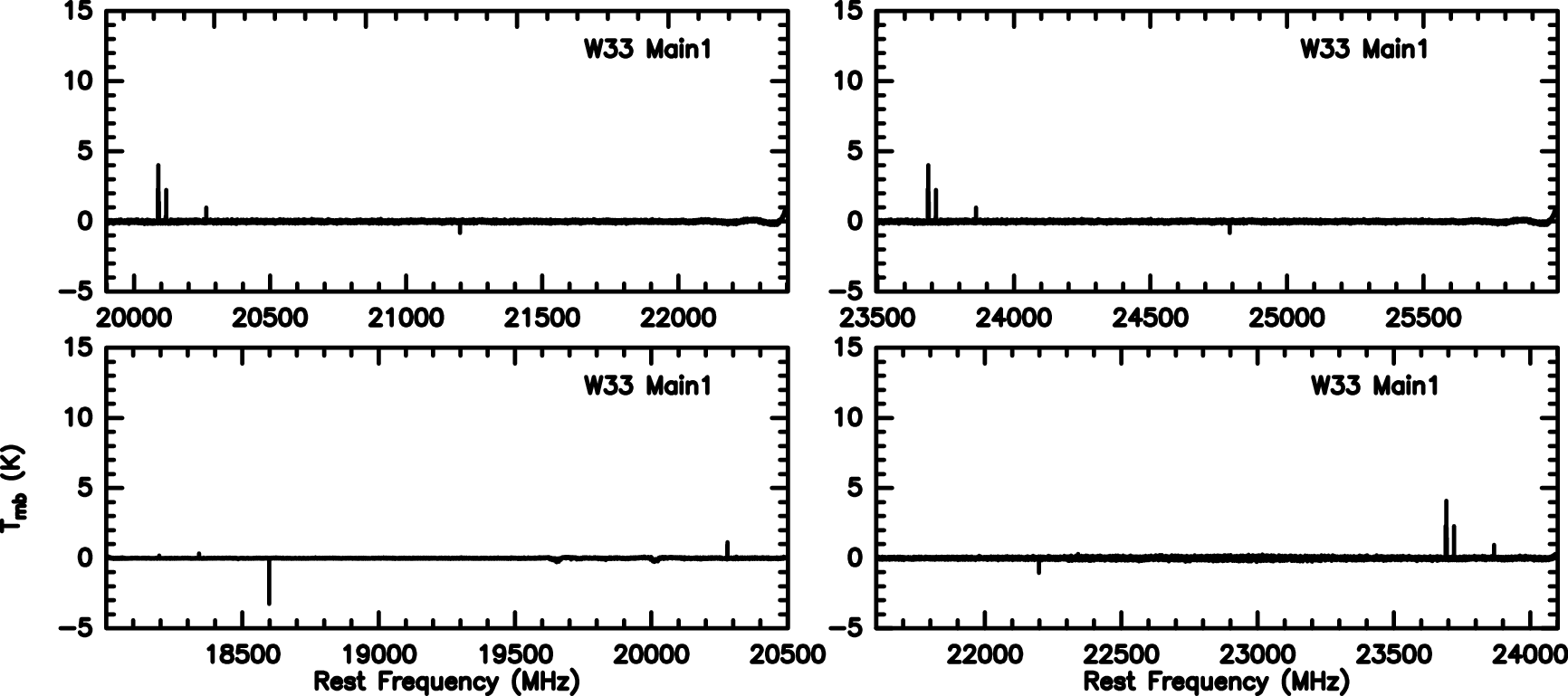}
\caption{Overview of the 1.3\,cm line survey toward W33\,Main1 with strong lines marked. For more details, see Fig.\,\ref{FigA.1}.}
\label{FigA.6}
\end{figure*}

\clearpage

\begin{figure*}[h]
\centering
\includegraphics[width=0.45\textwidth]{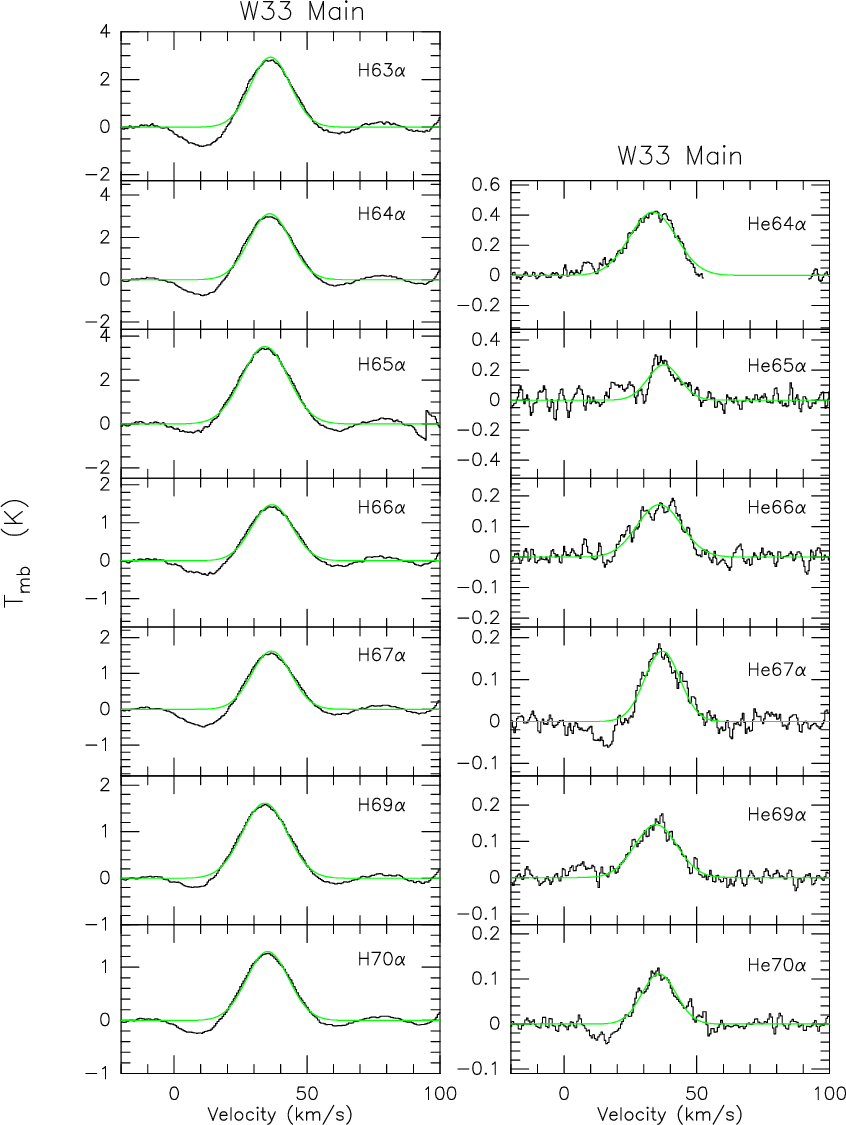}
\includegraphics[width=0.45\textwidth]{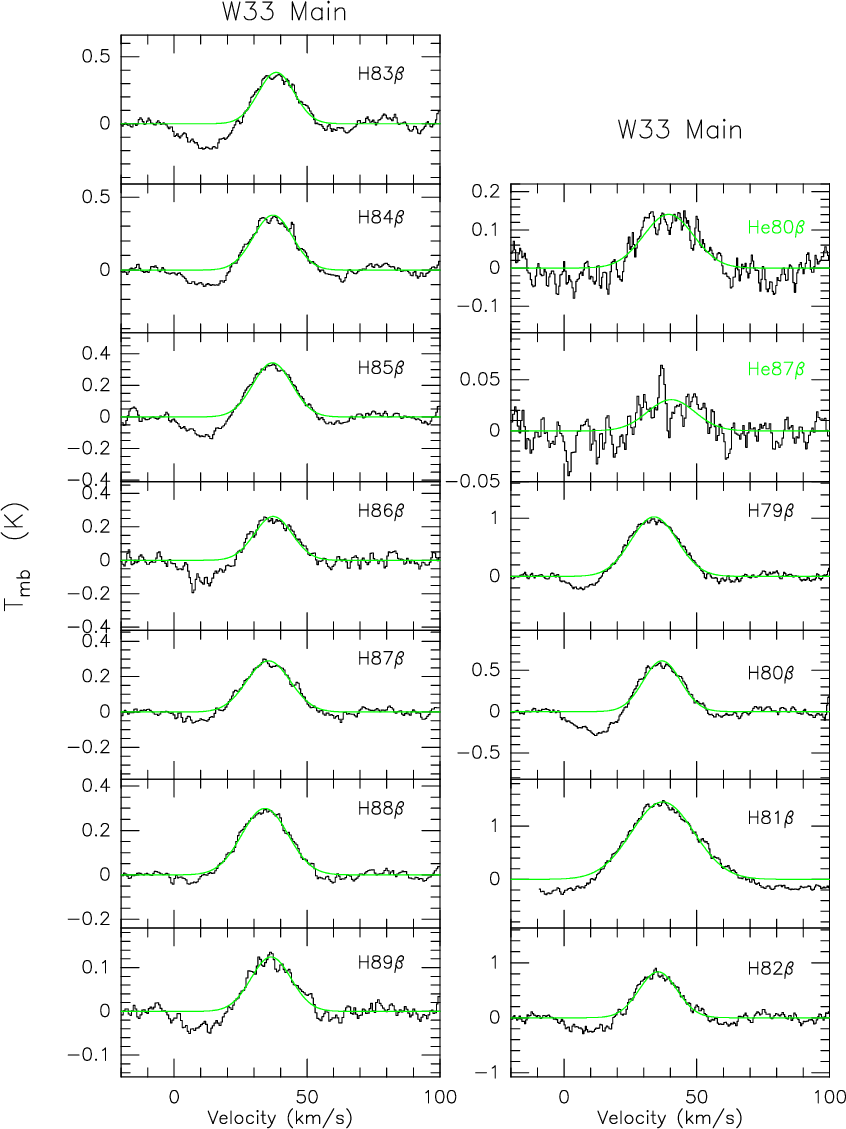}
\caption{Observed H$\alpha$ and He$\alpha$ transitions at offset ($0\arcsec, 0\arcsec$) with respect to the reference position for W33\,Main given in Table\,\ref{table:1} (\textit{left panels}). Similarly, the observed H$\beta$ and He$\beta$ transitions at offset ($0\arcsec, 0\arcsec$) with respect to the same reference position (\textit{right panels}). In each panel, the black solid line represents the observed spectrum, the green solid line indicates the one-component Gaussian fit. The velocity scale is LSR, here and elsewhere.}
\label{FigA.7}
\end{figure*}

\begin{figure*}[h]
\centering
\includegraphics[width=0.45\textwidth]{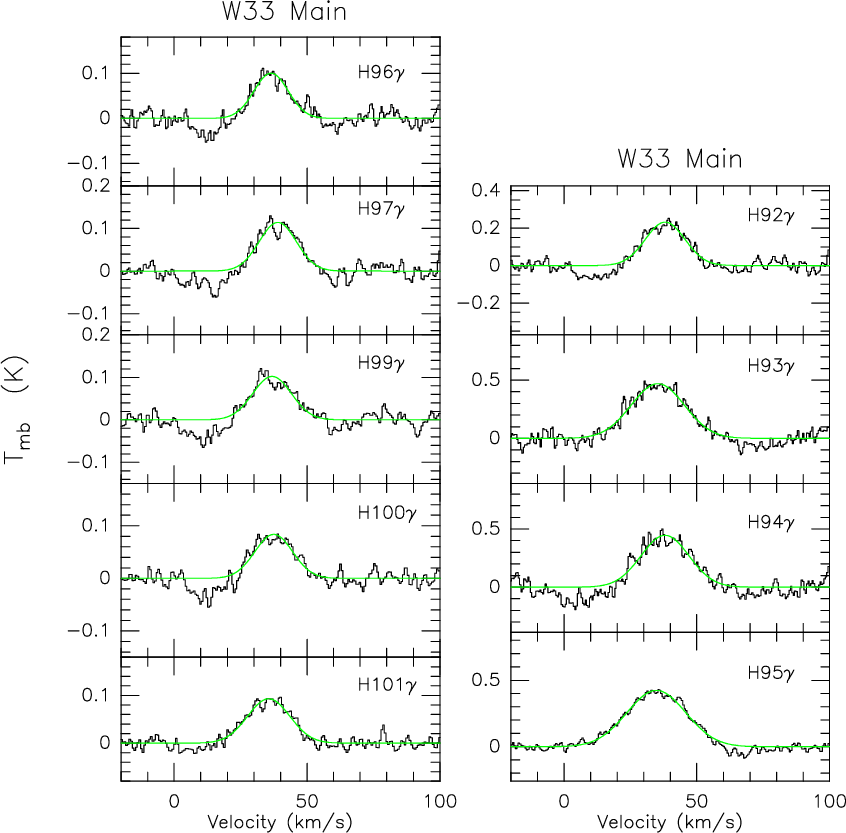}
\includegraphics[width=0.45\textwidth]{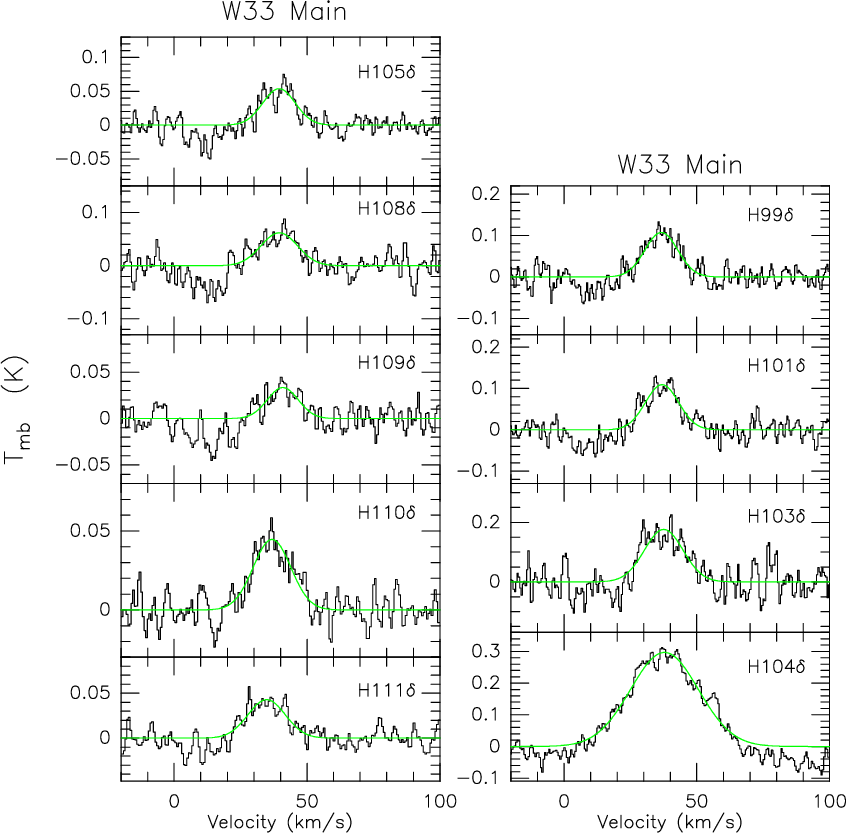}
\caption{Observed H$\gamma$ and He$\gamma$ transitions at offset ($0\arcsec, 0\arcsec$) with respect to the reference position for W33\,Main given in Table\,\ref{table:1} (\textit{left panels}). Similarly, the observed H$\delta$ and He$\delta$ transitions at offset ($0\arcsec, 0\arcsec$) with respect to the same reference position (\textit{right panels}). In each panel, the black solid line represents the observed spectrum, the green solid line indicates the one-component Gaussian fit.}
\label{FigA.8}
\end{figure*}

\begin{figure*}[h]
\centering
\includegraphics[width=1\textwidth]{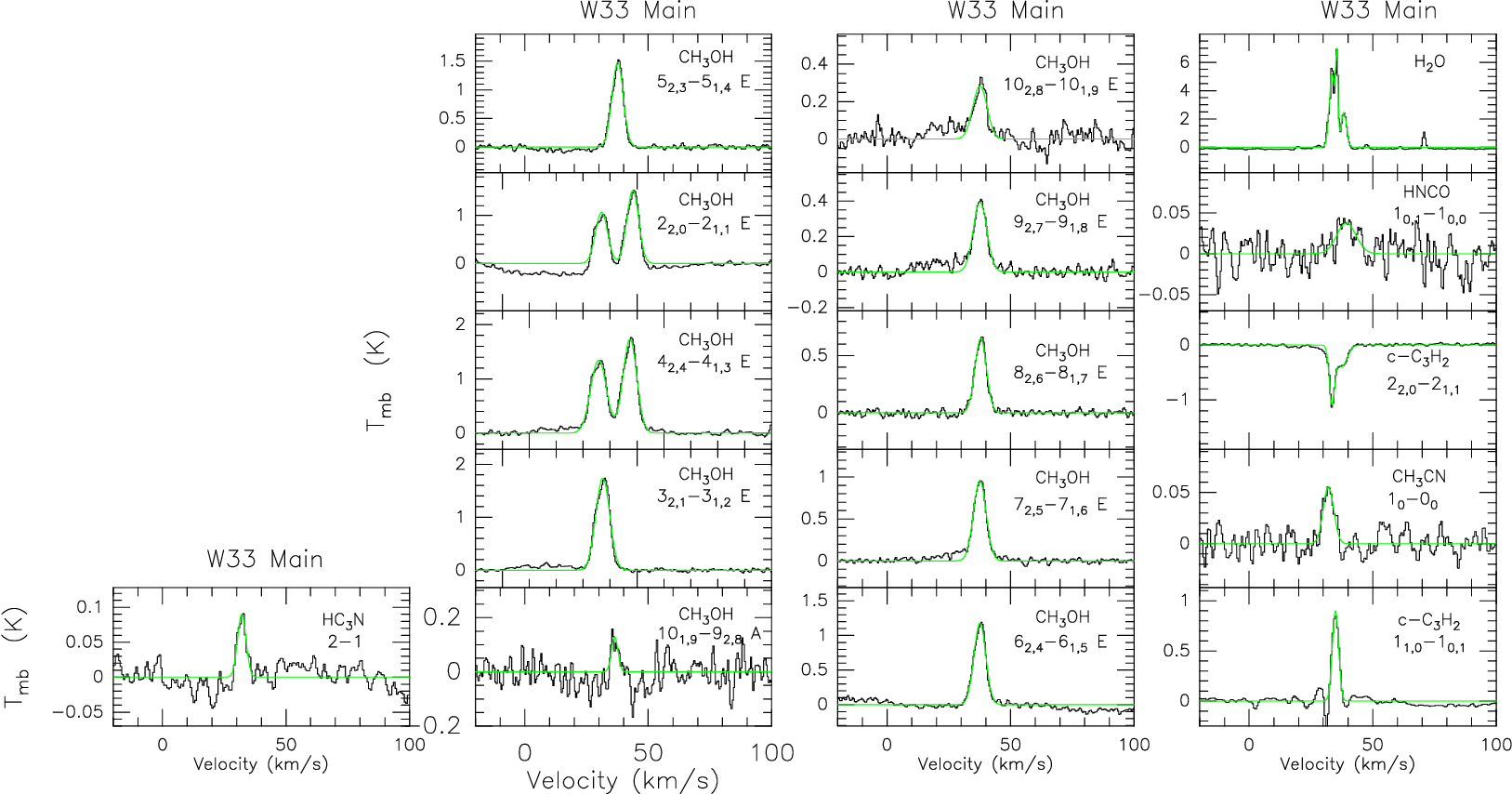}
\caption{Observed HC$_3$N, CH$_3$OH, c-C$_3$H$_2$, CH$_3$CN, HNCO, and H$_2$O transitions (black lines) with one or two-component Gaussian fits (green lines) in W33\,Main. Species and quantum numbers are given in the upper right of each panel.}
\label{FigA.9}
\end{figure*}

\begin{figure*}[h]
\centering
\includegraphics[width=1\textwidth]{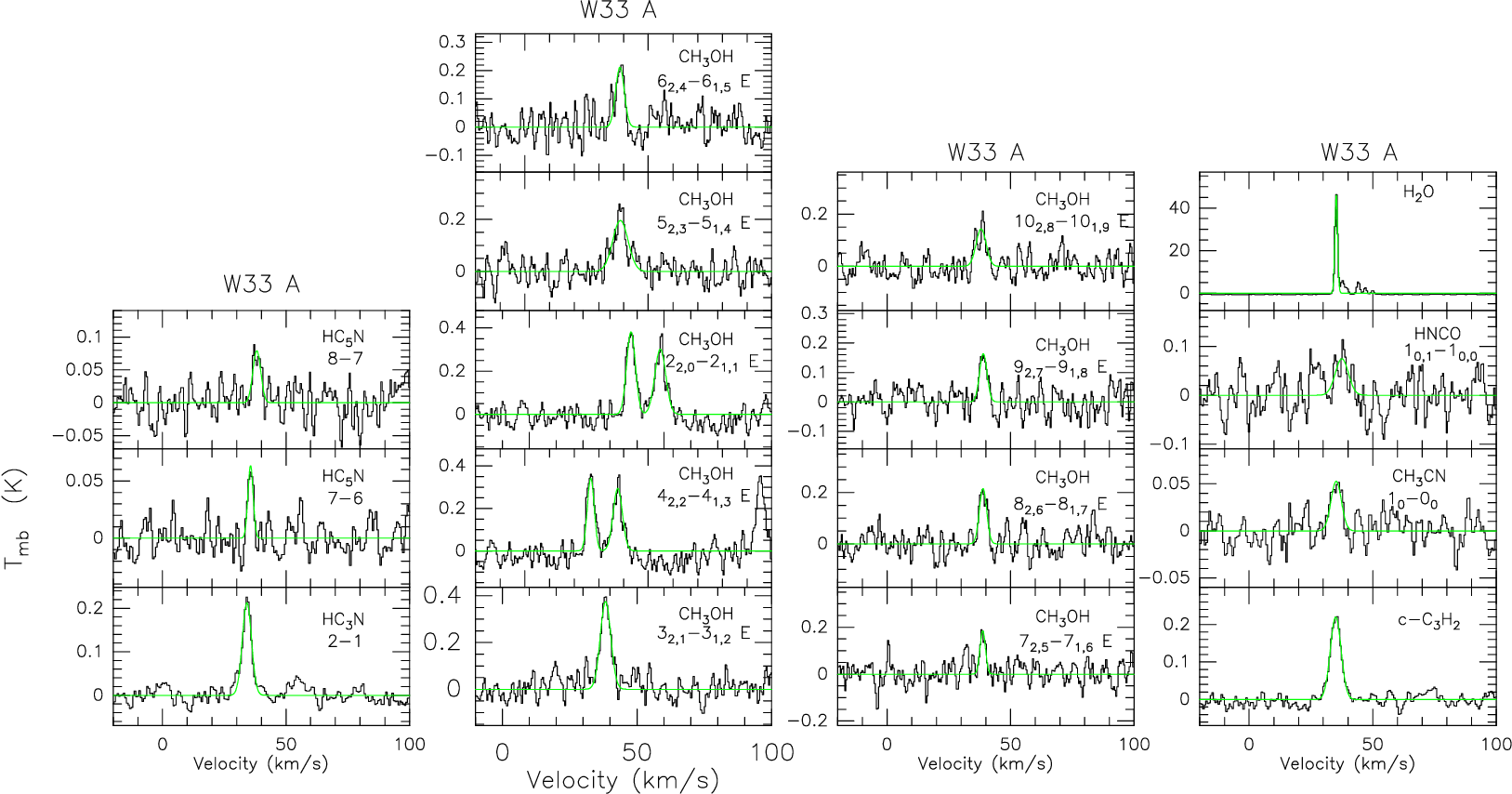}
\caption{Observed HC$_n$N, CH$_3$OH, c-C$_3$H$_2$, CH$_3$CN, HNCO, and H$_2$O transitions (black lines) with one or two-component Gaussian fits (green lines) in W33\,A. Species and quantum numbers are given in the upper right of each panel.}
\label{FigA.10}
\end{figure*}

\begin{figure*}[h]
\centering
\includegraphics[width=1\textwidth]{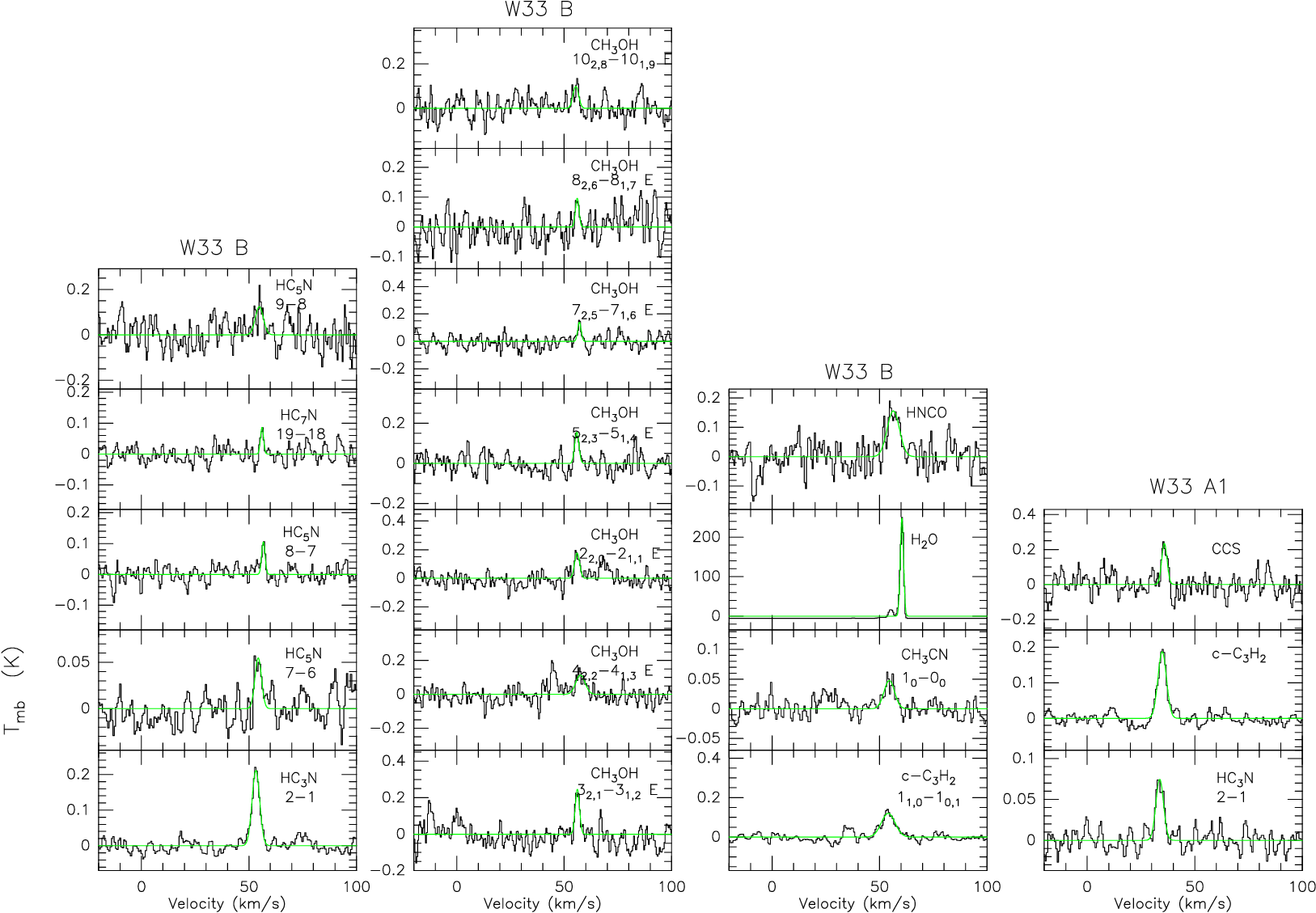}
\caption{Observed HC$_n$N, CH$_3$OH, c-C$_3$H$_2$, CH$_3$CN, H$_2$O, HNCO, and CCS transitions (black lines) with one-component Gaussian fits (green lines) in W33\,B and W33\,A1. Species and quantum numbers are given in the upper right of each panel.}
\label{FigA.11}
\end{figure*}

\begin{figure*}[h]
\centering
\includegraphics[width=1\textwidth]{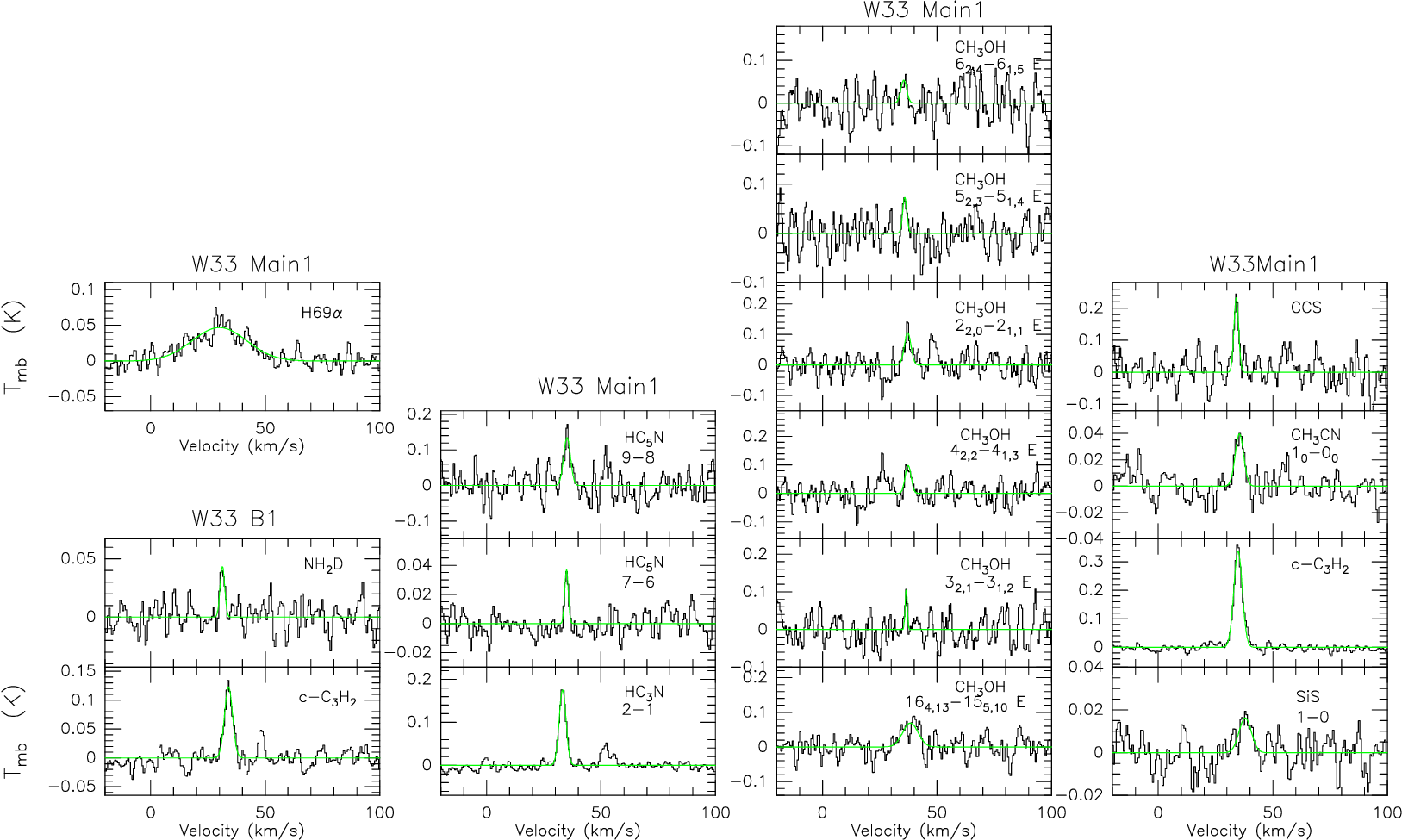}
\caption{Observed c-C$_3$H$_2$, NH$_2$D, H$\alpha$, HC$_n$N, CH$_3$OH, SiS, CH$_3$CN, and CCS transitions (black lines) with one-component Gaussian fits (green lines) in W33\,B1 and W33\,Main1. Species and quantum numbers are given in the upper right of each panel.}
\label{FigA.12}
\end{figure*}

\clearpage

\onecolumn
\section{Integrated intensity maps of NH$_3$ and c-C$_{3}$H$_{2}$}
\label{Appendix B}

\begin{figure*}[h]
\centerline{\hbox{
\includegraphics[width=6.5cm,height=6.2cm,angle=0]{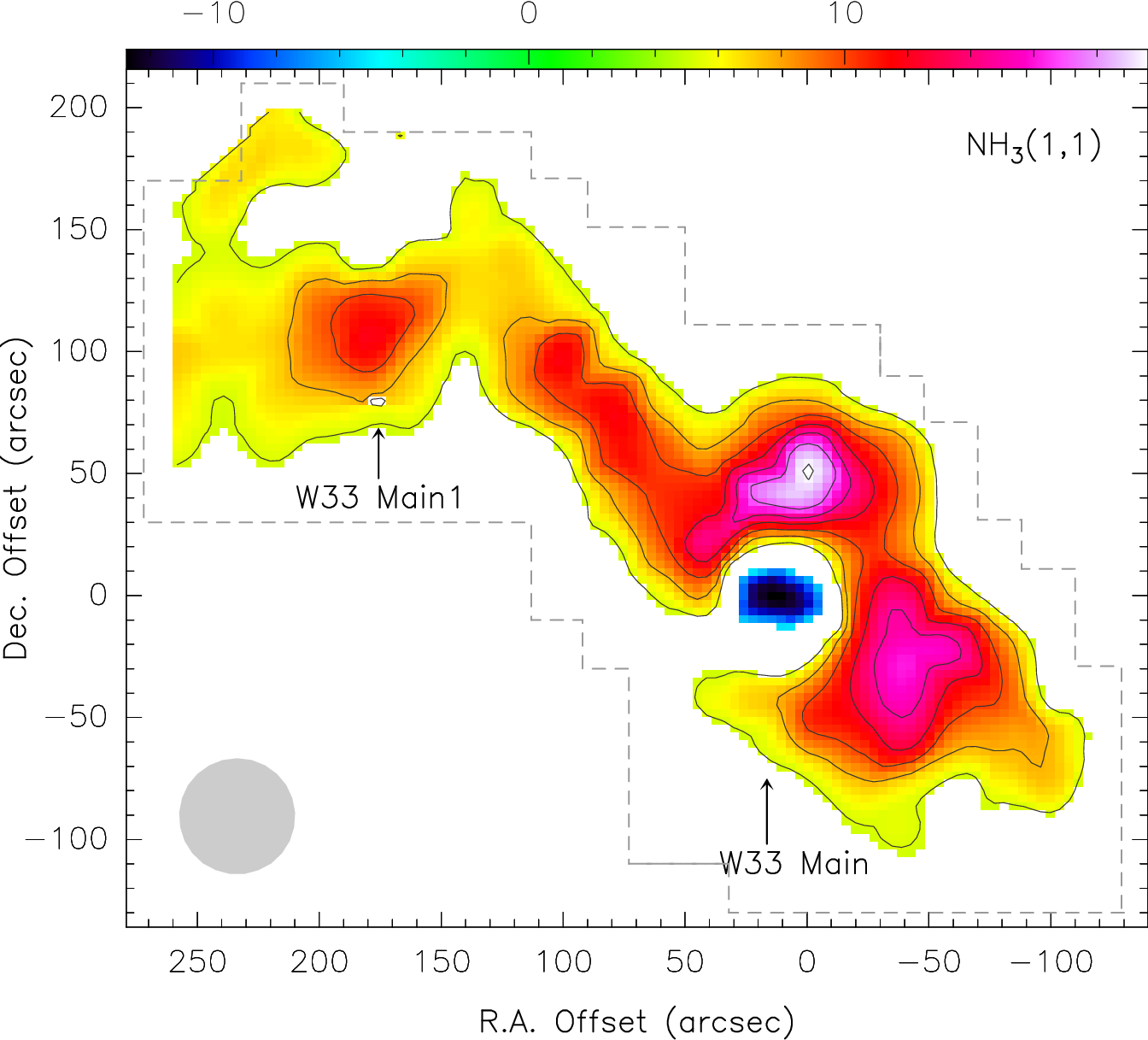}
\hspace{0.08cm}
\includegraphics[width=6.3cm,height=6.2cm,angle=0]{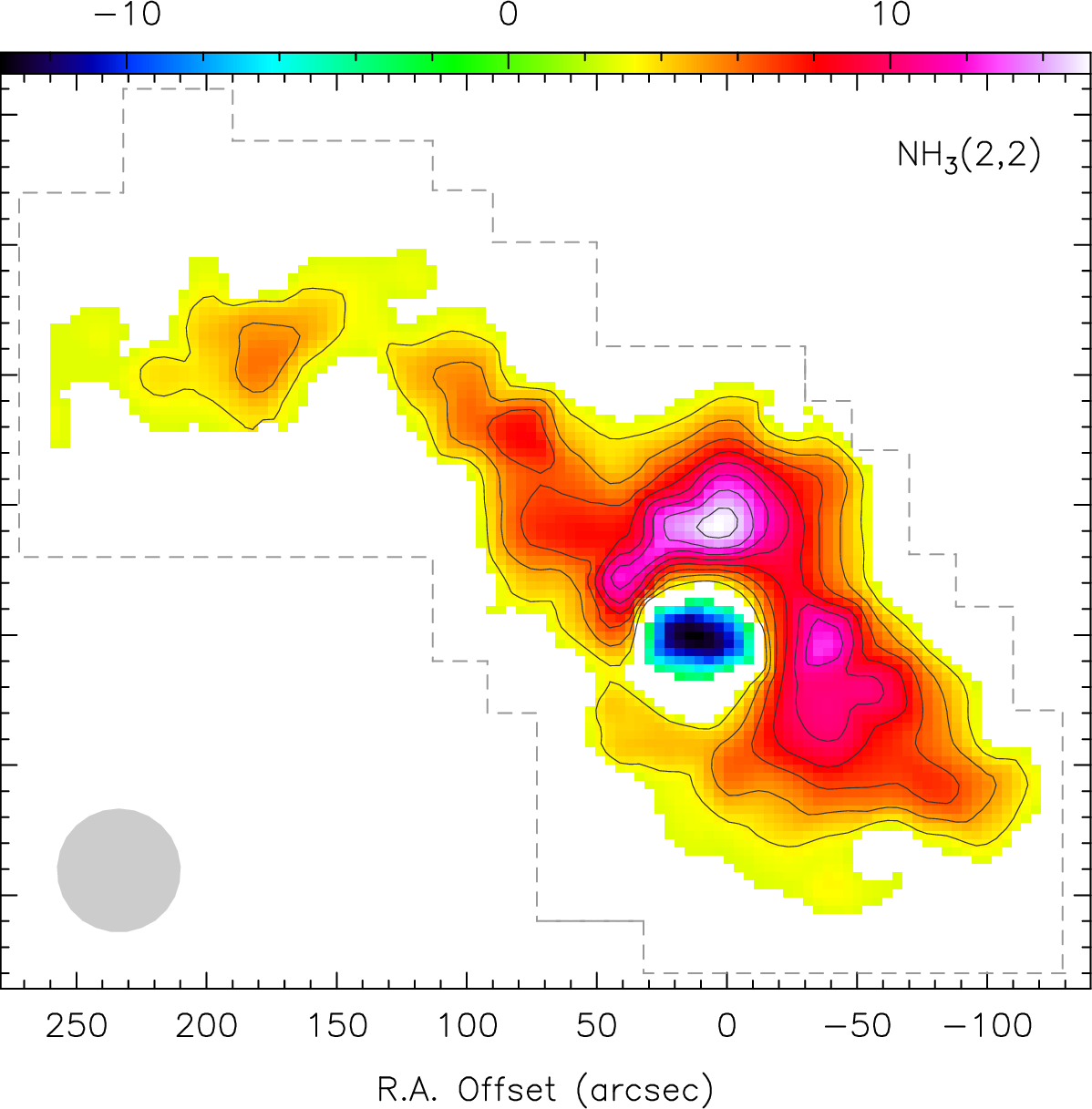}
\hspace{0.08cm}
\includegraphics[width=6.3cm,height=6.2cm,angle=0]{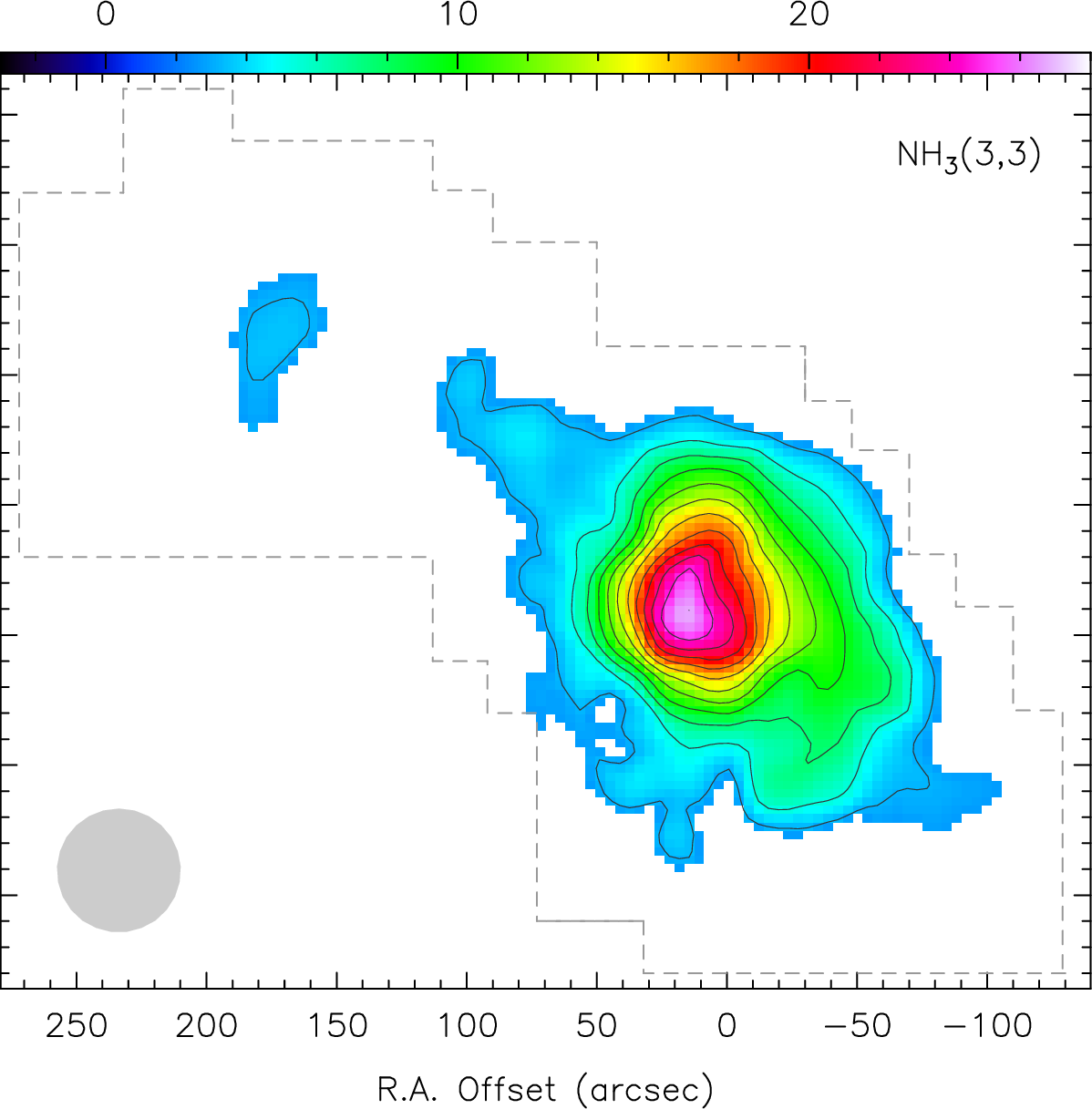}}}
\caption[]{Same as Figure\,\ref{Fig:4} but integrated intensities are presented for the NH$_3$\,(1,1) (\textit{left}), NH$_3$\,(2,2) (\textit{middle}), and NH$_3$\,(3,3) (\textit{right}) lines (adapted from \citealt{2022A&A...658A..34T}).}
\label{FigB.1}
\end{figure*}

\begin{figure}[h]
\centering
\includegraphics[width=0.45\textwidth]{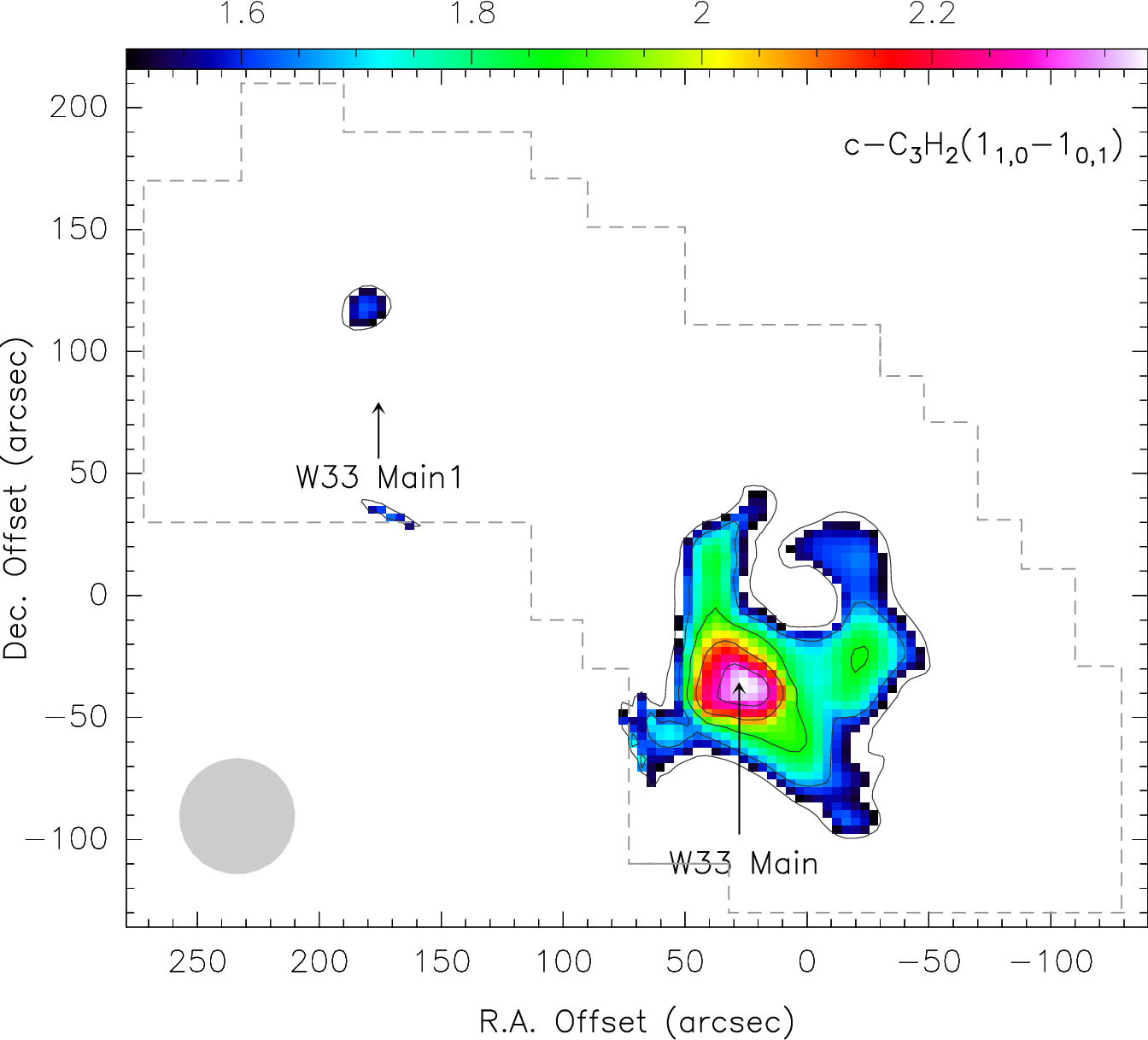}
\includegraphics[width=0.45\textwidth]{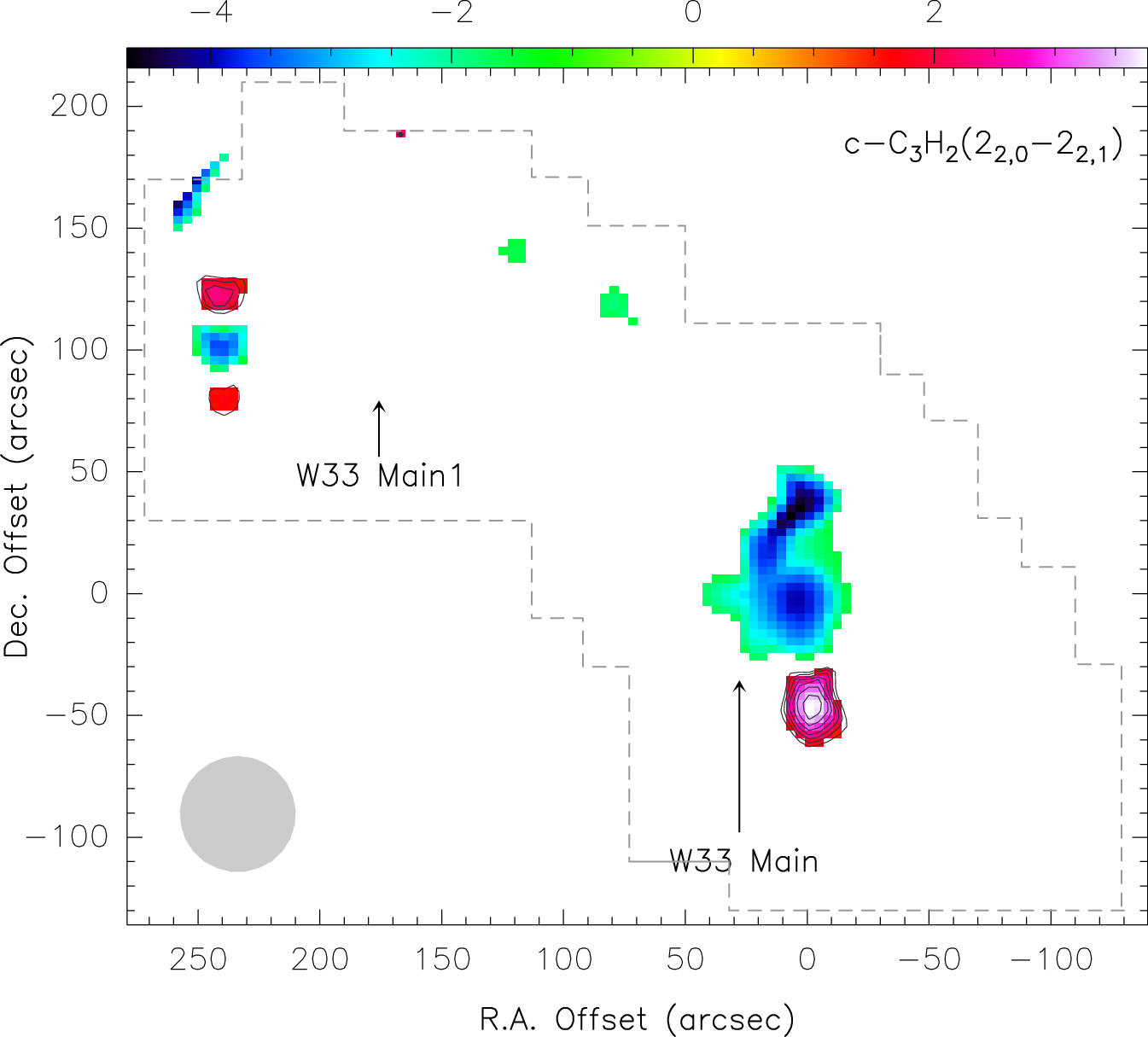}
\caption[]{Same as Figure\,\ref{Fig:4} but integrated intensities are presented for the c-C$_{3}$H$_{2}$\,(1$_{1,}$$_{0}$ -1$_{0,}$$_{1}$) (\textit{left}), and c-C$_{3}$H$_{2}$\,(2$_{2,}$$_{0}$ - 2$_{2,}$$_{1}$) (\textit{right}) lines.}
\label{FigB.2}
\end{figure}

\end{appendix}
\end{document}